\documentclass[pdflatex,sn-mathphys-num]{sn-jnl}

\usepackage{adjustbox} 
\usepackage{graphicx}%
\usepackage{multirow}%
\usepackage{amsmath,amssymb,amsfonts}%
\usepackage{amsthm}%
\usepackage{mathrsfs}%
\usepackage[title]{appendix}%
\usepackage{xcolor}%
\usepackage{textcomp}%
\usepackage{manyfoot}%
\usepackage{booktabs}%
\usepackage{algorithm}%
\usepackage{algorithmicx}%
\usepackage{algpseudocode}%
\usepackage{listings}%
\usepackage{makecell}
\usepackage{subcaption}
\usepackage{xcolor}
\usepackage{graphicx}
\usepackage{booktabs}
\usepackage{array}
\usepackage{tabularx}
\usepackage{adjustbox}
\usepackage{booktabs}
\usepackage{multirow}
\usepackage{flafter}
\usepackage{placeins}
\usepackage{float}
\usepackage{booktabs}
\usepackage{capt-of}

\theoremstyle{thmstyleone}%
\theoremstyle{thmstyletwo}%

\theoremstyle{thmstylethree}%

\begin{document}
\title[Article Title]{Feature‑-Adaptive Fusion in Hybrid Quantum--Classical Neural Networks for Robust Biomedical Image Classification}


\author[1]{\sur{Yan-Yan Hou}}\email{hyy@uzz.edu.cn}
\author[2]{\sur{Jian Li}}\email{lijian@bupt.edu.cn}
\author[3]{\sur{Chongqiang Ye}}\email{chongqiangye@hzcu.edu.cn}
\author[4]{\sur{Hengji Li}}\email{lihengji@henu.edu.cn}
\author[5]{\sur{Zhuo Wang}}\email{ZhuoWang@bupt.edu.cn}
\author*[6]{\sur{Qinghui Liu}}\email{qinghuiliu810716@126.com}
\affil[1]{\orgdiv{College of Information Science and Engineering}, \orgname{Zaozhuang University}, \orgaddress{\street{No. 1 Beian Road}, \city{Zaozhuang}, \postcode{277160}, \state{Shandong}, \country{China}}}
\affil[2]{\orgdiv{School of Cyberspace Security}, \orgname{Beijing University of Posts and Telecommunications}, \orgaddress{ \city{Beijing}, \postcode{100876}, \country{China}}}
\affil[3]{\orgname{Hangzhou City University}, \orgaddress{\city{Hangzhou}, \postcode{310015}, \country{China}}}
\affil[4]{\orgdiv{School of Artificial Intelligence}, \orgname{Henan University}, \orgaddress{\street{85 Nanyang Road}, \city{Zhengzhou}, \postcode{450000}, \state{Henan}, \country{China}}}
\affil[5]{\orgname{China Information Technology Security Evaluation Center}, \orgaddress{ \city{Beijing}, \postcode{100085}, \country{China}}}
\affil[6]{\orgdiv{Network Center}, \orgname{Zaozhuang University}, \orgaddress{\street{No. 1 Beian Road}, \city{Zaozhuang}, \postcode{277160}, \state{Shandong}, \country{China}}}

\abstract{Hybrid quantum--classical neural networks offer a promising framework for integrating quantum circuits into machine learning in the noisy intermediate-scale quantum era. However, fixed or globally shared fusion strategies used in many existing hybrid models may limit the effective integration of complementary classical and quantum information. To address this limitation, we propose a Feature-Adaptive Fusion Hybrid Quantum--Classical Neural Network (FAF-HQNN) for biomedical image classification. The proposed model combines a classical deep feature encoder with a variational quantum circuit (VQC) and employs a feature-adaptive fusion mechanism to dynamically weight classical and quantum predictions across samples and classes. Under the evaluated settings, FAF-HQNN achieves competitive accuracy on PathMNIST and BloodMNIST compared to internal variants and external hybrid quantum--classical baselines. On PathMNIST, FAF-HQNN achieves higher aggregate mean accuracy than all evaluated baselines under both Gaussian and Poisson noise. These results suggest that feature-adaptive fusion may help improve classification performance within the evaluated framework.}



\maketitle

\section{Introduction}\label{sec1}

Hybrid quantum-classical neural networks have emerged as a practical paradigm for quantum machine learning in the noisy intermediate-scale quantum (NISQ) era, where current quantum devices remain constrained by limited qubit counts, shallow circuit depth, and imperfect quantum operations \cite{alchieri2021introduction}. Under these hardware limitations, purely quantum models are difficult to scale to high-dimensional real-world tasks. Hybrid architectures provide a feasible alternative by combining the strong representation learning capabilities of classical neural networks with trainable quantum components, typically implemented as variational quantum circuits (VQCs), within end-to-end learning frameworks \cite{benedetti2019parameterized}.

Biomedical image classification plays an important role in computer-aided diagnosis and medical data analysis, supporting tasks such as tissue recognition, blood cell identification, and disease screening. Deep learning methods, especially convolutional neural networks (CNNs), have achieved strong performance in these applications by learning hierarchical visual representations directly from biomedical images \cite{litjens2017survey,esteva2017dermatologist,shen2017deep}. In recent years, there has been growing interest in hybrid quantum-classical CNN models for image analysis and biomedical data processing \cite{ajlouni2023medical,hdaib2024quantum,gujju2024quantum,chow2024quantum}. However, most studies have focused primarily on feasibility and clean-data performance \cite{huang2023image,liu2021hybrid,wei2023quantum}.

In practice, biomedical images are frequently affected by common image corruptions, particularly Gaussian and Poisson noise, which are closely related to sensor noise and acquisition variability in imaging systems and can shift input distributions and degrade predictive reliability~\cite{yang2022machine}. Robustness to such degradations is therefore an important requirement for practical biomedical image classification. Nevertheless, the behavior of hybrid quantum-classical neural networks under realistic input corruptions remains insufficiently understood. This issue is particularly important in biomedical applications, where robustness must be assessed under degradations commonly encountered in practical imaging workflows. In this work, our primary focus is on robustness to classical input corruptions applied to biomedical images.

A second limitation of existing hybrid quantum-classical neural networks lies in their typically fixed strategies for integrating classical and quantum components. Many existing architectures adopt fixed integration schemes, often taking sequential or replacement-based forms, in which the quantum module is appended after a classical encoder or replaces part of a classical classifier. Although these designs are simple and compatible with near-term hardware constraints, they assign largely predetermined roles to the classical and quantum branches across different inputs. This design choice may be restrictive under corrupted input conditions, where the relative usefulness of classical and quantum representations may vary with image content, corruption type, and corruption severity. As a result, a fixed integration strategy may fail to fully exploit the complementary strengths of the two branches. These considerations suggest that hybrid learning may benefit from a more adaptive fusion mechanism that dynamically weights branch contributions according to the input.

To address the limited flexibility of fixed fusion strategies in hybrid quantum–classical models, we propose a Feature-Adaptive Fusion Hybrid Quantum–Classical Neural Network (FAF-HQNN) for biomedical image classification. The proposed architecture combines a ResNet-18-based deep feature encoder with classical and variational quantum circuit (VQC) branches in an end-to-end trainable framework. Its central component is a feature-adaptive class-wise fusion mechanism that predicts sample-dependent, class-specific fusion weights from learned feature representations and dynamically combines the outputs of the two branches. This mechanism allows branch contributions to vary across inputs and classes rather than relying on fixed fusion weights.

We evaluate FAF-HQNN on two biomedical image benchmarks, PathMNIST and BloodMNIST, using both clean test data and synthetic Gaussian and Poisson input corruptions at multiple severity levels. The robustness evaluation concerns classical input corruptions rather than quantum hardware noise. To assess the contribution of adaptive fusion, we compare FAF-HQNN with seven internal variants, three external hybrid quantum–classical baselines, and four external CNN baselines in terms of clean-test accuracy and performance under input corruption.

The main contributions of this work are summarized as follows:
\begin{itemize}
    \item We propose FAF-HQNN, an end-to-end trainable architecture for biomedical image classification that integrates a ResNet-18-based feature encoder with classical and VQC-based branches.

    \item We introduce a feature-adaptive class-wise fusion mechanism that predicts sample-dependent, class-specific weights from learned feature representations to dynamically combine classical and quantum branch outputs.

   \item Experiments on PathMNIST and BloodMNIST demonstrate that FAF-HQNN achieves competitive accuracy compared to internal variants and external hybrid quantum--classical baselines under the evaluated settings. On PathMNIST, it achieves higher aggregate mean accuracy than all evaluated baselines under both Gaussian and Poisson noise.

\end{itemize}

The rest of this paper is organized as follows. Section 2 reviews related work on classical and hybrid quantum-classical approaches. Section 3 presents the proposed FAF-HQNN architecture. Section 4 describes the experimental setup and presents the results. Section 5 discusses the findings, limitations, and directions for future research. Finally, Section 6 concludes the paper.

\section{Related Work}\label{sec2}
This section reviews the prior research most relevant to the present study, focusing on robustness in classical biomedical image classification and hybrid quantum-classical learning for image-based tasks.

\subsection{Classical Deep Learning and Robustness in Biomedical Image Classification}
Deep learning has achieved strong performance in biomedical image classification, with applications in computer-aided diagnosis, digital pathology, and cellular image analysis. Among existing approaches, convolutional neural networks (CNNs) have been widely adopted for tasks such as histopathology classification, radiographic diagnosis, and blood cell recognition, owing to their ability to learn hierarchical visual representations directly from image data~\cite{huang2017densely,rajpurkar2018deep}. 
Residual networks, in particular, remain popular CNN backbones due to their optimization stability, scalability, and strong empirical performance, including in limited-data settings~\cite{he2016deep}. Transformer-based architectures and other global-context modeling approaches have also been explored in medical imaging~\cite{dosovitskiy2020image,liu2021swin}. However, despite their stronger capacity for modeling long-range dependencies, CNN-based encoders remain competitive in many biomedical classification tasks because of their computational efficiency and favorable inductive bias, particularly on compact benchmark datasets.

Despite strong performance on clean benchmark data, biomedical image classifiers remain vulnerable to image corruptions such as Gaussian and Poisson noise, which are relevant to sensor-related noise and acquisition variability in imaging systems~\cite{zech2018variable}. These perturbations can shift input distributions and compromise predictive reliability in real-world deployment settings. To improve robustness, prior studies have explored data augmentation, corruption-based training, adversarial training, and regularization-based strategies~\cite{shorten2019survey}. However, compared with standard clean-data evaluation, robustness evaluation across multiple corruption types and severity levels remains relatively limited in biomedical image classification.

Standardized benchmarks play an important role in robustness research by enabling controlled, reproducible evaluation under consistent data formats and protocols. MedMNIST provides a lightweight collection of biomedical image benchmarks designed for efficient experimentation and standardized evaluation, making it well suited for both classical and hybrid quantum-classical learning settings~\cite{yang2023medmnist}. Among its subsets, PathMNIST serves as a representative benchmark for multi-class histopathology classification, whereas BloodMNIST offers a complementary task centered on blood cell recognition. Their biomedical relevance, standardized formulation, and compact scale make them practical testbeds for assessing predictive performance under controlled image corruptions.

Overall, prior work in classical biomedical image classification has established strong baselines and highlighted the practical importance of robustness under realistic perturbations. In contrast, systematic robustness evaluation has been incorporated only to a limited extent in hybrid quantum-classical models. As a result, a clear gap remains at the intersection of biomedical image robustness evaluation and hybrid quantum-classical learning.

\subsection{Hybrid Quantum-Classical Learning for Image Classification}

Quantum machine learning has attracted growing attention as a framework for integrating quantum information processing into classical learning pipelines~\cite{biamonte2017quantum}. However, current quantum hardware remains in the NISQ regime and is constrained by limited qubit counts, shallow circuit depth, and imperfect quantum operations~\cite{chen2024nisq}. These constraints make it challenging to apply purely quantum models directly to high-dimensional image classification problems, where input representations are typically large and structurally complex. As a result, hybrid quantum-classical architectures have become a common approach in image-based tasks, where classical neural networks are used to extract compact and informative features, while trainable quantum modules, typically variational quantum circuits (VQCs), are applied to low-dimensional latent representations within an end-to-end framework.

A common design in hybrid quantum-classical learning for image tasks is to use a classical encoder to compress high-dimensional image representations before passing them to a quantum circuit~\cite{mari2020transfer}. This strategy reduces the dimensionality of the quantum input, improves compatibility with near-term hardware, and preserves the representational strength of classical deep models. 
Because such architectures are primarily motivated by the need to combine expressive classical feature extraction with hardware-constrained quantum processing, much of the existing literature has focused on proof-of-concept validation and benchmark performance under standard evaluation settings. 
Recent studies have explored these hybrid architectures in image analysis and biomedical learning~\cite{obayya2023hybrid,singh2021quantum,fan2023hybrid,li2022image,yan2025semi}, providing initial evidence that hybrid quantum-classical models are feasible for small- to medium-scale problems under current hardware constraints.

In image classification, existing hybrid quantum-classical models mainly adopt three types of integration patterns. A common design uses a classical backbone for feature extraction and places the quantum module at the classification stage, allowing the quantum circuit to operate on compressed representations rather than raw high-dimensional inputs~\cite{henderson2020quanvolutional,chen2021end,senokosov2024quantum,mahmud2024quantum}. A second line of work incorporates the quantum component into the prediction head, typically by replacing part of the final decision layers with a variational quantum circuit~\cite{skolik2021layerwise,arthur2022hybrid,wang2025hybrid}. A third approach introduces a parallel quantum branch and fuses its outputs with those of the classical pathway through shared trainable decision layers~\cite{gong2024quantum,bartolucci2023fusion,qu2023qnmf,gordienko2024multimodal}. Although these integration patterns provide different ways to combine classical and quantum processing, they do not necessarily provide explicit, input-dependent, class-wise weighting of the two branches’ predictions. Such weighting may be useful in biomedical image classification, where image quality, structural complexity, and corruption severity can vary substantially across samples.

Adaptive fusion and gating have been extensively studied in classical learning. Mixtures of experts use input-dependent gates to combine expert outputs~\cite{zhou2022adaptive}, while gated multimodal units regulate the contributions of different input representations~\cite{arevalo2017gated}. In multi-branch visual networks, selective kernel networks use input-dependent, channel-wise attention to combine branch features~\cite{li2019selective}. Our method builds on these established adaptive weighting ideas rather than introducing a new attention or gating principle. It uses a shared embedding to generate complementary weights for classical and quantum logits, with one coefficient for each sample–class pair. Unlike the representation-level fusion in gated multimodal units and selective kernel networks, this gate operates on class predictions. However, prediction-level fusion and class-wise weighting are not claimed as novel in themselves; our contribution lies in the specific integration and controlled evaluation of this embedding-conditioned logit gate within the proposed hybrid architecture.

This design is motivated by the structure and resource constraints of the proposed hybrid architecture. Although the classical and quantum branches use different internal representations, their logits correspond to the same output classes, providing a common fusion space without an additional cross-branch feature-alignment module. Computing the gate entirely on the classical side adds a classical gating computation but requires no additional qubits or quantum circuit depth relative to the same architecture without the gate. Given the branches' different representational capacities and resource constraints, sample-dependent, class-wise weighting allows their relative contributions to vary rather than imposing a fixed balance across inputs or classes. These properties motivate the design but do not by themselves establish improved robustness. We therefore evaluate whether this weighting structure improves robustness to image corruptions compared with less adaptive fusion strategies. The mechanism is not exclusive to quantum models.

\section{Methodology}\label{sec3}

This section presents the proposed Feature-Adaptive Fusion Hybrid Quantum-Classical Neural Network (FAF-HQNN) for biomedical image classification. We first introduce the overall architecture and its design rationale, and then describe its main components in detail, including classical feature extraction, the feature compression module, hybrid prediction and feature-adaptive fusion, and the training objective.

\subsection{Overall Architecture}
\begin{figure*}[t]
\centering
\includegraphics[scale=0.25]{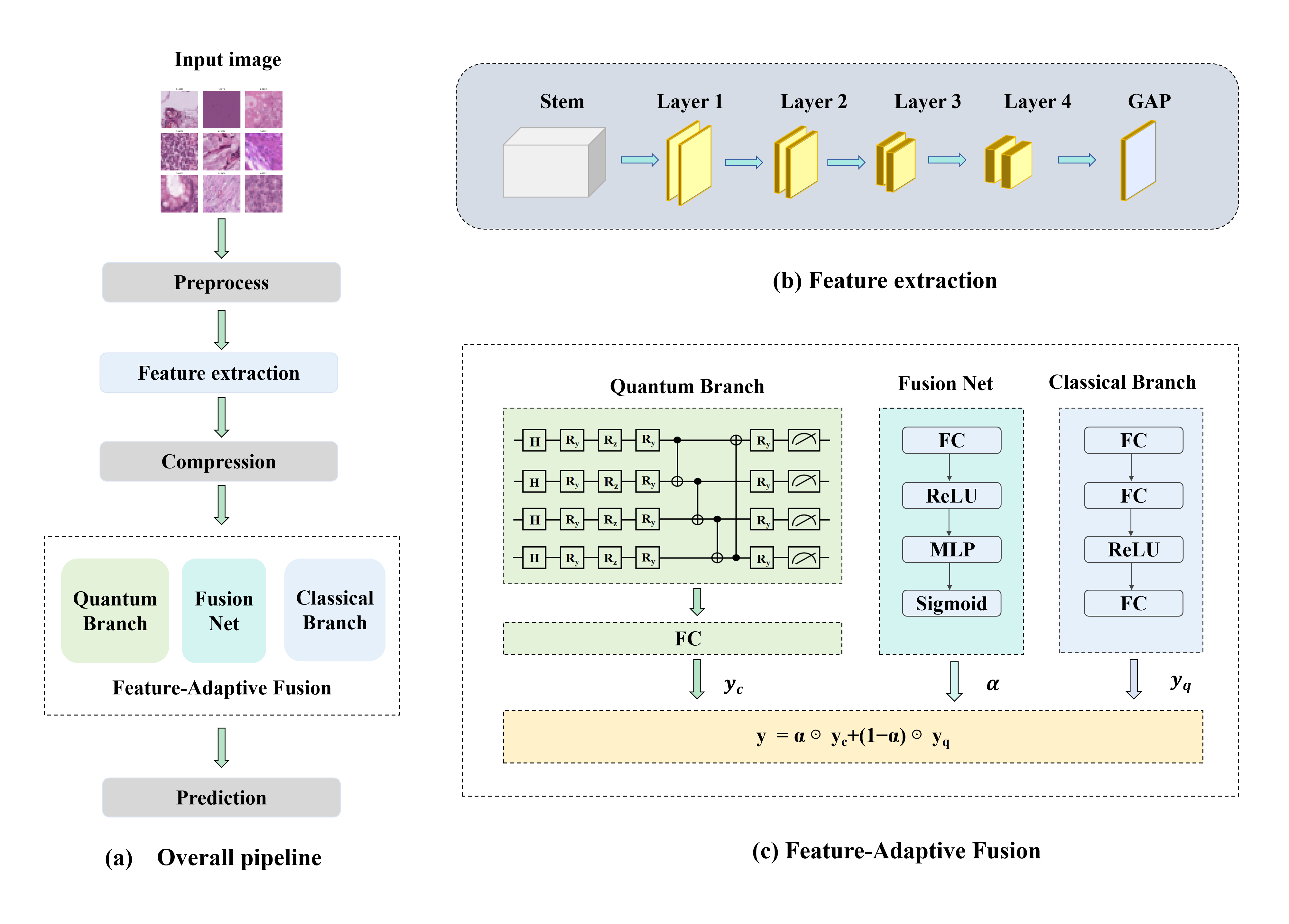}
\caption{Overall framework of the proposed FAF-HQNN model. 
(a) Pipeline of FAF-HQNN. Input biomedical images are first preprocessed and then fed into the feature extraction module, followed by a feature compression module for quantum-compatible embedding. The resulting shared embedding is subsequently passed to the feature-adaptive fusion module, including a quantum branch, fusion net, and a classical branch, whose outputs are combined to produce the final prediction. 
(b) Architecture of feature extraction, consisting of a stem block, four residual stages with two BasicBlocks in each stage, and a global average pooling (GAP) layer. 
(c) Structure of the feature-adaptive fusion. The embedding features are processed in parallel by a quantum branch implemented with a variational quantum circuit and a classical branch implemented as a multilayer perceptron. In parallel, a fusion net estimates class-wise adaptive fusion coefficients, which are used to combine the classical and quantum logits into the final output.}
\label{figure1}
\end{figure*}

Figure~\ref{figure1} presents the overall architecture of FAF-HQNN. As shown in Fig.~\ref{figure1}(a), the model follows a preprocessing--feature extraction--feature compression--feature-adaptive fusion pipeline for biomedical image classification.  In this pipeline, the feature compression module projects high-dimensional visual features into a low-dimensional shared embedding space, which serves as the interface between classical representation learning and downstream hybrid prediction.

Built upon this shared embedding, FAF-HQNN departs from conventional serial hybrid designs by adopting a parallel prediction strategy. Specifically, the shared embedding is fed into both the quantum and classical branches, while a lightweight fusion coefficient predictor generates sample-dependent, class-wise fusion coefficients to adaptively combine the logits produced by the two branches. In this way, the quantum branch serves as a complementary predictor rather than a direct replacement for the classical classifier, allowing the model to dynamically balance the relative contributions of the two branches across different input samples and categories. Overall, FAF-HQNN integrates classical deep feature extraction, VQC-based prediction, and feature-adaptive fusion within a unified end-to-end trainable framework.

\subsection{Feature Extraction}
This module extracts high-level visual representations from biomedical images for downstream hybrid prediction. In FAF-HQNN, feature extraction is implemented using ResNet-18 initialized with ImageNet-pretrained weights, which offers a suitable trade-off between representational capacity and computational complexity.

Specifically, the feature extractor consists of the convolutional stem and four residual stages of ResNet-18, while the original fully connected classification head is removed. Since the backbone is used solely for representation learning, it outputs deep visual features rather than dataset-specific classification logits. To obtain a fixed-length representation, the output of the final residual stage is further processed by a global adaptive average pooling layer. As a result, each input image is mapped to a 512-dimensional feature vector.

This extracted 512-dimensional feature vector is then passed to the subsequent feature compression module. By using an ImageNet-pretrained ResNet-18 as the classical backbone, the model benefits from transferable visual representations while maintaining a relatively lightweight architecture suitable for the proposed hybrid quantum-classical framework.

\subsection{Feature Compression for Quantum-Compatible Embedding}
Although the ResNet-18 backbone provides informative deep features, the resulting 512-dimensional representation remains too high-dimensional for direct quantum processing under NISQ constraints. To address this issue, we introduce a feature compression module that maps the backbone output to a compact shared embedding suitable for both quantum and classical prediction.

Let $\mathbf{h}_i \in \mathbb{R}^{512}$ denote the feature vector extracted from the $i$-th input image. The compression module first applies a learnable nonlinear projection to reduce the dimensionality of $\mathbf{h}_i$. Specifically, it consists of two fully connected layers with batch normalization and ReLU activation, which transform the feature dimension from 512 to 64. This step suppresses redundancy in the backbone representation while preserving the information most relevant to classification. This compact feature is then passed through a learnable embedding layer to produce the final shared embedding for downstream hybrid prediction.

Let $\mathbf{h}_i' \in \mathbb{R}^{64}$ denote the compressed feature, and let $f_{\mathrm{emb}} : \mathbb{R}^{64} \rightarrow \mathbb{R}^{n_q}$ denote the learnable embedding mapping.
To ensure numerical compatibility with parameterized quantum gates, the embedding output is passed through a hyperbolic tangent activation and rescaled to the interval $[-\pi,\pi]$. The shared embedding is defined as
$
\mathbf{z}_i = \pi \cdot \tanh\!\left(f_{\mathrm{emb}}(\mathbf{h}_i')\right),
$
where $\mathbf{z}_i \in \mathbb{R}^{n_q}$ denotes the shared embedding.

This shared embedding $\mathbf{z}_i$ serves as the interface between the classical backbone and the downstream hybrid prediction module. In this way, the feature compression module alleviates the dimensional mismatch between high-dimensional visual features and quantum-compatible inputs.

\subsection{Hybrid Prediction and Feature-Adaptive Fusion}

Given the shared embedding $\mathbf{z}_i$, FAF-HQNN performs parallel hybrid prediction and adaptive fusion. Specifically, $\mathbf{z}_i$ is simultaneously fed into the quantum branch, the classical branch, and the feature-adaptive class-wise fusion mechanism. The quantum and classical branches produce separate predictive outputs for adaptive fusion, while the feature-adaptive class-wise fusion mechanism assigns sample-specific fusion weights for each class to combine them adaptively. This design allows the model to dynamically balance the contributions of the two branches across different input samples and classes.

\subsubsection{Quantum Branch}

\begin{figure*}[t]
\centering
\includegraphics[scale=0.5]{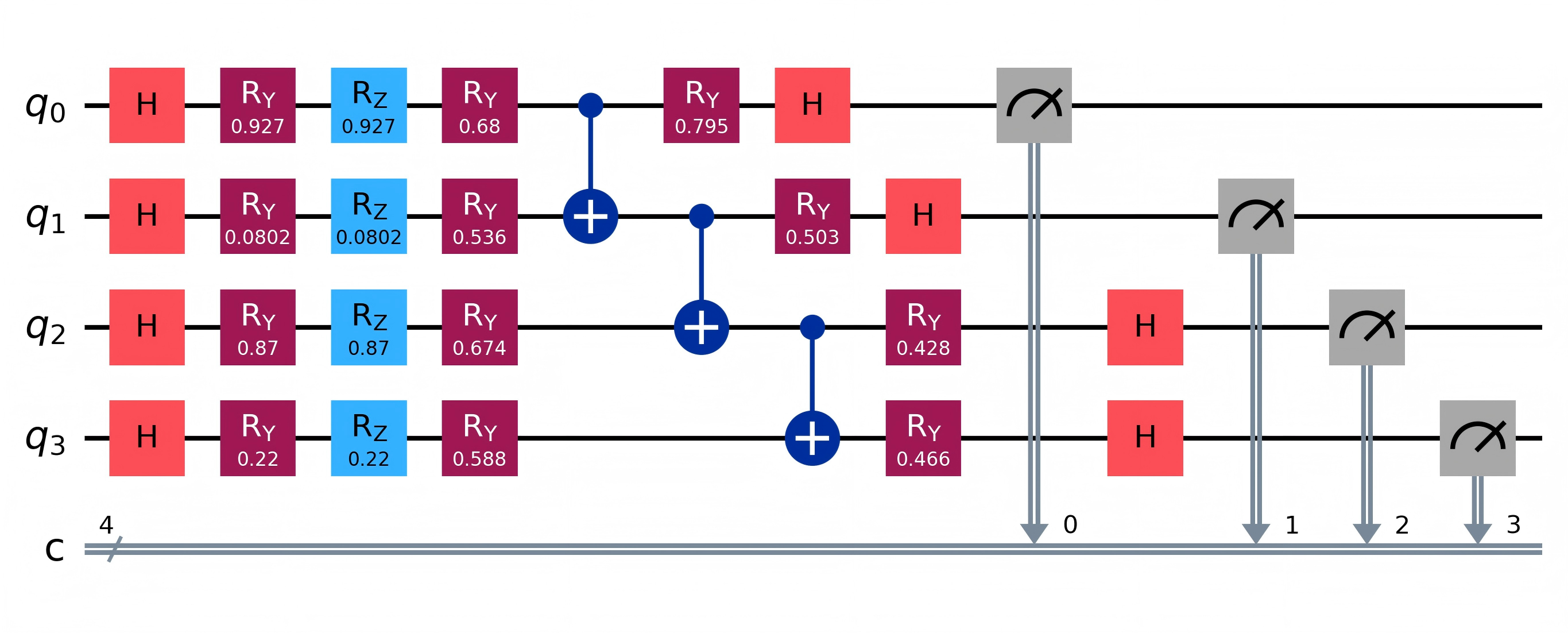}
\caption{Overall architecture of the quantum branch in FAF-HQNN. The branch consists of three stages: classical-to-quantum data encoding, variational quantum transformation, and measurement-based readout.}
\label{figure2}
\end{figure*}

The quantum branch processes the extracted features through three stages: classical-to-quantum data encoding, variational quantum transformation, and measurement-based readout. Its overall structure is illustrated in Fig.~\ref{figure2}.

\paragraph{Classical-to-quantum data encoding}
Let $\mathbf{z}_i = \left[z_i^0, z_i^1, \dots, z_i^{n_q-1}\right]^\top \in \mathbb{R}^{n_q}$ denote the shared embedding of the $i$-th sample used for classical-to-quantum data encoding, where $z_i^k$ denotes the $k$-th element of $\mathbf{z}_i$ for $k = 0,1,\dots,n_q-1$. The quantum register is first initialized in the computational basis state
$|\psi_0\rangle = |0\rangle^{\otimes n_q}$. 
To enrich the accessible state space while maintaining a shallow circuit structure, a layer of Hadamard gates is applied to all qubits, resulting in a uniform superposition state
\begin{equation}
\label{e4}
|\psi_H\rangle = H^{\otimes n_q} |\psi_0\rangle
= \bigotimes_{k=0}^{n_q-1} \frac{|0\rangle_k + |1\rangle_k}{\sqrt{2}}.
\end{equation}

The uniform superposition state $|\psi_H\rangle$ is then encoded through single-qubit $R_y$ rotations, 
\begin{equation}
\label{e5}
|\psi_{R_y}\rangle =
\left( \bigotimes_{k=0}^{n_q-1} R_y(z_i^k) \right) |\psi_H\rangle,
\end{equation}
where each component $z_i^k$ controls the rotation angle of qubit $k$. To further incorporate phase information, $R_z$ rotations are additionally applied to the state $|\psi_{R_y}\rangle$. The encoded quantum state is thus written as
\begin{equation}  
\label{e6}
|\psi_i\rangle =
\left[
\left( \bigotimes_{l=0}^{n_q-1} R_z(z_i^l) \right)
\right]
|\psi_{R_y}\rangle.
\end{equation}
This encoding scheme combines complementary angle and phase encoding operations to enrich the quantum representation while preserving a hardware-efficient and shallow circuit design.

\paragraph{Variational quantum transformation}

After data encoding, the quantum state is processed by an $L$-layer variational quantum circuit (VQC). Each layer contains
one trainable single-qubit rotation per qubit, followed by a nearest-neighbor CNOT entangling chain. Specifically, in layer
$l$, the rotation $R_y(\theta_{l,j})$ is applied to qubit $j$. The CNOT gates are then applied sequentially from qubit $0$
to qubit $n_q-1$. The overall VQC transformation is defined as
\begin{equation}
U_{\mathrm{VQC}} =
\prod_{l=0}^{L-1}
\left[
\left( \prod_{j=0}^{n_q-2} \mathrm{CNOT}(j,j+1) \right)
\left( \bigotimes_{j=0}^{n_q-1} R_y(\theta_{l,j}) \right)
\right].
\end{equation}
Here, operators act from right to left, and $\mathrm{CNOT}(j,j+1)$ uses qubit $j$ as the control and qubit $j+1$ as the target. Each layer therefore contains $n_q$ trainable rotation parameters and $n_q-1$ CNOT gates, giving $L n_q$ trainable circuit parameters in total. The four-qubit, single-layer configuration used in the main experiments has four trainable circuit parameters.

Together with data encoding and measurement, these trainable rotations and entangling operations map the shared classical
embedding to quantum readout features. The shallow circuit structure limits quantum resource requirements, although
its performance on physical quantum hardware remains to be evaluated.

\paragraph{Measurement-based readout}
After applying $U_{\mathrm{VQC}}$, the transformed quantum state is measured on each qubit to obtain a real-valued readout vector. In our implementation, the measurement outcome of the $k$-th qubit is defined as the expectation value of the Pauli-$X$ observable,
\begin{equation}
\label{e8}
q_i^k = \langle \psi_i | U_{\mathrm{VQC}}^\dagger X_k U_{\mathrm{VQC}} | \psi_i \rangle,
\quad k = 0, \dots, n_q - 1.
\end{equation}
Collecting the measurement outcomes from all qubits yields the quantum feature vector
\begin{equation}
\mathbf{q}_i = [q_i^0, \dots, q_i^{n_q-1}] \in \mathbb{R}^{n_q}.
\end{equation}
This vector is then mapped to the class space through a learnable linear layer,
\begin{equation}
\mathbf{y}_i^{q} = f_q(\mathbf{q}_i),
\end{equation}
where $f_q(\cdot)$ denotes the linear mapping of the quantum branch, and $\mathbf{y}_i^{q} \in \mathbb{R}^{C}$ represents the class logits produced by the quantum branch.

The quantum branch is constructed to remain compatible with NISQ devices. Each variational layer consists of parameterized single-qubit rotations and nearest-neighbor CNOT gates, resulting in a hardware-efficient circuit with limited depth. By combining local trainable operations with shallow entangling interactions, this design supports expressive quantum feature transformation while remaining practical for near-term quantum implementations.

\subsubsection{Classical Branch}

In parallel with the quantum branch, the shared embedding $\mathbf{z}_i \in \mathbb{R}^{n_q}$ is fed into a classical branch. This branch is implemented as a lightweight multilayer perceptron (MLP), which serves as an efficient classical predictor on the shared embedding.

Specifically, the classical branch consists of three fully connected layers with ReLU activations and dropout regularization. The output class logits are computed as
\begin{equation}
\mathbf{y}_i^{c} = f_c(\mathbf{z}_i),
\end{equation}
where $f_c(\cdot)$ denotes the classical MLP, and $\mathbf{y}_i^{c} \in \mathbb{R}^{C}$ represents the class logits produced by the classical branch.

Since this branch operates entirely in the classical domain, it provides a stable predictive reference within the hybrid framework. Together with the quantum branch, it forms a dual-branch prediction structure for subsequent adaptive fusion.

\subsubsection{Feature-Adaptive Class-Wise Fusion Mechanism}

Based on the shared embedding $\mathbf{z}_i$, the classical and quantum branches produce class logits $\mathbf{y}_i^{c}$ and $\mathbf{y}_i^{q}$, respectively. To adaptively combine these two outputs, a lightweight class-wise fusion predictor is applied to $\mathbf{z}_i$ to generate a fusion coefficient vector
\begin{equation}
\boldsymbol{\alpha}_i = f_{\alpha}(\mathbf{z}_i), \qquad \boldsymbol{\alpha}_i \in [0,1]^C,
\end{equation}
where $f_{\alpha}(\cdot)$ is implemented as a lightweight two-layer MLP with a sigmoid output layer, ensuring that each coefficient lies in $[0,1]$.

For each class $c$, the coefficient $\alpha_i^c$ controls the contribution of the classical branch, while $1-\alpha_i^c$ controls the contribution of the quantum branch. The fused logits are defined as
\begin{equation}
\hat{\mathbf{y}}_i =
\boldsymbol{\alpha}_i \odot \mathbf{y}_i^{c}
+
(\mathbf{1}-\boldsymbol{\alpha}_i) \odot \mathbf{y}_i^{q},
\end{equation}
where $\odot$ denotes element-wise multiplication and $\hat{\mathbf{y}}_i \in \mathbb{R}^{C}$ is the fused output vector. 

The fusion predictor implements an embedding-conditioned, prediction-level gate based on established adaptive weighting principles. It generates complementary branch weights for each sample-class pair, allowing the weights to vary across both inputs and output classes. This differs from fixed averaging and globally learned scalar fusion, which do not adapt their weights to individual inputs. It also differs from sample-wise scalar fusion, which shares a coefficient across classes within each sample, and globally learned class-wise fusion, which shares class-specific coefficients across samples. These alternatives are examined in the fusion ablation study to assess the effects of sample dependence and class-wise weighting.

Logit-level fusion is suited to the proposed hybrid architecture because the classical and quantum branches produce predictions over the same output classes despite using different internal representations. Their outputs can therefore be combined without explicitly aligning their internal feature representations. The gate adds classical parameters and computation but requires no additional qubits or quantum circuit depth relative to the same two-branch architecture without the gate. This design accommodates the quantum resource constraints of the hybrid architecture, while its effect on robustness to image corruptions is evaluated empirically.

\subsection{Training Objective}

FAF-HQNN is trained end-to-end in a supervised manner for multi-class biomedical image classification. Given an input image $\mathbf{x}_i$ with ground-truth label $y_i \in \{1, \dots, C\}$, the model produces fused class logits $\hat{\mathbf{y}}_i = [\hat{y}_{i,1}, \dots, \hat{y}_{i,C}] \in \mathbb{R}^{C}$ through the feature-adaptive class-wise fusion mechanism.

The model is optimized using the standard cross-entropy loss computed from the fused logits:
\begin{equation}
\mathcal{L}_{\mathrm{CE}}
=
-\frac{1}{N}
\sum_{i=1}^{N}
\log
\frac{
\exp(\hat{y}_{i,y_i})
}{
\sum_{c=1}^{C}\exp(\hat{y}_{i,c})
},
\end{equation}
where $N$ denotes the mini-batch size, and $\hat{y}_{i,c}$ denotes the $c$-th element of the fused logit vector $\hat{\mathbf{y}}_i$.

\section{Experiments}\label{sec4}

This section presents the empirical evaluation of the proposed FAF-HQNN. The experiments are designed to assess four aspects of the proposed method: classification performance under standard conditions, training behavior and the effect of quantum module design, robustness under common image corruptions, and generalization across datasets. We first conduct the main analysis on PathMNIST as the primary benchmark, and then examine whether the observed trends generalize to BloodMNIST.

\subsection{Experimental Setup}

This subsection describes the datasets and preprocessing procedures, compared methods, implementation details, and evaluation metrics and robustness protocol used throughout the experiments.

\subsubsection{Dataset and Preprocessing}

PathMNIST is a nine-class histopathology image classification dataset included in MedMNIST v2. Its official training and validation splits are derived from NCT-CRC-HE-100K, while its official test split is derived from CRC-VAL-HE-7K. The official splits contain 89,996 training images, 10,004 validation images, and 7,180 test images. In this study, we use the 28-by-28-pixel RGB version.

We use PathMNIST as the primary benchmark because colorectal histopathology image classification is a clinically relevant biomedical image analysis task, and the dataset is sufficiently large to support class-balanced subsampling and controlled multi-seed experiments. Histopathology images may be affected by variations in image acquisition and staining, motivating the study of robustness. However, the synthetic corruptions considered in this work provide a controlled evaluation of image-level perturbations rather than a comprehensive simulation of clinical domain shifts.

Due to the high computational cost of sample-wise quantum circuit simulation, we construct a fixed, class-balanced training subset by randomly sampling 1,000 images without replacement from each of the nine classes in the official training split, yielding 9,000 images in total. Subset sampling is performed once using a fixed sampling seed, separately from the random seeds used for model training. The resulting subset is kept unchanged across all compared models, ablation variants, and training seeds, ensuring that the comparisons are not confounded by differences in training subset composition. The full official validation and test splits are used for model selection and final evaluation, respectively. Accordingly, the reported PathMNIST results and associated conclusions are limited to this reduced-data training setting and do not establish whether the same benefits would be obtained when training on the complete official training set.

\subsubsection{Compared Methods}

We compare FAF-HQNN with two groups of methods. The first group consists of controlled baselines and internal ablation variants, which examine the prediction branches and fusion strategies within a common feature-extraction framework. The second group consists of external architecture baselines, which provide broader architectural comparisons rather than parameter-matched ablations.
\paragraph{Controlled baselines and internal ablations.}

All methods in this group use the same ImageNet-pretrained ResNet-18 backbone and feature-compression module, and each model is trained independently from the same pretrained initialization without weight sharing. The classical head takes the full four-dimensional shared embedding as input, while the four-qubit quantum head encodes all four components, assigning one component to each qubit.

\textbf{FAF-HQNN}, our proposed method, combines parallel classical and quantum prediction heads using embedding-dependent fusion weights that vary across both samples and classes.

This group comprises one single-branch classical control, one classical replacement control, and five variants that ablate the quantum branch and the fusion strategy. Together, they examine three aspects: single-branch prediction, replacement of the quantum branch with a classical alternative, and the fusion strategy, including fixed versus learnable weighting and the sample and class dependence of the fusion weights. Unless otherwise specified, each variant uses the same backbone and feature-compression module as FAF-HQNN.
\begin{itemize}
    \item \textbf{Classical}: This variant retains the classical prediction head while removing the quantum prediction branch and the adaptive fusion module. It serves as a single-branch classical baseline within the same backbone and feature-compression module architecture.
    \item \textbf{Q-Head}: This variant retains the quantum prediction head while removing the parallel classical prediction branch and the adaptive fusion module. It evaluates the predictive capability of the quantum head when used as the sole prediction branch, rather than in combination with the classical head.
    \item \textbf{C-AF}: This purely classical counterpart retains the original classical prediction head and the adaptive fusion module, but replaces the quantum prediction head with a lightweight classical multilayer perceptron (MLP). The replacement MLP takes the same compact embedding as the quantum head and produces logits of the same output dimension. This variant compares the quantum branch with its classical replacement within the same two-branch adaptive-fusion architecture.
    \item \textbf{H-Fix}: This variant retains both the classical and quantum prediction heads, but replaces adaptive fusion with fixed equal averaging of their logits. It provides a fixed-weight fusion baseline for assessing the effect of learning the relative contributions of the two branches.
    \item \textbf{H-Scalar}: This variant retains both prediction heads, but uses a single learnable global fusion scalar shared across all samples and classes. Compared with H-Fix, it assesses the effect of learning a global branch-weighting coefficient without introducing sample- or class-dependent fusion.
    \item \textbf{H-Sample}: This variant retains both prediction heads and uses an embedding-dependent fusion scalar for each sample, shared across all classes. Its comparison with FAF-HQNN assesses the effect of allowing the fusion weights to differ across classes in addition to varying across samples.
    \item \textbf{H-Class}: This variant retains both prediction heads and uses a learnable class-specific fusion vector shared across all samples. Its comparison with FAF-HQNN assesses the effect of conditioning the class-specific fusion weights on individual samples.
\end{itemize}

These variants assess single-branch performance, the choice of the second prediction branch, and fusion flexibility under a shared ResNet-18 backbone. \textbf{Classical} serves as both a purely classical control and an ablation of FAF-HQNN with the quantum head and adaptive fusion removed. Table~\ref{tab:model_definitions} summarizes the prediction heads, fusion mechanisms, and experimental roles of all models.

\begin{table*}[t]
\centering
\caption{FAF-HQNN and its controls and ablations. C-AF uses a replacement MLP alongside the original classical head.}
\label{tab:model_definitions}

\renewcommand{\arraystretch}{1.2}
\setlength{\tabcolsep}{3pt}
\resizebox{\textwidth}{!}{
\begin{tabular}{lcccl}
\hline
\textbf{Model} & \textbf{C-head(s)} & \textbf{Q-head} & \textbf{Fusion} & \textbf{Role} \\
\hline
Classical & Original       & No  & None                   & Control / ablation \\
Q-Head    & None           & Yes & None                   & Ablation \\
C-AF      & Original + MLP & No  & Sample--class adaptive & Replacement control \\
H-Fix     & Original       & Yes & Equal average          & Ablation \\
H-Scalar  & Original       & Yes & Global scalar          & Ablation \\
H-Sample  & Original       & Yes & Sample-wise scalar     & Ablation \\
H-Class   & Original       & Yes & Class-wise vector      & Ablation \\
FAF-HQNN  & Original       & Yes & Sample--class adaptive & Proposed \\
\hline
\end{tabular}
}
\end{table*}

\paragraph{External architecture baselines.}
To further position FAF-HQNN against established hybrid quantum-classical image-classification architectures, we include three external baselines, denoted
EST-QCNN, EST-RUQNN, and EST-QCQ-CNN. The prefix EST indicates that each architecture is re-implemented and tuned under the same experimental setting as
FAF-HQNN; these models are therefore task-adapted realizations of previously proposed architectural paradigms rather than exact reproductions, because the
original works differ from our setting in dataset type, input resolution, feature dimensionality, circuit configuration, and training protocol.

To keep the comparison as consistent as possible, all three baselines share the following settings with FAF-HQNN: an ImageNet-pretrained ResNet-18 backbone that
is fine-tuned end-to-end, a feature projection module that maps the backbone output to a four-dimensional embedding, and a four-qubit quantum circuit with a
single variational layer. The baselines therefore differ from one another only in the internal structure of the quantum module and in how the classical and
quantum components are arranged.

\textbf{EST-QCNN} is inspired by quantum convolutional neural networks~\cite{cong2019quantum}. The four-dimensional embedding is
angle-embedded into the four-qubit circuit, which contains four independently parameterized two-qubit convolution blocks and two controlled-rotation
pooling-inspired blocks, without mid-circuit measurements.

\textbf{EST-RUQNN} follows the data re-uploading quantum classifier paradigm~\cite{perez2020data}. The four-dimensional embedding is angle-embedded
into the four-qubit circuit and then re-uploaded in a single re-uploading layer, so that the encoding operations are interleaved with trainable quantum
transformations rather than applied only once at the circuit input.

\textbf{EST-QCQ-CNN} is inspired by the sequential quantum--classical--quantum architecture proposed by Long et al.~\cite{long2025hybrid}. It first extracts features using a fine-tuned ResNet-18 backbone, then processes them through a non-trainable four-qubit quantum convolutional filter, a classical multilayer perceptron (MLP), and a trainable single-layer four-qubit variational quantum classifier, in that order.

All external baselines are evaluated on the same class-balanced training subset, test images, corruption types, and severity levels as FAF-HQNN. These
comparisons provide architectural context rather than controlled tests of individual components, as the models differ in the internal design of their
quantum modules and in parameter count.

\subsubsection{Implementation Details and Training Protocol}

All images are resized to $64 \times 64$  pixels and normalized using the ImageNet mean and standard deviation, without data augmentation. All models are trained using cross-entropy loss, with training and evaluation batch sizes of 16 and 32, respectively. 

On PathMNIST, we tune the initial learning rates for all models, including FAF-HQNN, the internal variants, and the external baselines, using a budget-limited random search. Depending on the architecture, separate learning rates are assigned to the pretrained backbone, the feature-processing layers, the quantum circuit parameters, and the readout-related modules, including trainable fusion parameters where applicable. The candidate learning rates are
$\{1\times 10^{-5},\, 3\times 10^{-5},\, 1\times 10^{-4}\}$ for the backbone, $\{1\times 10^{-4},\, 3\times 10^{-4},\, 1\times 10^{-3}\}$ for the feature and
readout groups, and $\{1\times 10^{-4},\, 1\times 10^{-3},\, 1\times 10^{-2}\}$ for the quantum circuit parameters. No fixed learning-rate ratio is imposed between parameter groups.

We randomly sample a shared list of eight candidate learning-rate configurations over the full set of parameter groups, and each model evaluates these candidates using only its applicable parameter groups. Each candidate
configuration is trained with two tuning seeds, \{100, 101\}, for 50 epochs per run, yielding 16 tuning runs per model.
For each run, we record the highest validation accuracy achieved during training, and each model independently
selects the configuration with the highest mean best-validation accuracy across the two tuning seeds.
The shared candidate set, tuning seeds, and budget reduce differences in hyperparameter search effort across models.

Each model is then retrained with its selected configuration using five additional seeds, $\{0,1,2,3,4\}$, which are disjoint from the tuning seeds. Each seed is trained for 50 epochs. For every run, we retain the
checkpoint with the highest validation accuracy for testing. Results are reported as the mean and standard deviation across the five final runs, and the selected initial learning rates for each model are reported in Table~\ref{tab:selected_hyperparameters}.
 The test set is used only for final evaluation, and never for hyperparameter or checkpoint selection.
\begin{table*}[t]
\centering
\caption{Validation-selected initial learning rates for FAF-HQNN, internal controlled baselines and ablation variants, representative hybrid quantum-classical baselines, and external classical CNN baselines.}
\label{tab:selected_hyperparameters}
\begingroup
\small
\renewcommand{\arraystretch}{1.2}
\setlength{\tabcolsep}{3pt}
\begin{tabular*}{\textwidth}{@{\extracolsep{\fill}}lcccc@{}}
\toprule
\textbf{Model}
& \textbf{Backbone LR}
& \textbf{Feature LR}
& \textbf{Circuit LR}
& \textbf{Readout LR} \\
\midrule
Classical
& 3e-5 & 1e-3 & N/A  & 1e-4 \\
Q-Head
& 1e-4 & 1e-4 & 1e-3 & 1e-3 \\
C-AF
& 1e-4 & 1e-3 & N/A  & 1e-3 \\
H-Fix
& 3e-5 & 3e-4 & 1e-3 & 1e-4 \\
H-Scalar
& 3e-5 & 1e-3 & 1e-4 & 1e-4 \\
H-Sample
& 3e-5 & 1e-4 & 1e-3 & 1e-3 \\
H-Class
& 1e-4 & 1e-3 & 1e-3 & 1e-4 \\
EST-QCNN
& 3e-5 & 3e-4 & 1e-3 & 3e-4 \\
EST-RUQNN
& 3e-5 & 3e-4 & 1e-3 & 3e-4 \\
EST-QCQ-CNN
& 3e-5 & 3e-4 & 1e-3 & 3e-4 \\
FAF-HQNN
& 3e-5 & 3e-4 & 1e-3 & 1e-4 \\
ResNet-18
& 1e-4 & 1e-3  & N/A & 3e-4 \\
ResNet-50
& 3e-5 & 1e-3  & N/A & 3e-4 \\
EfficientNet-B0
& 1e-4 & 1e-3  & N/A & 1e-4 \\
MobileNetV3-Small
& 3e-5 & 1e-3  & N/A & 1e-3 \\
\bottomrule
\end{tabular*}
\par\medskip
\noindent
\begin{minipage}{\textwidth}
\footnotesize
\raggedright
Notes: LR denotes the initial learning rate. Backbone LR applies to the pretrained backbone. Feature LR applies to the feature-compression and embedding-projection layers. Circuit LR applies to the trainable quantum circuit parameters. Readout LR applies to the classical classification head, post-quantum linear readout, and trainable fusion parameters. N/A indicates that the corresponding parameter group is not applicable.
\end{minipage}
\par
\endgroup
\end{table*}

\subsubsection{Evaluation Metrics and Robustness Protocol}
Classification performance is evaluated using accuracy (ACC), support-weighted precision, support-weighted F1-score, and macro-averaged one-vs-rest ROC-AUC (AUC). ACC measures the proportion of correctly classified test samples. Precision and F1-score are computed for each class and averaged using weights proportional to class support. AUC is computed from predicted class probabilities using a one-vs-rest strategy and averaged equally across classes. Results are reported as the mean and sample standard deviation across five training seeds (0--4). The checkpoint with the highest validation accuracy is predefined as the primary checkpoint; the final-epoch and best-validation-AUC checkpoints are evaluated separately as auxiliary analyses.

Robustness is assessed under Gaussian and Poisson noise. All models are trained exclusively on clean data, and corruptions are applied only to test images in the pixel domain, followed by clipping to [0, 1] and renormalization. Noise severity is progressively increased, with identical noise realizations used across the three checkpoints within each seed. This train-clean/test-corrupted protocol isolates the effect of test-time perturbations from changes in the training distribution. The three checkpoints share a single training trajectory per seed and are not treated as independent training runs.

\subsection{Main Results on PathMNIST}
\paragraph{Clean-Test Performance Comparisons.}
We evaluate classification performance on PathMNIST  through two complementary comparisons. Table~\ref{tab:path_clean_internal} compares FAF-HQNN with seven internal controlled variants and ablation models, as well as three task-adapted external hybrid quantum--classical baselines. Table~\ref{tab:path_clean_external} provides a separate comparison with representative external classical CNN baselines.
Results are reported as means and standard deviations over five runs.

As shown in Table~\ref{tab:path_clean_internal}, FAF-HQNN achieves the highest mean ACC, precision, F1-score, and AUC among all methods listed, with an ACC of 0.8919. Its mean performance exceeds that of C-AF, the classical replacement variant, on all four metrics. 

FAF-HQNN also achieves higher mean performance on all four metrics than the evaluated fusion ablations. H-Fix yields the highest mean ACC among these ablations, at 0.8872, compared with 0.8919 for FAF-HQNN. These descriptive results favor the jointly sample- and class-dependent fusion configuration under the evaluated protocol.

Compared with the task-adapted external hybrid quantum--classical baselines, FAF-HQNN achieves higher mean performance on all four metrics. Based on the reported means, its ACC exceeds those of EST-QCNN, EST-RUQNN, and EST-QCQ-CNN by 1.81, 1.77, and 0.79 percentage points, respectively. Differences in model configurations, however, limit attribution of these gains solely to the fusion mechanism.

As shown in Table~\ref{tab:path_clean_external}, FAF-HQNN ranks third in mean ACC and F1-score among the five evaluated methods, behind EfficientNet-B0 and ResNet-50. EfficientNet-B0 achieves the highest mean ACC, precision, and F1-score, whereas ResNet-50 achieves the highest mean AUC. Overall, FAF-HQNN demonstrates competitive clean-test performance, although it does not achieve the highest mean value on any of the reported metrics.

\begin{table*}[t]
\centering
\caption{Performance comparison of FAF-HQNN with internal controlled variants, fusion ablations, and task-adapted external hybrid quantum--classical baselines on the clean PathMNIST test set. Results are reported as mean \ensuremath{\pm} standard deviation over five runs. Precision and F1-score are support-weighted. Bold metric values indicate the highest mean in each column, and the proposed method is highlighted in bold.}
\label{tab:path_clean_internal}
\small
\renewcommand{\arraystretch}{1.2}
\setlength{\tabcolsep}{3pt}

\resizebox{\linewidth}{!}{%
\begin{tabular}{@{}lcccc@{}}
\toprule
\textbf{Method}
& \textbf{ACC}
& \textbf{Precision}
& \textbf{F1-score}
& \textbf{AUC} \\
\midrule
Classical
& 0.8800 \textpm{} 0.0043
& 0.8831 \textpm{} 0.0048
& 0.8799 \textpm{} 0.0045
& 0.9754 \textpm{} 0.0029 \\
Q-Head
& 0.8777 \ensuremath{\pm} 0.0168
& 0.8797 \ensuremath{\pm} 0.0125
& 0.8763 \ensuremath{\pm} 0.0142
& 0.9490 \ensuremath{\pm} 0.0294 \\
C-AF 
& 0.8806 \ensuremath{\pm} 0.0196
& 0.8895 \ensuremath{\pm} 0.0158
& 0.8804 \ensuremath{\pm} 0.0181
& 0.9796 \ensuremath{\pm} 0.0034 \\
H-Fix
& 0.8872 \ensuremath{\pm} 0.0097
& 0.8895 \ensuremath{\pm} 0.0079
& 0.8868 \ensuremath{\pm} 0.0093
& 0.9760 \ensuremath{\pm} 0.0058 \\
H-Scalar 
& 0.8868 \ensuremath{\pm} 0.0085
& 0.8913 \ensuremath{\pm} 0.0068
& 0.8856 \ensuremath{\pm} 0.0083
& 0.9777 \ensuremath{\pm} 0.0048 \\
H-Sample
& 0.8747 \ensuremath{\pm} 0.0187
& 0.8776 \ensuremath{\pm} 0.0184
& 0.8725 \ensuremath{\pm} 0.0190
& 0.9738 \ensuremath{\pm} 0.0157 \\
H-Class
& 0.8763 \ensuremath{\pm} 0.0209
& 0.8799 \ensuremath{\pm} 0.0208
& 0.8746 \ensuremath{\pm} 0.0205
& 0.9792 \ensuremath{\pm} 0.0064 \\
EST-QCNN 
& 0.8738 \ensuremath{\pm} 0.0073
& 0.8749 \ensuremath{\pm} 0.0071
& 0.8712 \ensuremath{\pm} 0.0077
& 0.9490 \ensuremath{\pm} 0.0114 \\
EST-RUQNN
& 0.8742 \ensuremath{\pm} 0.0157
& 0.8771 \ensuremath{\pm} 0.0144
& 0.8718 \ensuremath{\pm} 0.0160
& 0.9571 \ensuremath{\pm} 0.0132 \\
EST-QCQ-CNN 
& 0.8840 \ensuremath{\pm} 0.0071
& 0.8852 \ensuremath{\pm} 0.0136
& 0.8830  \ensuremath{\pm} 0.0073
& 0.9590\ensuremath{\pm} 0.0082\\
\textbf{FAF-HQNN} 
& \textbf{0.8919} \ensuremath{\pm} 0.0078
& \textbf{0.8923} \ensuremath{\pm} 0.0075
&\textbf{0.8907} \ensuremath{\pm} 0.0077
& \textbf{0.9825} \ensuremath{\pm} 0.0027 \\
\bottomrule
\end{tabular}%
}
\end{table*}

\begin{table*}[t]
\centering
\caption{Performance comparison of FAF-HQNN with representative external classical CNN baselines on the clean PathMNIST test set. Results are reported as mean \ensuremath{\pm} standard deviation over five runs. Precision and F1-score are support-weighted. Bold metric values indicate the highest mean in each column, and the proposed method is highlighted in bold.}
\label{tab:path_clean_external}
\small
\renewcommand{\arraystretch}{1.2}
\setlength{\tabcolsep}{3pt}
\resizebox{\linewidth}{!}{%
\begin{tabular}{@{}lcccc@{}}
\toprule
\textbf{Method}& \textbf{ACC}& \textbf{Precision}& \textbf{F1-score}& \textbf{AUC} \\
\midrule
ResNet-18
& 0.8797 \ensuremath{\pm} 0.0104
& 0.8845 \ensuremath{\pm} 0.0099
& 0.8792 \ensuremath{\pm} 0.0107
& 0.9832 \ensuremath{\pm} 0.0058 \\
ResNet-50
& 0.8961 \ensuremath{\pm} 0.0090
& 0.8988 \ensuremath{\pm} 0.0071
& 0.8949 \ensuremath{\pm} 0.0088
& \textbf{0.9867} \ensuremath{\pm} 0.0021 \\
EfficientNet-B0
& \textbf{0.8974} \ensuremath{\pm} 0.0083
& \textbf{0.8999} \ensuremath{\pm} 0.0079
& \textbf{0.8952} \ensuremath{\pm} 0.0071
& 0.9822 \ensuremath{\pm} 0.0039 \\
MobileNetV3-Small
& 0.8627 \ensuremath{\pm} 0.0098
& 0.8641 \ensuremath{\pm} 0.0083
& 0.8592 \ensuremath{\pm} 0.0108
& 0.9836 \ensuremath{\pm} 0.0026 \\
\textbf{FAF-HQNN}
& 0.8919 \ensuremath{\pm} 0.0078
& 0.8923 \ensuremath{\pm} 0.0075
& 0.8907 \ensuremath{\pm} 0.0077
& 0.9825 \ensuremath{\pm} 0.0027 \\
\bottomrule
\end{tabular}%
}
\end{table*}
\paragraph{Exploratory Statistical Comparisons.}

Table~\ref{tab:path_significance_clean} reports two-sided paired \textit{t}-tests and paired-sample effect sizes across five seed-matched runs. FAF-HQNN shows nominally significant improvements in ACC over Classical, H-Fix, H-Sample, and MobileNetV3-Small, and in AUC over Classical, H-Fix, EST-QCNN, EST-RUQNN, and EST-QCQ-CNN. However, its AUC is nominally significantly lower than that of ResNet-50. None of the remaining comparisons reaches nominal significance at the 0.05 level. These results are exploratory, with \textit{p}-values unadjusted for multiple comparisons.

\begin{table*}[t]
\centering
\caption{Statistical comparisons on the clean PathMNIST test set across five seed-matched paired runs. Diff. denotes the mean paired difference (FAF-HQNN minus the baseline) on the original metric scale. Comparisons with ResNet-18 and ResNet-50 are computed from run-level results rounded to four decimal places; all other comparisons are retained from the original analysis. The paired-sample effect size is the mean paired difference divided by its sample standard deviation. All tests are two-sided paired t-tests. The analyses are exploratory, with p-values unadjusted for multiple comparisons; bold p-values indicate nominal significance at the 0.05 level.}
\label{tab:path_significance_clean}
\small
\renewcommand{\arraystretch}{1.2}
\setlength{\tabcolsep}{3pt}
\resizebox{\linewidth}{!}{%
\begin{tabular}{@{}lcccccc@{}}
\toprule
\textbf{Baseline}
& \textbf{ACC Diff.}
& \textbf{ACC \ensuremath{p}}
& \textbf{ACC \ensuremath{d_z}}
& \textbf{AUC Diff.}
& \textbf{AUC \ensuremath{p}}
& \textbf{AUC \ensuremath{d_z}} \\
\midrule
Classical
& 0.0119 & \textbf{0.0325} & 1.4368
& 0.0071 & \textbf{0.0010} & 3.8433 \\
Q-Head
& 0.0141 & 0.1726 & 0.7416
& 0.0335 & 0.0713 & 1.0906 \\
C-AF
& 0.0112 & 0.3250 & 0.5014
& 0.0029 & 0.1808 & 0.7240 \\
H-Fix
& 0.0046 & \textbf{0.0330} & 1.4300
& 0.0064 & \textbf{0.0427} & 1.3111 \\
H-Scalar
& 0.0051 & 0.4973 & 0.3335
& 0.0048 & 0.0746 & 1.0720 \\
H-Sample
& 0.0172 & \textbf{0.0454} & 1.2844
& 0.0087 & 0.2337 & 0.6267 \\
H-Class
& 0.0155 & 0.1071 & 0.9264
& 0.0033 & 0.2195 & 0.6505 \\
EST-QCNN
& 0.0181 & 0.0534 & 1.2130
& 0.0335 & \textbf{0.0030} & 2.8794 \\
EST-RUQNN
& 0.0177 & 0.0687 & 1.1060
& 0.0254 & \textbf{0.0143} & 1.8526 \\
EST-QCQ-CNN
& 0.0078 & 0.1570 & 0.7778
& 0.0234 & \textbf{0.0064} & 2.3365 \\
\midrule
ResNet-18
& 0.0122 & 0.1915 & 0.7020
& -0.0007 & 0.7521 & -0.1513 \\
ResNet-50
& -0.0043 & 0.4485 & -0.3754
& -0.0042 & \textbf{0.0042} & -2.6205 \\
EfficientNet-B0
& -0.0056 & 0.4068 & -0.4141
& 0.0003 & 0.9225 & 0.0463 \\
MobileNetV3-Small
& 0.0292 & \textbf{0.0125} & 1.9288
& -0.0011 & 0.6147 & -0.2438 \\
\bottomrule
\end{tabular}%
}
\end{table*}

\begin{figure*}[!t]

    \centering
    \includegraphics[width=\textwidth]{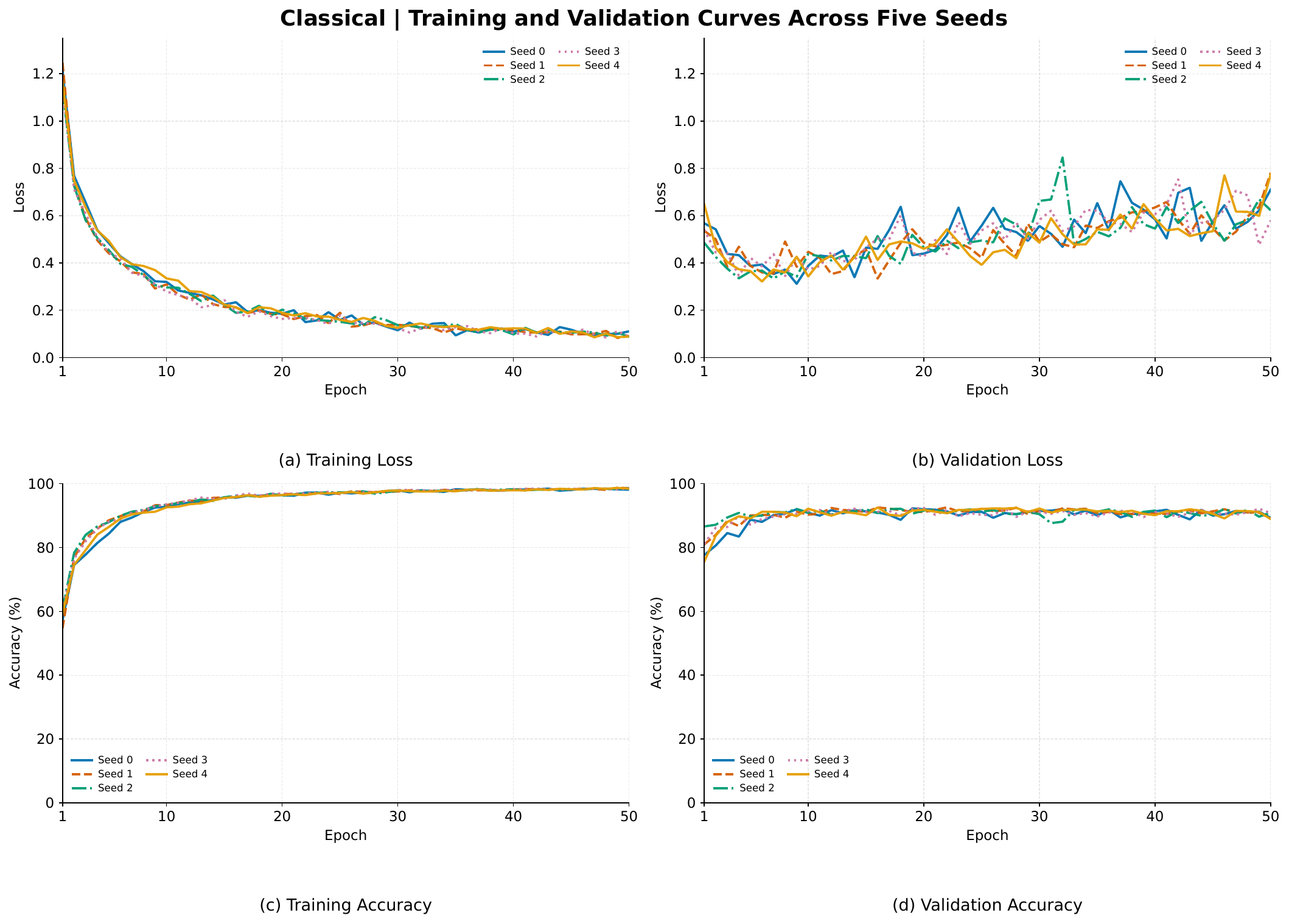}
    \caption{Training and validation curves for Classical on PathMNIST across five random seeds. (a) Training loss; (b) validation loss;
    (c) training accuracy; and (d) validation accuracy. Colors and line styles denote seeds 0-4. Each curve
    represents an independent 50-epoch training run, shown without smoothing.}
    \label{fig:classical_training}
\end{figure*}

\begin{figure*}[!t]

    \centering
    \includegraphics[width=\textwidth]{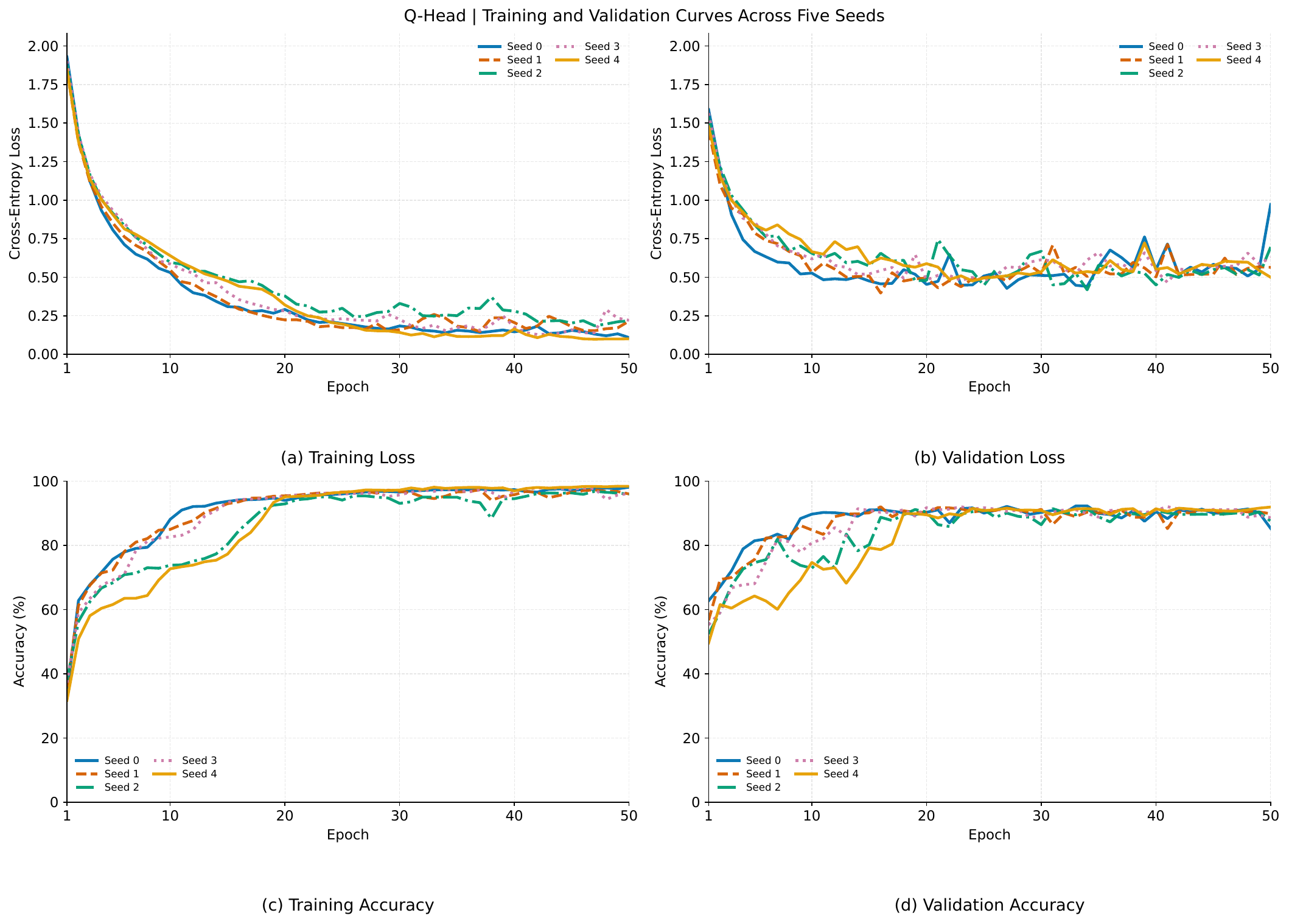}
    \caption{Training and validation curves for Q-Head on PathMNIST across five random seeds. (a) Training loss; (b) validation loss;
    (c) training accuracy; and (d) validation accuracy. Colors and line styles denote seeds 0-4. Each curve
    represents an independent 50-epoch training run, shown without smoothing.}
    \label{fig:qhead_training}
\end{figure*}

\begin{figure*}[!t]
    \centering
    \includegraphics[width=\textwidth]{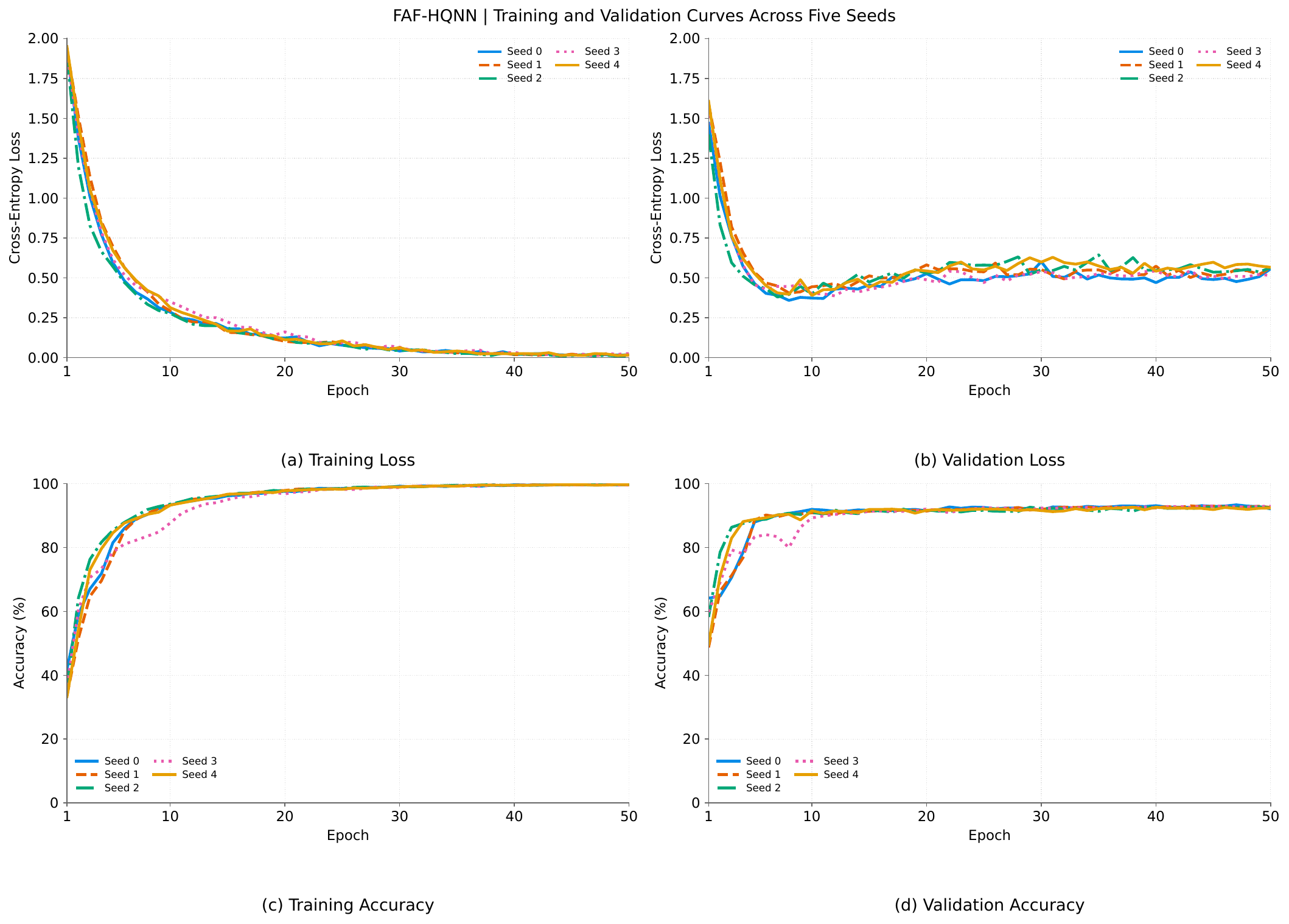}
    \caption{Training and validation curves for FAF-HQNN on PathMNIST across five random seeds.  (a) Training loss; (b) validation loss; (c) training accuracy; and (d) validation accuracy. Colors and line styles denote seeds 0-4. Each curve represents an independent 50-epoch training run, shown without smoothing.}
    \label{fig:faf_hqnn_training}
\end{figure*}

\paragraph{Training and Validation Dynamics.}
Figures~\ref{fig:classical_training}, \ref{fig:qhead_training}, and~\ref{fig:faf_hqnn_training} show the training and validation curves of Classical, Q-Head, and FAF-HQNN over 50 epochs across five random seeds. All three models exhibit an overall decrease in training loss and an increase in training accuracy. Under the evaluated training settings, FAF-HQNN shows more closely aligned training curves across seeds during early-to-middle training than Q-Head. Relative to Classical, its late-stage validation accuracy is visually slightly higher, and its validation loss exhibits less pronounced fluctuations. Nevertheless, validation loss rises after an initial decline while training loss continues to decrease, particularly for Classical and FAF-HQNN, suggesting overfitting despite relatively stable validation accuracy. Test performance is evaluated using the checkpoint with the highest validation accuracy in each run, rather than the final checkpoint.

\paragraph{Inference-Time Quantum-Logit Interventions.}

With trained parameters and the original gates fixed, zeroing the quantum logits (Q-Zero) reduces accuracy by 0.0099 \textpm{} 0.0003 relative to the full model, whereas permuting quantum logit vectors across test samples without recomputing the gates (Q-Shuffle) yields a much smaller mean reduction of 0.0002 \textpm{} 0.0004 (Table~\ref{tab:quantum_interventions}). This contrast indicates that the quantum-output term contributes to predictions under fixed gating, but provides limited evidence that its sample-specific correspondence explains the fusion gains. Q-Zero retains most of the full model's accuracy gain over C-Only while preserving class-dependent gating, which is absent from C-Only. These interventions therefore do not isolate the source of the fusion gains or the quantum branch's role in gate learning and joint optimization.

\begin{table}[t]
\centering
\caption{erformance of the full model, its internal branches (C-Only and Q-Only), and inference-time quantum-logit interventions on PathMNIST across five training seeds. Values are mean \textpm{} SD on the 0--1 scale; accuracy decreases are paired within-seed differences (Full minus each condition), computed before rounding. Q-Zero sets the quantum logits to zero; Q-Shuffle permutes complete quantum-output vectors across test samples. Both interventions retain the trained parameters and original gating coefficients, without retraining or gate recomputation.}
\label{tab:quantum_interventions}
\small
\renewcommand{\arraystretch}{1.2}
\setlength{\tabcolsep}{3pt}
\begin{tabularx}{\linewidth}{@{}lr>{\raggedleft\arraybackslash}X@{}}
\toprule
Condition & ACC & Accuracy decrease relative to Full \\
\midrule
Full & \textbf{0.8919} \textpm{} 0.0078 & --- \\
C-Only & 0.8170 \textpm{} 0.0577 & 0.0749 \textpm{} 0.0575 \\
Q-Only & 0.1102 \textpm{} 0.0262 & 0.7816 \textpm{} 0.0315 \\
Q-Zero & 0.8820 \textpm{} 0.0078 & 0.0099 \textpm{} 0.0003 \\
Q-Shuffle & 0.8917 \textpm{} 0.0076 & 0.0002 \textpm{} 0.0004 \\
\bottomrule
\end{tabularx}
\par\smallskip
\parbox{\linewidth}{\footnotesize
\raggedright
Note: For Q-Shuffle, results are first averaged over multiple random permutations within each training seed, and the reported mean and SD are computed across the five seed-level averages. C-Only and Q-Only are pre-fusion branches of the jointly trained FAF-HQNN, not the independently trained Classical and Q-Head baselines.
\par
}
\end{table}

\paragraph{Fusion-weight and sample-level prediction analyses.}

Table~\ref{tab:fusion_weight_statistics} summarizes the quantum fusion weights on the PathMNIST test set across five training seeds. The overall mean weight was 0.4505 \textpm{} 0.0982, and the proportion of weights below 0.05 was 0.0706 \textpm{} 0.0275. These statistics show that the quantum weights were not uniformly
suppressed across sample--output-class pairs, but nonzero weights alone do not establish a predictive contribution from the quantum branch. The mean weights assigned to the ground-truth and full-model-predicted classes were 0.1202 \textpm{} 0.0173 and 0.0865 \textpm{} 0.0146, respectively. These class-specific
statistics further characterize the learned weighting pattern, rather than directly measuring the utility of the quantum outputs.

\begin{table}[t]
\centering
\caption{Quantum fusion-weight statistics on the PathMNIST test set
across five training seeds. Quantum fusion weights are defined as
one minus the corresponding classical gating coefficients.
Within each seed, the overall mean weight and the proportion of
weights below 0.05 are computed over all test sample--output-class
pairs. The weights assigned to the ground-truth and
full-model-predicted classes are averaged over test samples.
Values are mean \textpm{} SD across the five seed-level statistics;
all quantities are on the 0--1 scale.}
\label{tab:fusion_weight_statistics}
\small
\renewcommand{\arraystretch}{1.2}
\setlength{\tabcolsep}{3pt}
\begin{tabularx}{\linewidth}{@{}>{\raggedright\arraybackslash}Xr@{}}
\toprule
Quantum fusion-weight statistic & Mean \textpm{} SD \\
\midrule
Overall mean weight
  & 0.4505 \textpm{} 0.0982 \\
Proportion of weights below 0.05
  & 0.0706 \textpm{} 0.0275 \\
Mean weight for the ground-truth class
  & 0.1202 \textpm{} 0.0173 \\
Mean weight for the full-model-predicted class
  & 0.0865 \textpm{} 0.0146 \\
\bottomrule
\end{tabularx}
\end{table}

Table~\ref{tab:branch_correction} summarizes sample-level prediction
events across the five training seeds, each evaluated on the same
7,180 test samples. Relative to C-Only, the full model corrected
errors on 8.501 \textpm{} 6.105\% of all test samples but introduced
errors on 1.017 \textpm{} 0.734\%. The resulting paired net accuracy
improvement, computed within each seed before rounding, was
7.485 \textpm{} 5.751 percentage points. On
7.050 \textpm{} 6.800\% of all test samples, both C-Only and Q-Only
were incorrect, whereas the full model was correct. Thus, fusion
can yield a correct prediction even when neither internal branch
assigns the highest logit to the ground-truth class.

These events describe the prediction behavior of the jointly
trained fusion model; they do not, by themselves, establish that
the corrections are attributable to sample-specific information
from the quantum branch. The fixed-gate interventions in
Table~\ref{tab:quantum_interventions} complement this analysis by
assessing the role of the quantum-output term in the trained fusion
computation. They do not resolve potential contributions through
gating learning or joint optimization.

\begin{table}[t]
\centering
\caption{Sample-level prediction analysis for the full model and
its internal branches on PathMNIST across five training seeds.
C-Only and Q-Only denote predictions from the pre-fusion outputs
of the jointly trained classical and quantum branches, respectively;
Full denotes the fused model prediction. All event percentages
use the same 7,180 test samples per seed as the denominator.
Values are mean \textpm{} SD across the five seed-level statistics.
Net accuracy improvement over C-Only is computed within each seed
as the correction rate (C-Only incorrect, Full correct) minus the
introduced-error rate (C-Only correct, Full incorrect), before
rounding, and is reported in percentage points.}
\label{tab:branch_correction}
\small
\renewcommand{\arraystretch}{1.2}
\setlength{\tabcolsep}{3pt}
\begin{tabularx}{\linewidth}{@{}>{\raggedright\arraybackslash}Xr@{}}
\toprule
Prediction event or statistic & Mean \textpm{} SD \\
\midrule
C-Only incorrect, Q-Only correct (\%)
  & 2.440 \textpm{} 2.849 \\
C-Only incorrect, Full correct (\%)
  & 8.501 \textpm{} 6.105 \\
C-Only correct, Full incorrect (\%)
  & 1.017 \textpm{} 0.734 \\
C-Only incorrect, Q-Only correct, Full correct (\%)
  & 1.451 \textpm{} 2.394 \\
C-Only incorrect, Q-Only incorrect, Full correct (\%)
  & 7.050 \textpm{} 6.800 \\
C-Only--Q-Only prediction disagreement (\%)
  & 90.925 \textpm{} 5.446 \\
\midrule
Net accuracy improvement over C-Only (percentage points)
  & 7.485 \textpm{} 5.751 \\
\bottomrule
\end{tabularx}
\end{table}

\subsection{Parameter and Computational Cost Analysis}

Table~\ref{tab:fusion_cost} summarizes trainable parameter counts and measured training and inference times for FAF-HQNN and the comparison models.  Total Params includes all trainable parameters in the classical backbone, feature-processing modules, quantum circuits, and readout and fusion modules, where applicable. Quantum Params and Fusion Params report the corresponding subsets. 

Experiments were conducted using Windows Server 2019, Python 3.12.3, an Intel CPU with 64 logical processors, eight
PyTorch threads, and PennyLane's \texttt{default.qubit} simulator. Timing was performed using \texttt{time.perf\_counter()}, with
CUDA synchronization where applicable. Times are averaged across five seeds. Training time is the per-epoch mean over 50 epochs per seed, including data loading and training but excluding validation, checkpoint saving, and testing. Inference time is measured per image for full-model forward passes on the clean test set at batch size 32, using a consistent checkpoint-selection
strategy and excluding data loading, preprocessing, device transfer, and metric computation.

Across the comparison models, FAF-HQNN's training time is higher than those of Classical and C-AF, close to those of
Q-Head and H-Fix, and lower than those of all three task-adapted hybrid quantum--classical baselines. Its inference time is higher than that of EST-QCNN but lower than those of EST-RUQNN and EST-QCQ-CNN. Among the external CNN baselines, FAF-HQNN trains more slowly than ResNet-18 and MobileNetV3-Small but faster than ResNet-50 and EfficientNet-B0. Its inference time is close to that of ResNet-50 and higher than those of the other three CNN baselines.

Overall, FAF-HQNN adds only 506 trainable parameters relative to Classical, while incurring measured training and inference
overheads of 26.41\% and 48.74\%, respectively. These results highlight that a small parameter increase does not necessarily
imply a small runtime overhead. 

As shown in Table~\ref{tab:fusion_cost}, adding the quantum prediction branch to classical baseline
(Classical to H-Fix) introduces 49 trainable parameters (4 quantum-circuit and 45 classical readout parameters),
increasing training and inference times by 12.78 s/epoch (24.71\%) and 0.75 ms/image (63.03\%), respectively.
Replacing fixed averaging with adaptive fusion (H-Fix to FAF-HQNN) adds 457 classical parameters and
increases training time by 0.88 s/epoch (1.36\%), without increasing the qubit count or circuit depth. The measured full-model forward-pass time changes from 1.94 to 1.77 ms/image. Overall, FAF-HQNN adds 506 parameters relative to Classical,
with training and inference overheads of 26.41\% and 48.74\%, respectively. These runtimes reflect classical circuit
simulation rather than execution on quantum hardware.

\begin{table}[t]
\centering
\caption{Trainable parameters and computational costs of FAF-HQNN, internal variants, task-adapted hybrid quantum--classical baselines, and external CNN baselines.}
\label{tab:fusion_cost}
\small
\renewcommand{\arraystretch}{1.2}
\setlength{\tabcolsep}{2pt}
\begin{tabular*}{\linewidth}{@{\extracolsep{\fill}}lrrrrr@{}}
\toprule
\textbf{Method} &
\shortstack{\textbf{Total}\\\textbf{Params}}  &
\shortstack{\textbf{Quantum}\\\textbf{Params}}  &
\shortstack{\textbf{Fusion }\\\textbf{Params}}  &
\shortstack{\textbf{Training Time}\\\textbf{(s/epoch)}} &
\shortstack{\textbf{Inference Time}\\\textbf{(ms/image)}} \\
\midrule
Classical
& 11,332,045 & 0 & 0 & 51.72 & 1.19\\
Q-Head
& 11,329,397 & 4 & 0 & 65.30 & 1.52\\
C-AF
& 11,333,567 & 0 & 457 & 59.39 &1.66\\
H-Fix
& 11,332,094 & 4 & 0 & 64.50 & 1.94\\
H-Scalar
& 11,332,095 & 4 & 1 & 72.78 & 1.78\\
H-Sample
& 11,332,287 & 4 & 193 & 61.69 & 1.60\\
H-Class
& 11,332,103 & 4 & 9 & 67.69 & 1.64 \\
EST-QCNN
& 11,329,421 & 28 & 0 & 75.46 & 1.17\\
EST-RUQNN
& 11,329,397 & 4 & 0 & 75.19 & 2.09\\
EST-QCQ-CNN
& 11,331,405 & 12 & 0 & 92.73 & 1.99\\
\midrule
\textbf{FAF-HQNN}
& 11,332,551 & 4 & 457 & 65.38 & 1.77 \\
\midrule
ResNet-18
& 11,181,129 & -- & -- & 40.90 & 0.72\\
ResNet-50
& 23,526,473 & -- & -- & 109.59 &1.78\\
EfficientNet-B0
& 4,019,077 & -- & -- & 92.67 & 1.03 \\
MobileNetV3-Small
& 1,527,081 & -- & -- & 41.77 & 0.54\\
\bottomrule
\end{tabular*}
\end{table}

\subsection{Analysis of the Quantum Module}
To better understand the role of the quantum component in FAF-HQNN, we analyze how the design of the variational quantum circuit (VQC) affects the overall performance of the model. Specifically, we examine three factors: circuit layout, measurement basis, and circuit depth.

\subsubsection{Effect of Circuit Layout and Measurement Basis}

\begin{figure*}[t]
\centering
\includegraphics[scale=0.23]{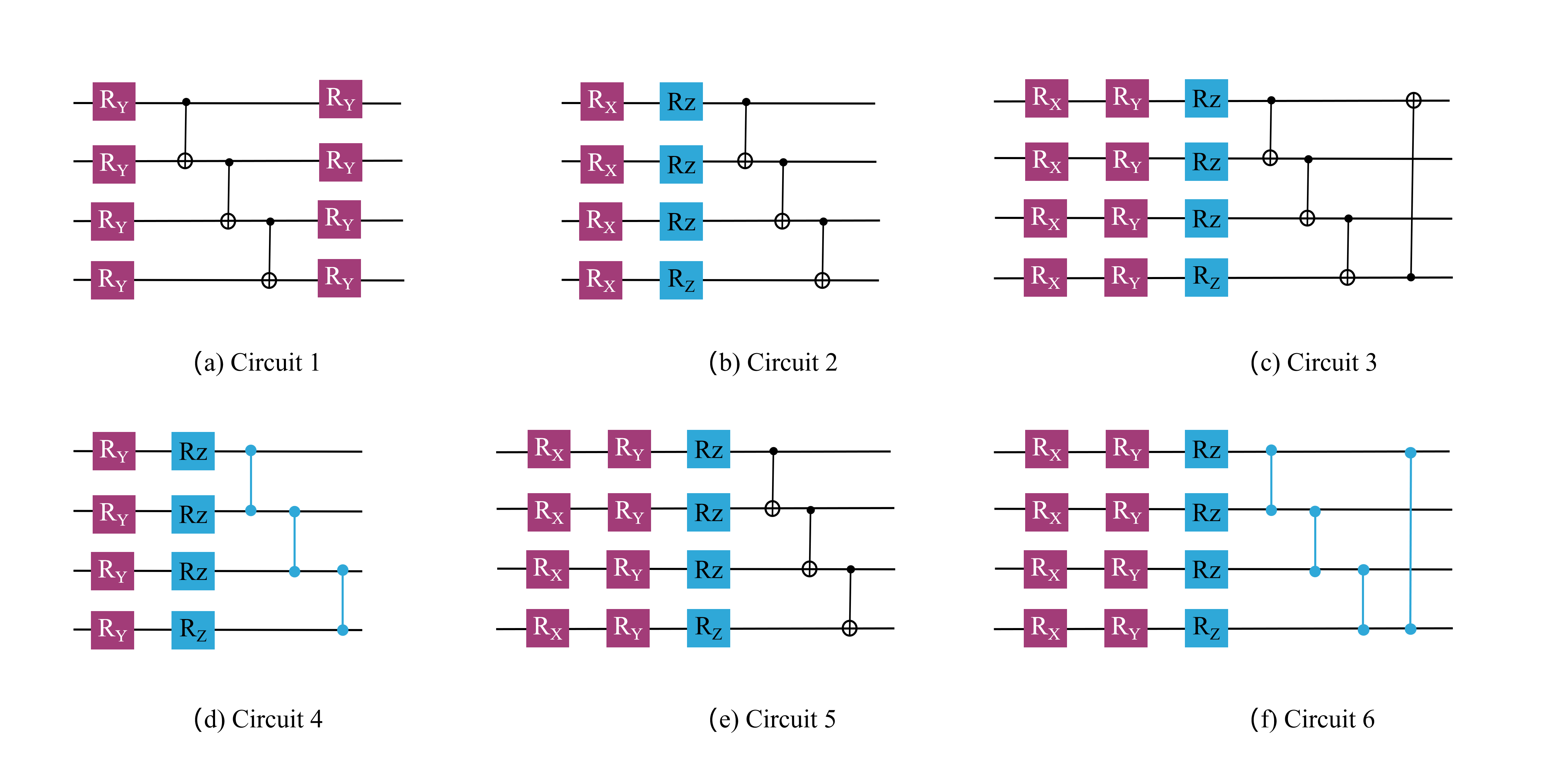}
\caption{Six variational quantum circuit layouts considered in this work.}
\label{fig:circuit_layouts}
\end{figure*}

We compare six VQC layouts that differ in their combinations of single-qubit rotation gates and entangling operations, as illustrated in Fig.~\ref{fig:circuit_layouts}. For each layout, we first consider three single-basis measurement settings, namely Pauli-X, Pauli-Y, and Pauli-Z. In each case, the quantum features are given by the expectation values of the corresponding Pauli observable for all qubits.  Unless otherwise specified, all other model components and training settings are kept fixed so that the comparison isolates the effect of VQC design choices. The resulting classification accuracies on PathMNIST are summarized in Table~\ref{tab:circuit_basis}.

The results show that the performance of FAF-HQNN depends on both the circuit layout and the measurement setting. Although the overall variation is moderate, the observed differences indicate that the design of the quantum branch can affect the final hybrid performance. No single measurement setting consistently performs best across all circuit layouts, suggesting that the usefulness of the quantum readout depends on its interaction with the underlying circuit structure. Among all configurations, the best result is obtained with Circuit 1 under Pauli-X measurement, achieving an ACC of $0.8919 \pm 0.0078$.

\begin{table}[t]
\centering
\caption{Classification accuracy of FAF-HQNN on PathMNIST under different VQC layouts and measurement settings. Values are reported as mean $\pm$ standard deviation over five runs.}
\label{tab:circuit_basis}
\setlength{\tabcolsep}{3pt}
\renewcommand{\arraystretch}{3}
\begin{tabular}{lcccccc}
\toprule
\textbf{Measurement} & \textbf{Circuit 1} & \textbf{Circuit 2} & \textbf{Circuit 3} & \textbf{Circuit 4} & \textbf{Circuit 5} & \textbf{Circuit 6} \\
\midrule
Pauli-X   & \makecell{\textbf{0.8919} \\ \textbf{$\pm$0.0078}} & \makecell{0.8801 \\ $\pm$ 0.0121} & \makecell{0.8917\\ $\pm$
0.0070 } & \makecell{0.8902\\ $\pm$0.0052} & \makecell{0.8869 \\ $\pm$0.0046} & \makecell{0.8627\\ $\pm$0.0109} \\

Pauli-Y   & \makecell{0.8769 \\ $\pm$0.0165} & \makecell{0.8854  \\ $\pm$0.0051} & \makecell{0.8873\\ $\pm$0.0037 } & \makecell{0.8872\\ $\pm$0.0069} & \makecell{0.8869\\ $\pm$0.0047} & \makecell{0.8845\\ $\pm$0.0035} \\

Pauli-Z   & \makecell{0.8801 \\ $\pm$0.0121} & \makecell{0.8901 \\ $\pm$0.0069} & \makecell{0.8858\\ $\pm$0.0012 } & \makecell{0.8877\\ $\pm$0.0062} & \makecell{0.8865\\ $\pm$0.0044 } & \makecell{0.8875\\ $\pm$0.0049} \\
\bottomrule
\end{tabular}
\end{table}
\subsubsection{Effect of Circuit Depth}

We evaluate FAF-HQNN with one, two, and three VQC layers to assess the effect of circuit depth on PathMNIST. As shown in Table~\ref{tab:circuit_depth}, the single-layer configuration achieves the highest mean performance across all four reported metrics. Based on the reported means, the two- and three-layer configurations have lower ACC by 0.38 and 0.24 percentage points, respectively, relative to the single-layer configuration. Both deeper configurations have a mean AUC of 0.9790 at the reported precision. The three-layer configuration achieves slightly higher mean ACC, precision, and F1-score than the two-layer configuration, but neither deeper configuration exceeds the single-layer configuration on any reported metric.

These descriptive results favor the single-layer configuration in the evaluated setting: increasing circuit depth from one to two or three layers does not improve mean predictive performance. However, the reported means and standard deviations alone do not establish statistically significant differences between the configurations. The findings therefore do not rule out potential benefits from deeper circuits under other training or optimization settings.

\begin{table*}[t]
\centering
\caption{Performance of FAF-HQNN on PathMNIST with different numbers of VQC layers. Values are mean \textpm{} SD across five runs, with all metrics reported on the 0--1 scale. Boldface indicates the highest mean for each metric.}
\label{tab:circuit_depth}
\begingroup
\footnotesize
\renewcommand{\arraystretch}{1.2}
\setlength{\tabcolsep}{3pt}
\begin{tabular*}{\textwidth}{@{\extracolsep{\fill}}lcccc@{}}
\toprule
\textbf{VQC layers}
& \textbf{ACC}
& \textbf{Precision}
& \textbf{F1-score}
& \textbf{AUC} \\
\midrule
1
& \textbf{0.8919} \textpm{} 0.0078
& \textbf{0.8923} \textpm{} 0.0075
& \textbf{0.8907} \textpm{} 0.0077
& \textbf{0.9825} \textpm{} 0.0027 \\
2
& 0.8881 \textpm{} 0.0056
& 0.8903 \textpm{} 0.0040
& 0.8876 \textpm{} 0.0045
& 0.9790 \textpm{} 0.0036 \\
3
& 0.8895 \textpm{} 0.0063
& 0.8914 \textpm{} 0.0053
& 0.8888 \textpm{} 0.0061
& 0.9790 \textpm{} 0.0032 \\
\bottomrule
\end{tabular*}
\endgroup
\end{table*}

\subsection{Robustness to Gaussian and Poisson Noise on PathMNIST}

We assess robustness to Gaussian and Poisson noise on PathMNIST using models trained on clean images. Clean-test ACC serves as a reference for interpreting performance under corruption. The analysis reports ACC at each noise level and paired comparisons
of aggregate corrupted-test ACC. Comparisons at individual noise levels are descriptive unless accompanied by a statistical test.

\subsubsection{Gaussian Noise}
Table~\ref{tab:path_gaussian_acc} reports ACC under Gaussian noise with standard deviations from 0.00 to 0.03, where 0.00 denotes the clean reference. Mean ACC decreases monotonically as noise intensity increases for every evaluated method.

FAF-HQNN achieves higher mean ACC than all three external hybrid quantum--classical baselines at every evaluated noise level.
It also outperforms all seven internal variants, except at a standard deviation of 0.02, where Classical achieves a slightly higher mean ACC. Overall, FAF-HQNN ranks among the top three methods at every evaluated level. EfficientNet-B0 achieves the highest mean ACC on the clean test set and at the lowest nonzero noise level, whereas ResNet-18 leads at a standard deviation of 0.02 and FAF-HQNN leads at 0.01 and 0.03.

\begin{figure*}[t]
    \centering
    \begin{subfigure}[t]{0.49\textwidth}
        \centering
        \includegraphics[width=\linewidth]{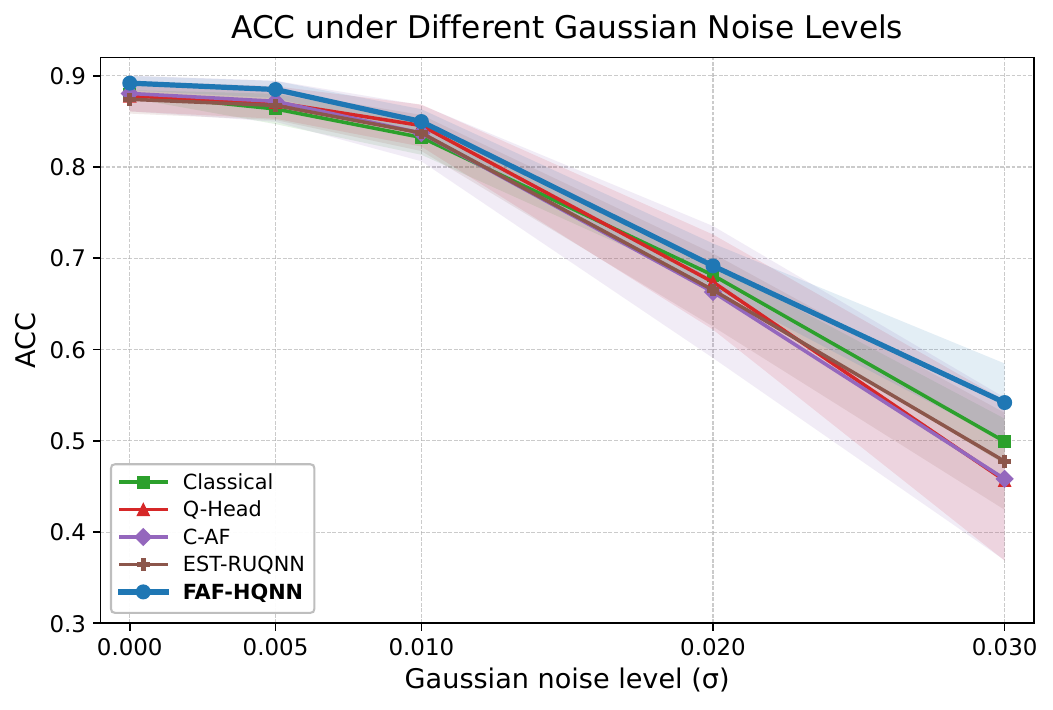}
        \caption{Gaussian noise.}
        \label{fig:path_gaussian}
    \end{subfigure}
    \hfill
    \begin{subfigure}[t]{0.49\textwidth}
        \centering
        \includegraphics[width=\linewidth]{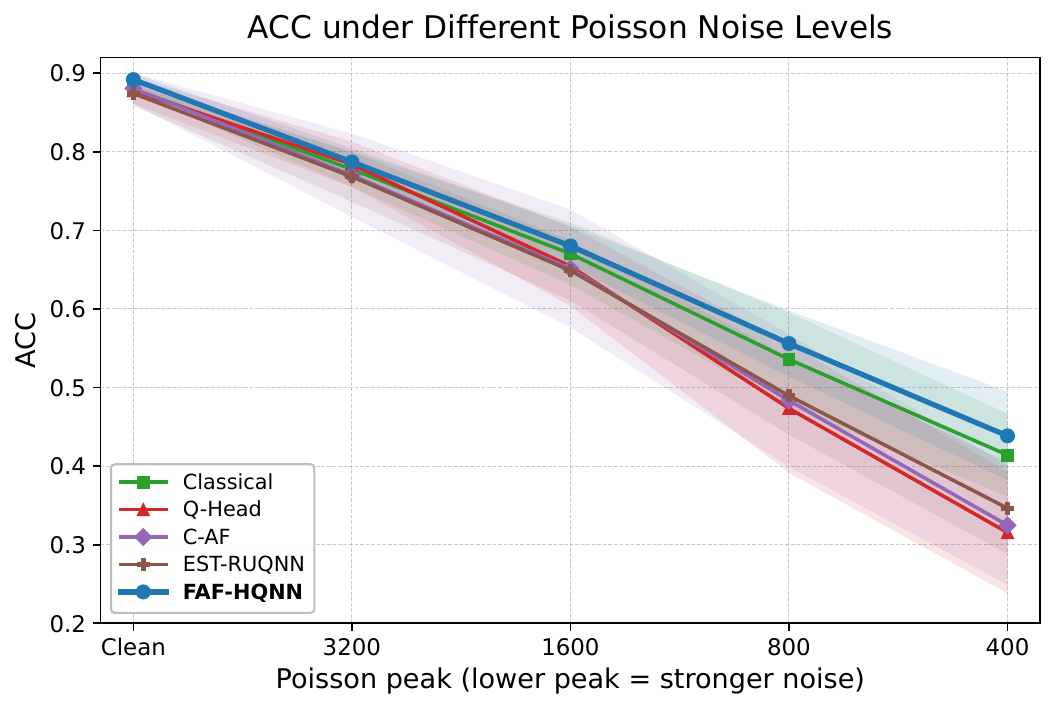}
        \caption{Poisson noise.}
        \label{fig:path_sp}
    \end{subfigure}
    \caption{Mean ACC of FAF-HQNN and selected baselines on PathMNIST under (a) Gaussian noise and (b) Poisson noise.
    Gaussian noise intensity increases with the standard deviation, whereas Poisson noise intensity increases
    as the peak parameter decreases. Only selected methods are displayed for readability; complete results for all
    evaluated methods are reported in Tables~\ref{tab:path_gaussian_acc} and~\ref{tab:path_poisson_acc}, respectively. Shaded bands    
    indicate \textpm{}1 SD across five training seeds.}
    \label{fig:path_noise}
\end{figure*}

These results show that FAF-HQNN maintains competitive ACC under the evaluated Gaussian noise conditions, achieving the highest
mean ACC at two nonzero noise levels, including the strongest corruption setting. These comparisons are based on mean ACC and do not establish statistical significance. Fig.~\ref{fig:path_gaussian} illustrates the mean ACC trends of FAF-HQNN and selected baselines; comparisons across all methods are provided in Table~\ref{tab:path_gaussian_acc}.

\begin{table*}[t]
\centering
\caption{ACC on PathMNIST under Gaussian noise (standard deviation \ensuremath{\sigma}; \ensuremath{\sigma = 0} denotes clean data). Results are mean \ensuremath{\pm} SD across five runs. Bold values indicate the highest mean ACC at each noise level.}
\label{tab:path_gaussian_acc}
\scriptsize
\setlength{\tabcolsep}{3pt}
\renewcommand{\arraystretch}{1.2}
\resizebox{\textwidth}{!}{%
\begin{tabular}{lccccc}
\toprule
\textbf{Method}
& \ensuremath{\boldsymbol{\sigma=0.00}}
& \ensuremath{\boldsymbol{\sigma=0.005}}
& \ensuremath{\boldsymbol{\sigma=0.01}}
& \ensuremath{\boldsymbol{\sigma=0.02}}
& \ensuremath{\boldsymbol{\sigma=0.03}} \\
\midrule
Classical
& \ensuremath{0.8800\pm0.0043}
& \ensuremath{0.8704\pm0.0068}
& \ensuremath{0.8370\pm0.0168}
& \ensuremath{0.6920\pm0.0376}
& \ensuremath{0.5299\pm0.0653} \\
Q-Head
& \ensuremath{0.8777\pm0.0168}
& \ensuremath{0.8707\pm0.0178}
& \ensuremath{0.8454\pm0.0230}
& \ensuremath{0.6739\pm0.0521}
& \ensuremath{0.4565\pm0.0878} \\
C-AF
& \ensuremath{0.8806\pm0.0196}
& \ensuremath{0.8720\pm0.0225}
& \ensuremath{0.8371\pm0.0308}
& \ensuremath{0.6630\pm0.0723}
& \ensuremath{0.4582\pm0.0891} \\
H-Fix
& \ensuremath{0.8872\pm0.0097}
& \ensuremath{0.8785\pm0.0103}
& \ensuremath{0.8385\pm0.0151}
& \ensuremath{0.6863\pm0.0337}
& \ensuremath{0.5360\pm0.0641} \\
H-Scalar
& \ensuremath{0.8868\pm0.0085}
& \ensuremath{0.8776\pm0.0093}
& \ensuremath{0.8417\pm0.0142}
& \ensuremath{0.6360\pm0.0588}
& \ensuremath{0.4063\pm0.1233} \\

H-Sample
& \ensuremath{0.8747\pm0.0187}
& \ensuremath{0.8695\pm0.0206}
& \ensuremath{0.8429\pm0.0260}
& \ensuremath{0.6769\pm0.0496}
& \ensuremath{0.4609\pm0.1073} \\

H-Class
& \ensuremath{0.8763\pm0.0209}
& \ensuremath{0.8690\pm0.0196}
& \ensuremath{0.8412\pm0.0226}
& \ensuremath{0.6888\pm0.0733}
& \ensuremath{0.4847\pm0.1024} \\

EST-QCNN
& \ensuremath{0.8738\pm0.0073}
& \ensuremath{0.8668\pm0.0053}
& \ensuremath{0.8370\pm0.0112}
& \ensuremath{0.6847\pm0.0507}
& \ensuremath{0.4944\pm0.0897} \\

EST-RUQNN
& \ensuremath{0.8742\pm0.0157}
& \ensuremath{0.8675\pm0.0160}
& \ensuremath{0.8373\pm0.0198}
& \ensuremath{0.6652\pm0.0400}
& \ensuremath{0.4775\pm0.0531} \\

EST-QCQ-CNN
& \ensuremath{0.8840\pm0.0071}
& \ensuremath{0.8729\pm0.0098}
& \ensuremath{0.8273\pm0.0176}
& \ensuremath{0.6301\pm0.0432}
& \ensuremath{0.4452\pm0.0640} \\
\midrule
\textbf{FAF-HQNN}
& \ensuremath{0.8919\pm0.0078}
& \ensuremath{0.8849\pm0.0094}
& \ensuremath{0.8498\pm0.0135}
& \ensuremath{0.6916\pm0.0246}
& \ensuremath{0.5419\pm0.0428} \\
\midrule
ResNet-18
& \ensuremath{0.8797\pm0.0104}
& \ensuremath{0.8722\pm0.0146}
& \ensuremath{0.8415\pm0.0271}
& \ensuremath{\boldsymbol{0.6967\pm0.0469}}
& \ensuremath{0.5236\pm0.0558} \\
ResNet-50
& \ensuremath{0.8961\pm0.0090}
& \ensuremath{0.8777\pm0.0100}
& \ensuremath{0.8118\pm0.0080}
& \ensuremath{0.6304\pm0.0167}
& \ensuremath{0.4728\pm0.0404} \\

EfficientNet-B0
& \ensuremath{\boldsymbol{0.8974\pm0.0083}}
& \ensuremath{\boldsymbol{0.8856\pm0.0063}}
& \ensuremath{0.8365\pm0.0171}
& \ensuremath{0.6186\pm0.0409}
& \ensuremath{0.2569\pm0.0406} \\

MobileNetV3-Small
& \ensuremath{0.8627\pm0.0098}
& \ensuremath{0.8276\pm0.0161}
& \ensuremath{0.7532\pm0.0273}
& \ensuremath{0.5857\pm0.0648}
& \ensuremath{0.4570\pm0.0768} \\
\bottomrule
\end{tabular}%
}
\end{table*}

\subsubsection{Poisson Noise}
Table~\ref{tab:path_poisson_acc} reports classification accuracy (ACC) on PathMNIST under Poisson noise at different peak values,
where lower peaks indicate stronger corruption. Mean ACC decreases monotonically with increasing corruption severity
for all methods. FAF-HQNN achieves higher mean ACC than all seven internal comparison variants and three external hybrid
quantum--classical baselines across the evaluated noise settings, except for H-Class at peak 3200.

Among the external CNN baselines, ResNet-18 achieves the highest mean ACC at peaks 3200, 1600, and 800, whereas MobileNetV3-Small performs best at peak 400. FAF-HQNN performs nearly identically to ResNet-18 under milder corruption, with mean ACC values of 0.7871 versus 0.7869 at peak 3200 and 0.6802 versus 0.6831 at peak 1600. Under stronger corruption, FAF-HQNN achieves the highest mean ACC among all evaluated methods, reaching 0.5560 at peak 800 and 0.4385 at peak 400. These values exceed those of the strongest external CNN at each respective level by 2.27 and 5.10 percentage points. Fig.~\ref{fig:path_sp} illustrates the performance trends for selected methods.

\begin{table*}[t]
\centering
\caption{ACC on PathMNIST under clean conditions and Poisson noise (lower peak values indicate stronger corruption). Results are mean \textpm{} sample standard deviation over five independent training runs; bold values denote the highest mean ACC per column. Models include FAF-HQNN, seven internal variants, three external hybrid quantum--classical baselines, and four external CNN baselines.}

\label{tab:path_poisson_acc}
\scriptsize
\setlength{\tabcolsep}{3.0pt}
\renewcommand{\arraystretch}{1.2}
\resizebox{\textwidth}{!}{%
\begin{tabular}{lccccc}
\toprule
\textbf{Method}
& \textbf{Clean}
& \textbf{Peak=3200}
& \textbf{Peak=1600}
& \textbf{Peak=800}
& \textbf{Peak=400} \\
\midrule

Classical
& 0.8800 \textpm{} 0.0043
& 0.7776 \textpm{} 0.0225
& 0.6703 \textpm{} 0.0395
& 0.5359 \textpm{} 0.0595
& 0.4141 \textpm{} 0.0525 \\

Q-Head
& \ensuremath{0.8777\pm0.0168}
& \ensuremath{0.7840\pm0.0289}
& \ensuremath{0.6543\pm0.0503}
& \ensuremath{0.4735\pm0.0820}
& \ensuremath{0.3155\pm0.0773} \\

C-AF
& \ensuremath{0.8806\pm0.0196}
& \ensuremath{0.7702\pm0.0530}
& \ensuremath{0.6514\pm0.0750}
& \ensuremath{0.4835\pm0.0842}
& \ensuremath{0.3248\pm0.0759} \\

H-Fix
& \ensuremath{0.8872\pm0.0097}
& \ensuremath{0.7761\pm0.0225}
& \ensuremath{0.6762\pm0.0360}
& \ensuremath{0.5500\pm0.0593}
& \ensuremath{0.4190\pm0.0847} \\

H-Scalar
& \ensuremath{0.8868\pm0.0085}
& \ensuremath{0.7622\pm0.0291}
& \ensuremath{0.6149\pm0.0634}
& \ensuremath{0.4341\pm0.0891}
& \ensuremath{0.2873\pm0.1061} \\

H-Sample
& \ensuremath{0.8747\pm0.0187}
& \ensuremath{0.7836\pm0.0333}
& \ensuremath{0.6616\pm0.0479}
& \ensuremath{0.4909\pm0.0693}
& \ensuremath{0.3367\pm0.1319} \\

H-Class
& \ensuremath{0.8763\pm0.0209}
& \ensuremath{0.7882\pm0.0427}
& \ensuremath{0.6768\pm0.0812}
& \ensuremath{0.5094\pm0.1071}
& \ensuremath{0.3592\pm0.0940} \\

EST-QCNN
& \ensuremath{0.8738\pm0.0073}
& \ensuremath{0.7760\pm0.0386}
& \ensuremath{0.6646\pm0.0577}
& \ensuremath{0.4991\pm0.0866}
& \ensuremath{0.3482\pm0.0855} \\

EST-RUQNN
& \ensuremath{0.8742\pm0.0157}
& \ensuremath{0.7683\pm0.0317}
& \ensuremath{0.6489\pm0.0381}
& \ensuremath{0.4896\pm0.0497}
& \ensuremath{0.3464\pm0.0583} \\

EST-QCQ-CNN
& \ensuremath{0.8840\pm0.0071}
& \ensuremath{0.7482\pm0.0293}
& \ensuremath{0.6130\pm0.0420}
& \ensuremath{0.4575\pm0.0466}
& \ensuremath{0.3337\pm0.0615} \\

\midrule

\textbf{FAF-HQNN}
& \ensuremath{0.8919\pm0.0078}
& \ensuremath{0.7871\pm0.0173}
& \ensuremath{0.6802\pm0.0257}
& \ensuremath{\mathbf{0.5560\pm0.0421}}
& \ensuremath{\mathbf{0.4385\pm0.0563}} \\

\midrule

ResNet-18
& \ensuremath{0.8797\pm0.0104}
& \ensuremath{0.7869\pm0.0332}
& \ensuremath{\boldsymbol{0.6831\pm0.0481}}
& \ensuremath{0.5333\pm0.0479}
& \ensuremath{0.3794\pm0.0255} \\
ResNet-50
& \ensuremath{0.8961\pm0.0090}
& \ensuremath{0.7269\pm0.0105}
& \ensuremath{0.6136\pm0.0229}
& \ensuremath{0.4660\pm0.0453}
& \ensuremath{0.3332\pm0.0285} \\

EfficientNet-B0
& \ensuremath{\mathbf{0.8974\pm0.0083}}
& \ensuremath{0.7597\pm0.0322}
& \ensuremath{0.6247\pm0.0456}
& \ensuremath{0.3336\pm0.0313}
& \ensuremath{0.1600\pm0.0188} \\

MobileNetV3-Small
& \ensuremath{0.8627\pm0.0098}
& \ensuremath{0.6983\pm0.0398}
& \ensuremath{0.6013\pm0.0603}
& \ensuremath{0.4878\pm0.0740}
& \ensuremath{0.3875\pm0.0805} \\

\bottomrule
\end{tabular}%
}
\end{table*}

\subsubsection{Aggregate Corrupted-Test Accuracy}

For each seed and corruption type, aggregate corrupted-test ACC is calculated as the equally weighted mean ACC across
the four non-clean levels. Table~\ref{tab:path_robustness_Aggregate} reports these aggregate scores over five matched seeds,
together with paired mean differences, effect sizes, 95\% confidence intervals, and two-sided paired
\textit{t}-tests.

FAF-HQNN achieves higher aggregate mean ACC than all seven internal comparison variants, all three external hybrid
quantum--classical baselines, and all four external CNN baselines under both Gaussian and Poisson noise.

At the nominal 0.05 level, the paired tests favor FAF-HQNN over H-Scalar, ResNet-50, and EfficientNet-B0 under both
corruption types, as well as MobileNetV3-Small under Gaussian noise and EST-QCQ-CNN under Poisson noise. The remaining
comparisons do not reach nominal significance.


\begin{table*}[!tp]
\centering
\caption{Aggregate corrupted-test accuracy (ACC) and paired comparisons between FAF-HQNN and internal comparison variants,
external hybrid quantum--classical baselines, and external CNN baselines on PathMNIST under Gaussian and Poisson noise.
Results are based on five matched seeds.}
\label{tab:path_robustness_Aggregate}
\begingroup
\scriptsize
\setlength{\tabcolsep}{3pt}
\renewcommand{\arraystretch}{1.2}

\begin{tabular*}{\textwidth}{@{\extracolsep{\fill}}lccccc@{}}
\toprule
\textbf{Method}
& \textbf{Mean Corr. ACC}
& \textbf{ACC Diff.}
& \textbf{ACC \textit{d}\textsubscript{z}}
& \textbf{95\% CI of Diff.}
& \textbf{Raw \textit{p}} \\
\midrule
\multicolumn{6}{@{}l}{\textbf{(a) Gaussian noise}} \\
\midrule

Classical
& 0.7323 \textpm{} 0.0302
& 0.0097
& 0.373
& [\textminus{}0.0226, 0.0420]
& 0.4516 \\

Q-Head
& 0.7116 \textpm{} 0.0396
& 0.0304
& 0.616
& [\textminus{}0.0308, 0.0917]
& 0.2400 \\

C-AF
& 0.7076 \textpm{} 0.0510
& 0.0345
& 0.572
& [\textminus{}0.0404, 0.1094]
& 0.2705 \\

H-Fix
& 0.7348 \textpm{} 0.0268
& 0.0072
& 0.456
& [\textminus{}0.0124, 0.0268]
& 0.3660 \\

H-Scalar
& 0.6904 \textpm{} 0.0488
& 0.0517
& 1.571
& [0.0108, 0.0925]
& \textbf{0.0247} \\

H-Sample
& 0.7125 \textpm{} 0.0425
& 0.0295
& 0.720
& [\textminus{}0.0213, 0.0804]
& 0.1824 \\

H-Class
& 0.7209 \textpm{} 0.0485
& 0.0211
& 0.466
& [\textminus{}0.0351, 0.0774]
& 0.3560 \\

EST-QCNN
& 0.7207 \textpm{} 0.0379
& 0.0213
& 0.402
& [\textminus{}0.0444, 0.0871]
& 0.4188 \\

EST-RUQNN
& 0.7119 \textpm{} 0.0276
& 0.0302
& 1.026
& [\textminus{}0.0064, 0.0667]
& 0.0840 \\

EST-QCQ-CNN
& 0.6939 \textpm{} 0.0326
& 0.0482
& 1.228
& [\textminus{}0.0005, 0.0969]
& 0.0516 \\

ResNet-18
& 0.7335 \textpm{} 0.0301
& 0.0085
& 0.380
& [\textminus{}0.0194, 0.0364]
& 0.4432 \\

ResNet-50
& 0.6983 \textpm{} 0.0162
& 0.0437
& 3.376
& [0.0276, 0.0598]
& \textbf{0.0016} \\

EfficientNet-B0
& 0.6494 \textpm{} 0.0234
& 0.0927
& 2.622
& [0.0488, 0.1366]
& \textbf{0.0042} \\

MobileNetV3-Small
& 0.6559 \textpm{} 0.0456
& 0.0862
& 1.510
& [0.0153, 0.1571]
& \textbf{0.0279} \\

\textbf{FAF-HQNN}
& \textbf{0.7420 \textpm{} 0.0160}
& --
& --
& --
& -- \\

\midrule
\multicolumn{6}{@{}l}{\textbf{(b) Poisson noise}} \\
\midrule

Classical
& 0.5995 \textpm{} 0.0427
& 0.0160
& 0.361
& [\textminus{}0.0389, 0.0708]
& 0.4649 \\

Q-Head
& 0.5568 \textpm{} 0.0580
& 0.0586
& 0.999
& [\textminus{}0.0143, 0.1314]
& 0.0890 \\

C-AF
& 0.5575 \textpm{} 0.0711
& 0.0580
& 0.606
& [\textminus{}0.0608, 0.1768]
& 0.2469 \\

H-Fix
& 0.6078 \textpm{} 0.0471
& 0.0076
& 0.191
& [\textminus{}0.0418, 0.0570]
& 0.6916 \\

H-Scalar
& 0.5246 \textpm{} 0.0689
& 0.0908
& 1.936
& [0.0326, 0.1491]
& \textbf{0.0124} \\

H-Sample
& 0.5682 \textpm{} 0.0631
& 0.0472
& 0.559
& [\textminus{}0.0576, 0.1520]
& 0.2790 \\

H-Class
& 0.5834 \textpm{} 0.0798
& 0.0320
& 0.333
& [\textminus{}0.0874, 0.1515]
& 0.4980 \\

EST-QCNN
& 0.5720 \textpm{} 0.0634
& 0.0435
& 0.505
& [\textminus{}0.0636, 0.1505]
& 0.3228 \\

EST-RUQNN
& 0.5633 \textpm{} 0.0379
& 0.0521
& 1.132
& [\textminus{}0.0050, 0.1093]
& 0.0640 \\

EST-QCQ-CNN
& 0.5381 \textpm{} 0.0440
& 0.0773
& 1.397
& [0.0086, 0.1461]
& \textbf{0.0354} \\

ResNet-18
& 0.5957 \textpm{} 0.0334
& 0.0198
& 1.096
& [\textminus{}0.0026, 0.0422]
& 0.0704 \\

ResNet-50
& 0.5349 \textpm{} 0.0250
& 0.0805
& 4.423
& [0.0579, 0.1031]
& \textbf{0.0006} \\

EfficientNet-B0
& 0.4695 \textpm{} 0.0269
& 0.1459
& 3.690
& [0.0968, 0.1950]
& \textbf{0.0012} \\

MobileNetV3-Small
& 0.5437 \textpm{} 0.0625
& 0.0717
& 0.856
& [\textminus{}0.0323, 0.1757]
& 0.1281 \\

\textbf{FAF-HQNN}
& \textbf{0.6154 \textpm{} 0.0285}
& --
& --
& --
& -- \\

\bottomrule
\end{tabular*}

\par\smallskip
\noindent
\begin{minipage}{\textwidth}
\scriptsize
\raggedright
Note:
Mean Corr. ACC is the per-seed equally weighted average ACC across the four non-clean settings for each noise type, reported as mean \textpm{} sample SD over five matched seeds.
ACC Diff. is FAF-HQNN minus the baseline; positive values favor FAF-HQNN.
Paired-sample Cohen's \textit{d}\textsubscript{z} is calculated directly as the mean paired difference divided by the sample SD of the paired differences.
The 95\% CIs refer to mean paired differences, and raw \textit{p}-values are from two-sided paired \textit{t}-tests on the five seed-matched aggregate scores.
CIs and \textit{p}-values are unadjusted for multiple comparisons; bold \textit{p}-values indicate nominal significance at the
0.05 level.
These exploratory tests assess aggregate accuracy, not performance at individual corruption levels.
Nonsignificant results do not establish equivalence. Bold FAF-HQNN entries identify the reference model rather
than the column maximum; dashes denote inapplicable self-comparisons.
\end{minipage}
\par
\endgroup
\end{table*}

\subsection{Cross-Dataset Validation on BloodMNIST}
To examine whether the findings on PathMNIST also hold on another dataset, we train and evaluate all compared methods
on BloodMNIST, an eight-class peripheral blood cell image classification benchmark in MedMNIST. BloodMNIST contains
17,092 RGB images at a resolution of 28-by-28 pixels, with 11,959 training, 1,712 validation, and 3,421 test samples
in the official split.

We follow the same preprocessing pipeline used for PathMNIST: images are resized to 64-by-64 pixels, converted to tensors,
and normalized using the ImageNet mean and standard deviation. For BloodMNIST, we use the full official training set rather
than applying the per-class subsampling strategy used for PathMNIST, and retain the official validation and test splits.
The hyperparameter configurations used for PathMNIST are directly reused for the corresponding methods on BloodMNIST,
without additional BloodMNIST-specific hyperparameter tuning or a separate search. All methods are trained and evaluated using the same BloodMNIST data splits and five matched random seeds to ensure a controlled comparison.

\subsubsection{Performance under Clean Conditions}

Table~\ref{tab:blood_clean} summarizes clean-test performance on BloodMNIST. FAF-HQNN achieves higher mean scores across
all four metrics than all seven internal comparison variants and all three external hybrid quantum--classical baselines.
Across all evaluated methods, it achieves the highest mean ACC, precision, and F1-score, although its mean AUC is lower
than that of all four external CNN baselines.

Table~\ref{tab:blood_significance_clean} reports two-sided paired \textit{t}-tests and paired-sample effect sizes over
five matched seeds. At the nominal 0.05 level, FAF-HQNN achieves significant ACC gains over all internal comparison
variants and external hybrid quantum--classical baselines, as well as ResNet-50 and MobileNetV3-Small. Its ACC differences
relative to ResNet-18 and EfficientNet-B0 do not reach nominal significance. In contrast, its lower AUC relative to each of
the four external CNN baselines is nominally significant.

\begin{table*}[!tp]
\centering
\caption{Clean-test performance on BloodMNIST for FAF-HQNN, seven internal comparison variants, three external hybrid
quantum--classical baselines, and four external CNN baselines. Values are reported as mean \textpm{} sample standard deviation
over five matched seeds. Bold numerical entries indicate the highest mean in each metric column.}
\label{tab:blood_clean}
\small
\renewcommand{\arraystretch}{1.2}
\setlength{\tabcolsep}{3pt}

\resizebox{\linewidth}{!}{%
\begin{tabular}{@{}lcccc@{}}
\toprule
\textbf{Method}
& \textbf{ACC}
& \textbf{Precision}
& \textbf{F1-score}
& \textbf{AUC} \\
\midrule
Classical
& 0.9569 \ensuremath{\pm} 0.0030
& 0.9584 \ensuremath{\pm} 0.0034
& 0.9571 \ensuremath{\pm} 0.0031
& 0.9949 \ensuremath{\pm} 0.0007 \\
Q-Head
& 0.9517 \ensuremath{\pm} 0.0053
& 0.9533 \ensuremath{\pm} 0.0054
& 0.9519 \ensuremath{\pm} 0.0053
& 0.9864 \ensuremath{\pm} 0.0022 \\
C-AF
& 0.9569 \ensuremath{\pm} 0.0028
& 0.9573 \ensuremath{\pm} 0.0027
& 0.9568 \ensuremath{\pm} 0.0028
& 0.9950 \ensuremath{\pm} 0.0011 \\
H-Fix
& 0.9566 \ensuremath{\pm} 0.0028
& 0.9573 \ensuremath{\pm} 0.0029
& 0.9567 \ensuremath{\pm} 0.0029
& 0.9938 \ensuremath{\pm} 0.0013 \\
H-Scalar
& 0.9545 \ensuremath{\pm} 0.0038
& 0.9554 \ensuremath{\pm} 0.0035
& 0.9547 \ensuremath{\pm} 0.0037
& 0.9934 \ensuremath{\pm} 0.0010 \\
H-Sample
& 0.9555 \ensuremath{\pm} 0.0028
& 0.9566 \ensuremath{\pm} 0.0027
& 0.9558 \ensuremath{\pm} 0.0028
& 0.9947 \ensuremath{\pm} 0.0008 \\
H-Class
& 0.9551 \ensuremath{\pm} 0.0018
& 0.9561 \ensuremath{\pm} 0.0018
& 0.9553 \ensuremath{\pm} 0.0019
& 0.9948 \ensuremath{\pm} 0.0006 \\
EST-QCNN
& 0.9509 \ensuremath{\pm} 0.0042
& 0.9514 \ensuremath{\pm} 0.0043
& 0.9508 \ensuremath{\pm} 0.0043
& 0.9879 \ensuremath{\pm} 0.0010 \\
EST-RUQNN
& 0.9508 \ensuremath{\pm} 0.0024
& 0.9521 \ensuremath{\pm} 0.0020
& 0.9509 \ensuremath{\pm} 0.0024
& 0.9869 \ensuremath{\pm} 0.0012 \\
EST-QCQ-CNN
& 0.9539 \ensuremath{\pm} 0.0031
& 0.9545 \ensuremath{\pm} 0.0034
& 0.9539 \ensuremath{\pm} 0.0032
& 0.9857 \ensuremath{\pm} 0.0027 \\
\midrule
\textbf{FAF-HQNN}
& \textbf{0.9654} \ensuremath{\pm} 0.0024
& \textbf{0.9656} \ensuremath{\pm} 0.0023
& \textbf{0.9654} \ensuremath{\pm} 0.0024
& 0.9956 \ensuremath{\pm} 0.0004 \\
\midrule
ResNet-18
& 0.9619 \ensuremath{\pm} 0.0032
& 0.9623 \ensuremath{\pm} 0.0028
& 0.9620 \ensuremath{\pm} 0.0031
& 0.9974 \ensuremath{\pm} 0.0002 \\

ResNet-50
& 0.9614 \ensuremath{\pm} 0.0013
& 0.9619 \ensuremath{\pm} 0.0011
& 0.9615 \ensuremath{\pm} 0.0012
& 0.9972 \ensuremath{\pm} 0.0004 \\

EfficientNet-B0
& 0.9639 \ensuremath{\pm} 0.0020
& 0.9642 \ensuremath{\pm} 0.0019
& 0.9639 \ensuremath{\pm} 0.0020
& \textbf{0.9976} \ensuremath{\pm} 0.0001 \\
MobileNetV3-Small
& 0.9497 \ensuremath{\pm} 0.0043
& 0.9505 \ensuremath{\pm} 0.0043
& 0.9498 \ensuremath{\pm} 0.0044
& 0.9964 \ensuremath{\pm} 0.0005 \\
\bottomrule
\end{tabular}%
}
\end{table*}

\begin{table*}[!tp]
\centering
\caption{Paired statistical comparisons of clean-test ACC and AUC between FAF-HQNN and the comparison methods on BloodMNIST over five matched seeds.}
\label{tab:blood_significance_clean}
\small
\renewcommand{\arraystretch}{1.2}
\setlength{\tabcolsep}{3pt}
\resizebox{\linewidth}{!}{%
\begin{tabular}{@{}lcccccc@{}}
\toprule
\textbf{Baseline}
& \textbf{ACC Diff.}
& \textbf{ACC \ensuremath{p}}
& \textbf{ACC \ensuremath{d_z}}
& \textbf{AUC Diff.}
& \textbf{AUC \ensuremath{p}}
& \textbf{AUC \ensuremath{d_z}} \\
\midrule
Classical & 0.0085 & \textbf{0.0001} & 7.3822 & 0.0008 & \textbf{0.0296} & 1.4820 \\
Q-Head & 0.0137 & \textbf{0.0128} & 1.9180 & 0.0092 & \textbf{0.0009} & 3.9794 \\
C-AF & 0.0085 & \textbf{0.0035} & 2.7661 & 0.0006 & 0.3412 & 0.4827 \\
H-Fix & 0.0088 & \textbf{0.0040} & 2.6673 & 0.0019 & \textbf{0.0359} & 1.3906 \\
H-Scalar & 0.0109 & \textbf{0.0016} & 3.4108 & 0.0022 & \textbf{0.0038} & 2.6998 \\
H-Sample & 0.0099 & \textbf{0.0035} & 2.7538 & 0.0009 & 0.1493 & 0.7971 \\
H-Class & 0.0103 & \textbf{0.0019} & 3.2434 & 0.0008 & \textbf{0.0196} & 1.6868 \\
EST-QCNN & 0.0145 & \textbf{0.0008} & 4.0768 & 0.0077 & \textbf{4.56e-5} & 8.4774 \\
EST-RUQNN & 0.0146 & \textbf{3.84e-7} & 28.0977 & 0.0087 & \textbf{2.75e-5} & 9.6341 \\
EST-QCQ-CNN & 0.0114 & \textbf{0.0008} & 4.1099 & 0.0099 & \textbf{0.0010} & 3.8838 \\
\midrule
ResNet-18 & 0.0034 & 0.2211 & 0.6478 & -0.0018 & \textbf{0.0013} & -3.5816 \\
ResNet-50 & 0.0040 & \textbf{0.0195} & 1.6881 & -0.0016 & \textbf{0.0027} & -2.9587 \\
EfficientNet-B0 & 0.0015 & 0.1407 & 0.8199 & -0.0020 & \textbf{9.02e-5} & -7.1354 \\
MobileNetV3-Small & 0.0157 & \textbf{0.0011} & 3.7694 & -0.0008 & \textbf{0.0180} & -1.7300 \\
\bottomrule
\end{tabular}%
}
\par\smallskip
\noindent
\begin{minipage}{\textwidth}
\scriptsize
\raggedright
Note: Diff. is the mean paired difference (FAF-HQNN minus the comparison method); positive values
favor FAF-HQNN. Paired-sample Cohen's \textit{d}\textsubscript{z} is the mean paired difference divided by the sample standard deviation of the paired differences. Raw \textit{p}-values are from two-sided paired \textit{t}-tests on the five seed-matched scores and are unadjusted for multiple comparisons. Bold \textit{p}-values indicate nominal significance at the
0.05 level, irrespective of the direction of the difference. These exploratory tests assess clean-test ACC and AUC.
Nonsignificant results do not establish equivalence.
\end{minipage}
\end{table*}

\subsubsection{Training and Validation Dynamics}
{Figures~\ref{fig:classical_training_blood}, \ref{fig:qhead_training_blood}, and~\ref{fig:faf_hqnn_training_blood} show training and validation curves on BloodMNIST over 50 epochs and five random seeds. Compared with Q-Head, FAF-HQNN exhibits
smoother training trajectories and less visible variability across seeds. Relative to Classical, it shows slightly higher late-stage validation accuracy and smaller validation-loss fluctuations.

FAF-HQNN's training loss approaches zero and its training accuracy approaches 100\%, while its validation loss rises
modestly after an early minimum before stabilizing. Despite the relatively stable validation accuracy, this divergence suggests some degree of overfitting.

\begin{figure*}[t]
    \centering
    \includegraphics[width=\textwidth]{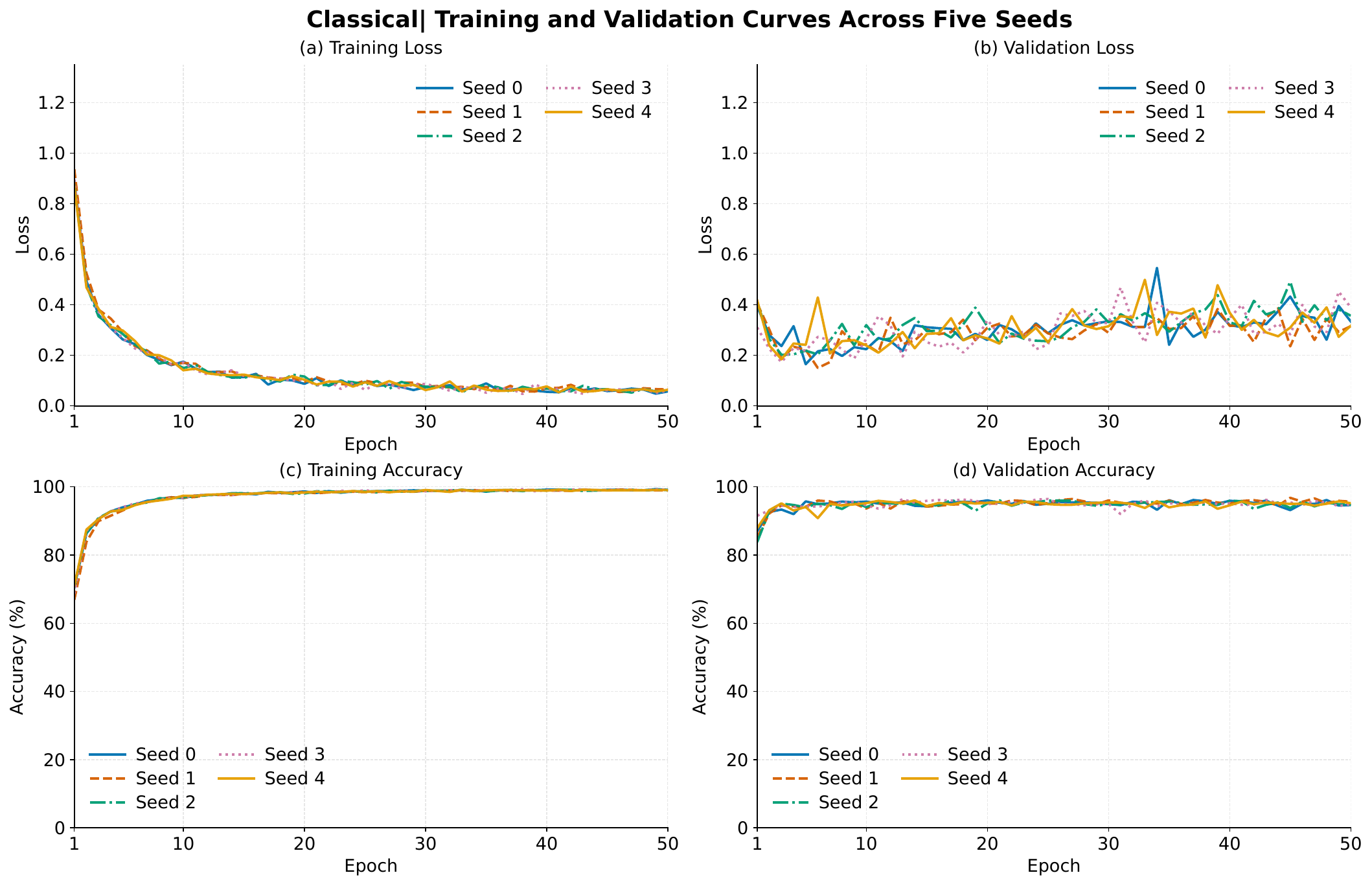}
    \caption{Training and validation curves for Classical on BloodMNIST across five random seeds. (a) Training loss; (b) validation loss;
    (c) training accuracy; and (d) validation accuracy. Colors and line styles denote seeds 0--4. Each curve represents an independent 50-epoch training run, shown without smoothing.}
    \label{fig:classical_training_blood}
\end{figure*}

\begin{figure*}[t]
    \centering
    \includegraphics[width=\textwidth]{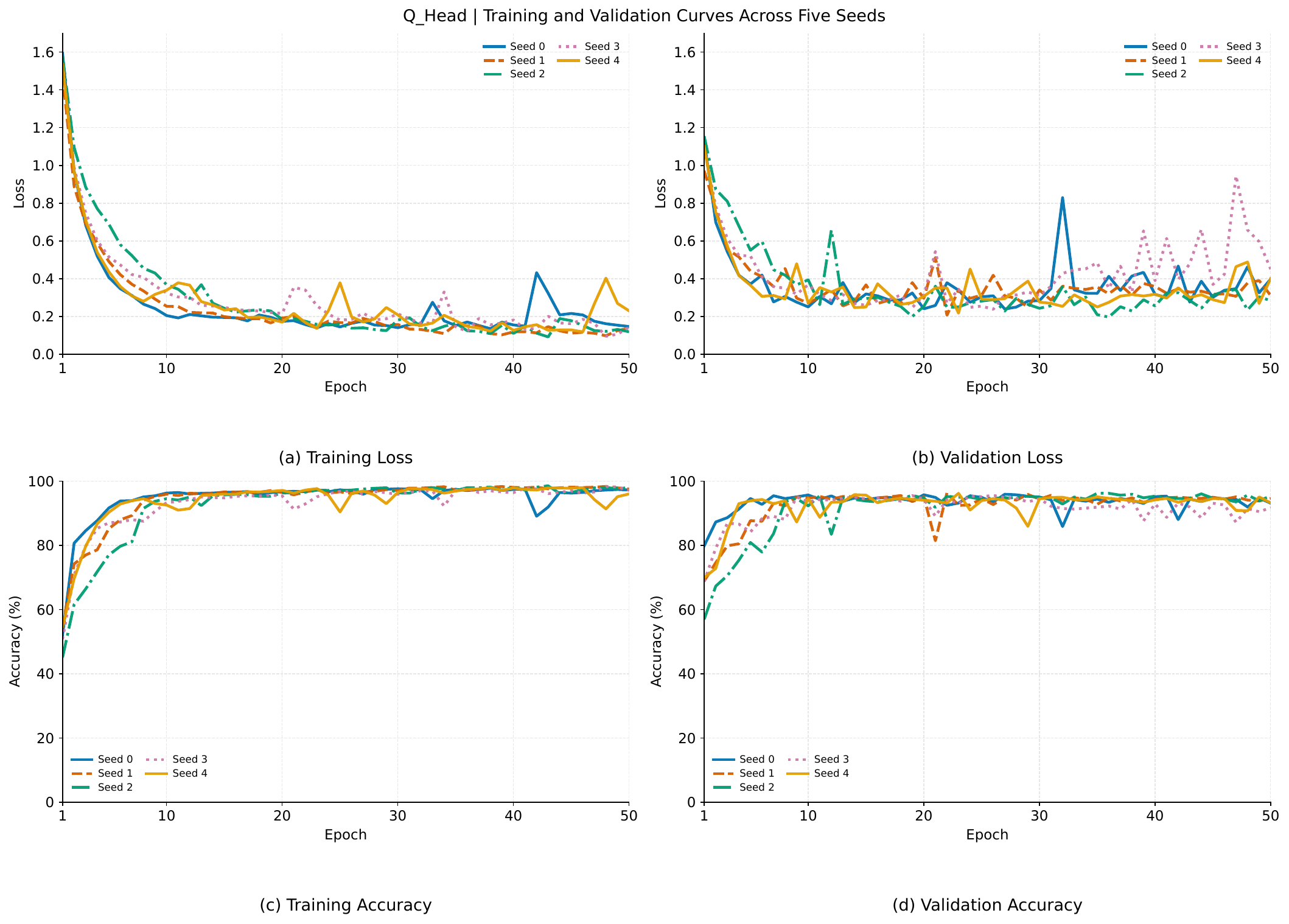}
    \caption{Training and validation curves for Q-Head on BloodMNIST across five random seeds.  (a) Training loss; (b) validation loss; (c) training accuracy; and (d) validation accuracy. Colors and line styles denote seeds 0--4. Each curve represents an independent 50-epoch training run, shown without smoothing.}
    \label{fig:qhead_training_blood}
\end{figure*}

\begin{figure*}[t]
    \centering
    \includegraphics[width=\textwidth]{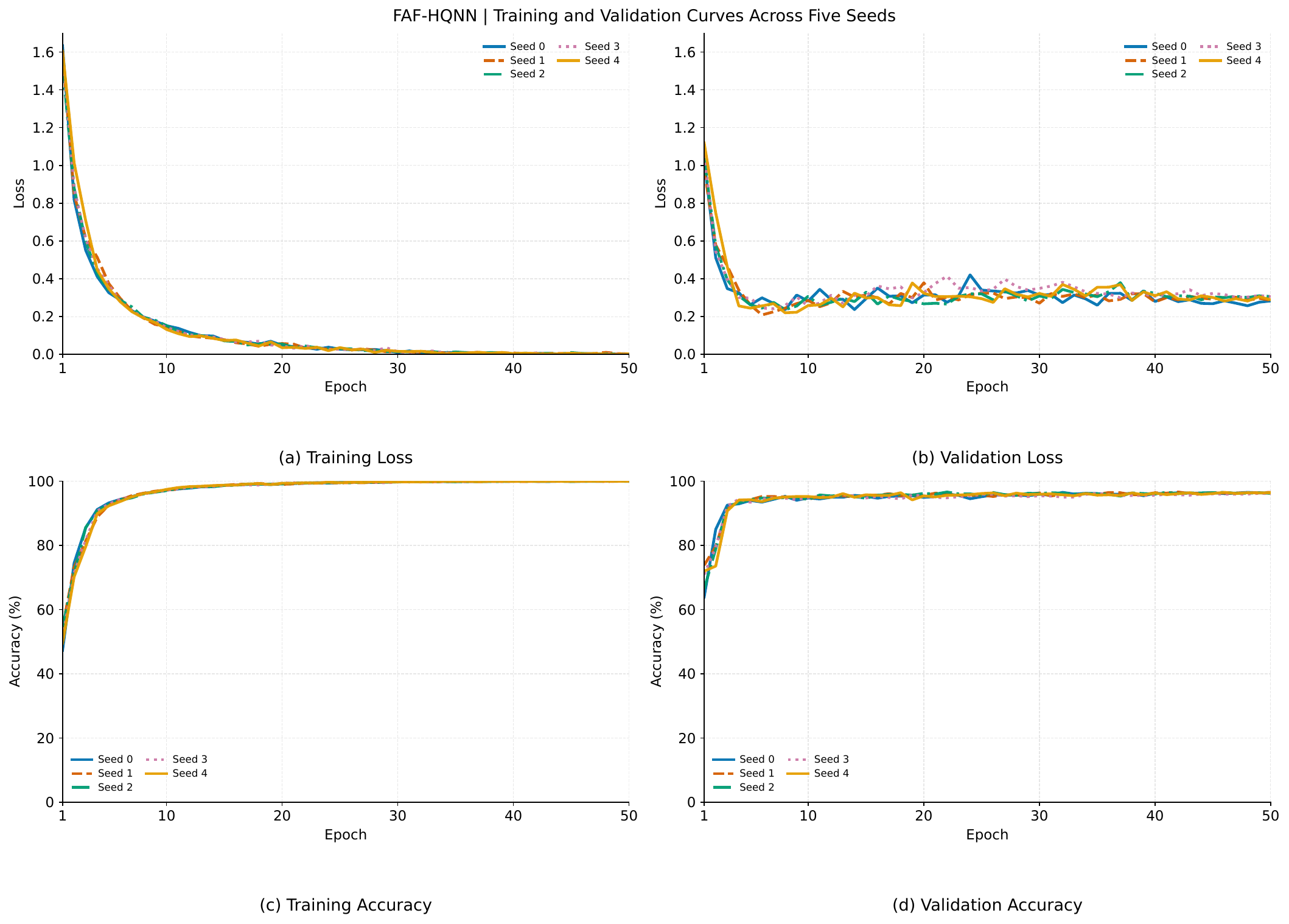}
    \caption{Training and validation curves for FAF-HQNN on BloodMNIST across five random seeds. (a) Training loss; (b) validation loss; (c) training accuracy; and (d) validation accuracy. Colors and line styles denote seeds 0-4. Each curve represents an independent 50-epoch training run, shown without smoothing.}
    \label{fig:faf_hqnn_training_blood}
\end{figure*}

\subsubsection{Robustness under Image Corruptions}

Tables~\ref{tab:blood_gaussian_acc} and~\ref{tab:blood_poisson_acc}, together with Figure~\ref{fig:blood_noise}, summarize performance on BloodMNIST under Gaussian and Poisson noise. Mean ACC generally declines as corruption severity increases, although small non-monotonic changes occur at low Gaussian noise levels. Table~\ref{tab:blood_robustness_acc} reports aggregate ACC across the four corrupted settings and paired comparisons over five matched random seeds.

Under Gaussian noise, FAF-HQNN achieves the highest aggregate mean ACC among all evaluated methods. Its gains over Classical, Q-Head, C-AF, H-Fix, EST-QCNN, EST-QCQ-CNN, ResNet-18, EfficientNet-B0, and MobileNetV3-Small reach nominal significance, whereas the remaining comparisons, including that with ResNet-50, do not.

Under Poisson noise, FAF-HQNN achieves higher aggregate mean ACC than all three external hybrid quantum--classical baselines, but ranks below Q-Head, H-Fix, H-Scalar, H-Sample, and ResNet-50; ResNet-50 ranks first overall. Gains reach nominal significance only over Classical, ResNet-18, EfficientNet-B0, and MobileNetV3-Small. The remaining comparisons, including that with ResNet-50, do not reach nominal significance. Under the present evaluation settings, FAF-HQNN's ranking in aggregate mean ACC differs between the two corruption types: it ranks first under Gaussian noise but not under Poisson noise.

\begin{table*}[!tp]
\centering
\caption{ACC on BloodMNIST under Gaussian noise for FAF-HQNN, internal comparison variants, external hybrid
quantum--classical baselines, and external CNN baselines. Values are reported as mean \textpm{} sample standard
deviation over five matched seeds. Bold entries indicate the highest mean ACC in each column; ties at the reported precision are also highlighted.}
\label{tab:blood_gaussian_acc}
\scriptsize
\setlength{\tabcolsep}{3pt}
\renewcommand{\arraystretch}{1.2}
\resizebox{\textwidth}{!}{%
\begin{tabular}{lccccc}
\toprule
\textbf{Method}
& \textbf{\ensuremath{\sigma=0.000}}
& \textbf{\ensuremath{\sigma=0.005}}
& \textbf{\ensuremath{\sigma=0.010}}
& \textbf{\ensuremath{\sigma=0.020}}
& \textbf{\ensuremath{\sigma=0.030}} \\
\midrule

Classical
& \ensuremath{0.9569\pm0.0030}
& \ensuremath{0.9567\pm0.0033}
& \ensuremath{0.9548\pm0.0038}
& \ensuremath{0.9433\pm0.0064}
& \ensuremath{0.9194\pm0.0092} \\

Q-Head
& \ensuremath{0.9517\pm0.0053}
& \ensuremath{0.9512\pm0.0059}
& \ensuremath{0.9504\pm0.0047}
& \ensuremath{0.9459\pm0.0074}
& \ensuremath{0.9311\pm0.0170} \\

C-AF
& \ensuremath{0.9569\pm0.0028}
& \ensuremath{0.9567\pm0.0026}
& \ensuremath{0.9536\pm0.0023}
& \ensuremath{0.9413\pm0.0047}
& \ensuremath{0.9138\pm0.0092} \\

H-Fix 
& \ensuremath{0.9566\pm0.0028}
& \ensuremath{0.9570\pm0.0027}
& \ensuremath{0.9563\pm0.0024}
& \ensuremath{0.9496\pm0.0049}
& \ensuremath{0.9286\pm0.0142} \\

H-Scalar
& \ensuremath{0.9545\pm0.0038}
& \ensuremath{0.9545\pm0.0037}
& \ensuremath{0.9551\pm0.0040}
& \ensuremath{0.9491\pm0.0041}
& \ensuremath{0.9311\pm0.0094} \\

H-Sample
& \ensuremath{0.9555\pm0.0028}
& \ensuremath{0.9552\pm0.0030}
& \ensuremath{0.9552\pm0.0030}
& \ensuremath{0.9486\pm0.0065}
& \ensuremath{0.9275\pm0.0081} \\

H-Class 
& \ensuremath{0.9551\pm0.0018}
& \ensuremath{0.9557\pm0.0025}
& \ensuremath{0.9549\pm0.0039}
& \ensuremath{0.9483\pm0.0058}
& \ensuremath{0.9251\pm0.0058} \\

EST-QCNN
& \ensuremath{0.9509\pm0.0042}
& \ensuremath{0.9501\pm0.0050}
& \ensuremath{0.9488\pm0.0047}
& \ensuremath{0.9380\pm0.0112}
& \ensuremath{0.9092\pm0.0248} \\

EST-RUQNN
& \ensuremath{0.9508\pm0.0024}
& \ensuremath{0.9511\pm0.0020}
& \ensuremath{0.9500\pm0.0042}
& \ensuremath{0.9441\pm0.0128}
& \ensuremath{0.9286\pm0.0218} \\
EST-QCQ-CNN
& \ensuremath{0.9539\pm0.0031}
& \ensuremath{0.9533\pm0.0036}
& \ensuremath{0.9533\pm0.0032}
& \ensuremath{0.9438\pm0.0056}
& \ensuremath{0.9132\pm0.0187} \\
\midrule

\textbf{FAF-HQNN}
& \ensuremath{\mathbf{0.9654\pm0.0024}}
& \ensuremath{\mathbf{0.9652\pm0.0033}}
& \ensuremath{\mathbf{0.9644\pm0.0028}}
& \ensuremath{0.9526\pm0.0063}
& \ensuremath{0.9291\pm0.0095} \\

\midrule

ResNet-18
& \ensuremath{0.9619\pm0.0032}
& \ensuremath{0.9619\pm0.0034}
& \ensuremath{0.9593\pm0.0031}
& \ensuremath{0.9429\pm0.0076}
& \ensuremath{0.9042\pm0.0119} \\

ResNet-50
& \ensuremath{0.9614\pm0.0013}
& \ensuremath{0.9615\pm0.0012}
& \ensuremath{0.9606\pm0.0013}
& \ensuremath{\mathbf{0.9542\pm0.0028}}
& \ensuremath{\mathbf{0.9334\pm0.0028}} \\

EfficientNet-B0
& \ensuremath{0.9639\pm0.0020}
& \ensuremath{\mathbf{0.9652\pm0.0010}}
& \ensuremath{0.9585\pm0.0031}
& \ensuremath{0.8665\pm0.0194}
& \ensuremath{0.3998\pm0.1459} \\

MobileNetV3-Small
& \ensuremath{0.9497\pm0.0043}
& \ensuremath{0.9421\pm0.0045}
& \ensuremath{0.9060\pm0.0097}
& \ensuremath{0.8054\pm0.0289}
& \ensuremath{0.7268\pm0.0380} \\

\bottomrule
\end{tabular}%
}
\end{table*}

\begin{table*}[!tp]
\centering
\caption{ACC on BloodMNIST under Poisson noise for FAF-HQNN, internal comparison variants, external hybrid quantum--classical baselines, and external CNN baselines. Lower peak values correspond to stronger corruption. Values are reported as mean \textpm{} sample standard deviation over five matched seeds. Bold entries indicate the highest mean ACC in each column; ties at the reported precision are also highlighted.}

\label{tab:blood_poisson_acc}
\scriptsize
\setlength{\tabcolsep}{3pt}
\renewcommand{\arraystretch}{1.2}

\resizebox{\textwidth}{!}{%
\begin{tabular}{lccccc}
\toprule
\textbf{Method}
& \textbf{Clean}
& \textbf{Peak=3200}
& \textbf{Peak=1600}
& \textbf{Peak=800}
& \textbf{Peak=400} \\
\midrule

Classical
& \ensuremath{0.9569\pm0.0030}
& \ensuremath{0.9521\pm0.0029}
& \ensuremath{0.9448\pm0.0057}
& \ensuremath{0.9301\pm0.0097}
& \ensuremath{0.8947\pm0.0188} \\

Q-Head
& \ensuremath{0.9517\pm0.0053}
& \ensuremath{0.9498\pm0.0057}
& \ensuremath{0.9471\pm0.0054}
& \ensuremath{0.9389\pm0.0093}
& \ensuremath{\mathbf{0.9116\pm0.0319}} \\

C-AF
& \ensuremath{0.9569\pm0.0028}
& \ensuremath{0.9505\pm0.0026}
& \ensuremath{0.9419\pm0.0035}
& \ensuremath{0.9269\pm0.0030}
& \ensuremath{0.8889\pm0.0162} \\

H-Fix
& \ensuremath{0.9566\pm0.0028}
& \ensuremath{0.9543\pm0.0031}
& \ensuremath{0.9505\pm0.0034}
& \ensuremath{0.9386\pm0.0081}
& \ensuremath{0.9090\pm0.0227} \\

H-Scalar
& \ensuremath{0.9545\pm0.0038}
& \ensuremath{0.9536\pm0.0032}
& \ensuremath{0.9502\pm0.0023}
& \ensuremath{0.9384\pm0.0041}
& \ensuremath{0.9091\pm0.0101} \\

H-Sample
& \ensuremath{0.9555\pm0.0028}
& \ensuremath{0.9542\pm0.0041}
& \ensuremath{0.9508\pm0.0044}
& \ensuremath{0.9383\pm0.0044}
& \ensuremath{0.9050\pm0.0083} \\

H-Class
& \ensuremath{0.9551\pm0.0018}
& \ensuremath{0.9547\pm0.0048}
& \ensuremath{0.9494\pm0.0046}
& \ensuremath{0.9377\pm0.0064}
& \ensuremath{0.9008\pm0.0046} \\

EST-QCNN
& \ensuremath{0.9509\pm0.0042}
& \ensuremath{0.9463\pm0.0049}
& \ensuremath{0.9411\pm0.0091}
& \ensuremath{0.9258\pm0.0201}
& \ensuremath{0.8862\pm0.0395} \\

EST-RUQNN
& \ensuremath{0.9508\pm0.0024}
& \ensuremath{0.9487\pm0.0056}
& \ensuremath{0.9453\pm0.0091}
& \ensuremath{0.9352\pm0.0170}
& \ensuremath{0.9096\pm0.0320} \\

EST-QCQ-CNN
& \ensuremath{0.9539\pm0.0031}
& \ensuremath{0.9522\pm0.0043}
& \ensuremath{0.9457\pm0.0061}
& \ensuremath{0.9307\pm0.0102}
& \ensuremath{0.8819\pm0.0304} \\
\midrule

\textbf{FAF-HQNN}
& \ensuremath{\mathbf{0.9654\pm0.0024}}
& \ensuremath{\mathbf{0.9607\pm0.0034}}
& \ensuremath{0.9507\pm0.0081}
& \ensuremath{0.9338\pm0.0121}
& \ensuremath{0.8993\pm0.0171} \\

\midrule

ResNet-18
& \ensuremath{0.9619\pm0.0032}
& \ensuremath{0.9557\pm0.0057}
& \ensuremath{0.9451\pm0.0098}
& \ensuremath{0.9196\pm0.0138}
& \ensuremath{0.8680\pm0.0169} \\

ResNet-50
& \ensuremath{0.9614\pm0.0013}
& \ensuremath{0.9595\pm0.0014}
& \ensuremath{\mathbf{0.9557\pm0.0027}}
& \ensuremath{\mathbf{0.9427\pm0.0033}}
& \ensuremath{0.9053\pm0.0068} \\

EfficientNet-B0
& \ensuremath{0.9639\pm0.0020}
& \ensuremath{0.9425\pm0.0023}
& \ensuremath{0.8670\pm0.0267}
& \ensuremath{0.5242\pm0.1922}
& \ensuremath{0.2588\pm0.0480} \\

MobileNetV3-Small
& \ensuremath{0.9497\pm0.0043}
& \ensuremath{0.8662\pm0.0200}
& \ensuremath{0.8042\pm0.0298}
& \ensuremath{0.7405\pm0.0327}
& \ensuremath{0.6578\pm0.0400} \\
\bottomrule
\end{tabular}%
}
\end{table*}

\begin{figure*}[t]
    \centering
    \begin{subfigure}[t]{0.49\textwidth}
        \centering
        \includegraphics[width=\linewidth]{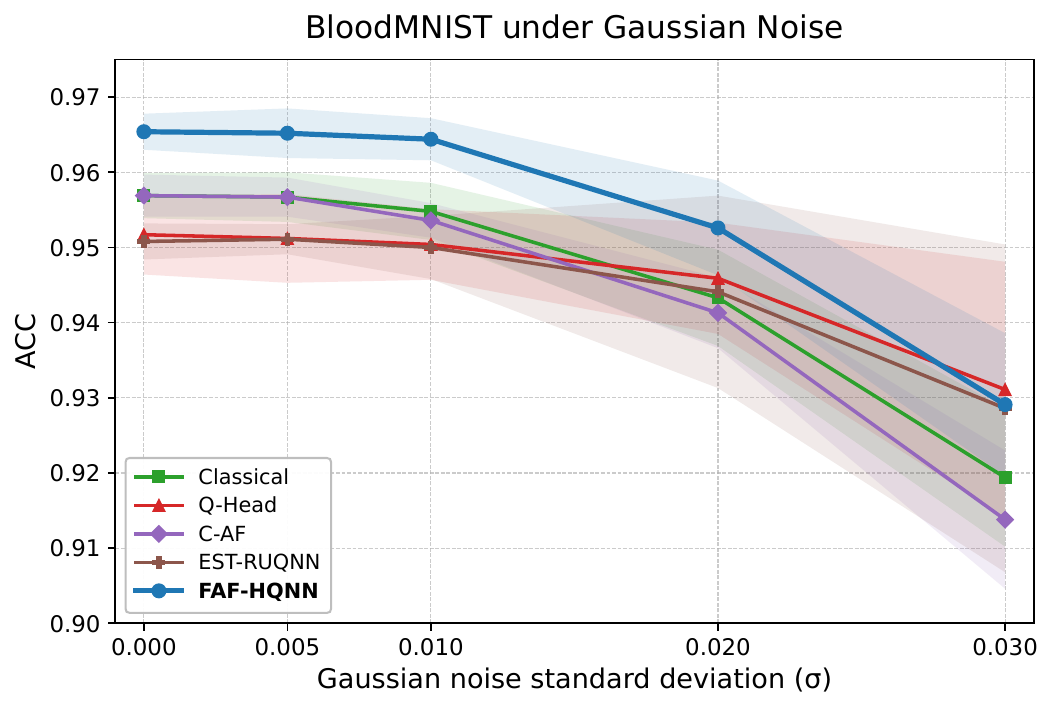}
        \caption{Gaussian noise.}
        \label{fig:blood_gaussian_acc}
    \end{subfigure}
    \hfill
    \begin{subfigure}[t]{0.49\textwidth}
        \centering
        \includegraphics[width=\linewidth]{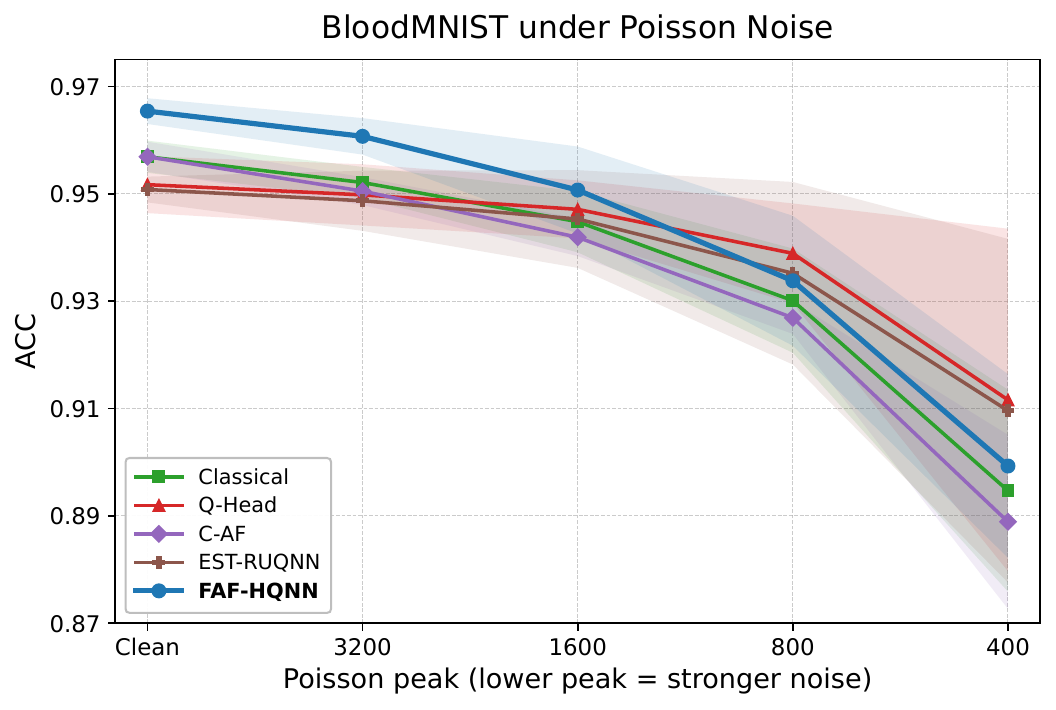}
        \caption{Poisson noise.}
        \label{fig:blood_poisson_acc}
    \end{subfigure}
    \caption{Mean ACC of FAF-HQNN and selected baselines on BloodMNIST under (a) Gaussian noise and (b) Poisson noise.
Corruption severity increases with Gaussian noise standard deviation and decreases with Poisson peak value. Shaded regions indicate mean \textpm{} one sample standard deviation over five matched seeds.}
    \label{fig:blood_noise}
\end{figure*}

\begin{table*}[!tp]
\centering
\caption{Aggregate corrupted-test ACC and paired comparisons between FAF-HQNN and internal comparison variants, external
hybrid quantum--classical baselines, and external CNN baselines on BloodMNIST over five matched seeds.}
\label{tab:blood_robustness_acc}
\begingroup
\small
\setlength{\tabcolsep}{3pt}
\renewcommand{\arraystretch}{1}
\begin{adjustbox}{max width=\textwidth,center}
\begin{tabular}{@{}lccccc@{}}
\toprule
\textbf{Method}
& \textbf{Mean Corr. ACC}
& \textbf{ACC Diff.}
& \textbf{ACC \textit{d}\textsubscript{z}}
& \textbf{95\% CI}
& \textbf{Raw \textit{p}} \\
\midrule
\multicolumn{6}{@{}l}{\textbf{(a) Gaussian noise}} \\
\midrule
Classical
& 0.9436 \textpm{} 0.0049
& 0.0093
& 2.457
& [0.0046, 0.0140]
& \textbf{0.0054} \\
Q-Head
& 0.9447 \textpm{} 0.0063
& 0.0082
& 1.349
& [0.0006, 0.0157]
& \textbf{0.0402} \\
C-AF
& 0.9414 \textpm{} 0.0032
& 0.0115
& 1.595
& [0.0025, 0.0204]
& \textbf{0.0235} \\
H-Fix
& 0.9479 \textpm{} 0.0044
& 0.0050
& 1.774
& [0.0015, 0.0085]
& \textbf{0.0171} \\
H-Scalar
& 0.9475 \textpm{} 0.0023
& 0.0054
& 0.944
& [\textminus{}0.0017, 0.0125]
& 0.1040 \\
H-Sample
& 0.9466 \textpm{} 0.0049
& 0.0062
& 0.770
& [\textminus{}0.0038, 0.0162]
& 0.1592 \\
H-Class
& 0.9460 \textpm{} 0.0042
& 0.0068
& 0.965
& [\textminus{}0.0019, 0.0156]
& 0.0953 \\
EST-QCNN
& 0.9365 \textpm{} 0.0101
& 0.0163
& 1.823
& [0.0052, 0.0274]
& \textbf{0.0150} \\
EST-RUQNN
& 0.9435 \textpm{} 0.0101
& 0.0094
& 0.741
& [\textminus{}0.0064, 0.0251]
& 0.174 \\
EST-QCQ-CNN
& 0.9409 \textpm{} 0.0069
& 0.0119
& 1.606
& [0.0027, 0.0211]
& \textbf{0.023} \\
ResNet-18
& 0.9421 \textpm{} 0.0054
& 0.0108
& 2.416
& [0.0052, 0.0163]
& \textbf{0.0058} \\
ResNet-50
& 0.9524 \textpm{} 0.0015
& 0.0005
& 0.121
& [\textminus{}0.0047, 0.0056]
& 0.8177 \\
EfficientNet-B0
& 0.7975 \textpm{} 0.0332
& 0.1553
& 4.304
& [0.1105, 0.2001]
& \textbf{6.53e-4} \\
MobileNetV3-Small
& 0.8451 \textpm{} 0.0172
& 0.1078
& 5.474
& [0.0833, 0.1322]
& \textbf{2.57e-4} \\
\textbf{FAF-HQNN}
& \textbf{0.9528 \textpm{} 0.0047}
& --
& --
& --
& -- \\
\midrule
\multicolumn{6}{@{}l}{\textbf{(b) Poisson noise}} \\
\midrule
Classical
& 0.9304 \textpm{} 0.0085
& 0.0056
& 1.656
& [0.0014, 0.0098]
& \textbf{0.0206} \\
Q-Head
& 0.9368 \textpm{} 0.0110
& \textminus{}0.0008
& \textminus{}0.118
& [\textminus{}0.0093, 0.0076]
& 0.7981 \\
C-AF
& 0.9271 \textpm{} 0.0050
& 0.0089
& 0.648
& [\textminus{}0.0081, 0.0260]
& 0.2184 \\
H-Fix
& 0.9381 \textpm{} 0.0078
& \textminus{}0.0021
& \textminus{}0.478
& [\textminus{}0.0075, 0.0034]
& 0.3493 \\
H-Scalar
& 0.9378 \textpm{} 0.0035
& \textminus{}0.0018
& \textminus{}0.154
& [\textminus{}0.0163, 0.0127]
& 0.7463 \\
H-Sample
& 0.9371 \textpm{} 0.0048
& \textminus{}0.0011
& \textminus{}0.090
& [\textminus{}0.0163, 0.0142]
& 0.8558 \\
H-Class
& 0.9357 \textpm{} 0.0042
& 0.0003
& 0.026
& [\textminus{}0.0142, 0.0149]
& 0.9528 \\
EST-QCNN
& 0.9248 \textpm{} 0.0180
& 0.0112
& 0.685
& [\textminus{}0.0091, 0.0315]
& 0.2015 \\
EST-RUQNN
& 0.9347 \textpm{} 0.0159
& 0.0013
& 0.058
& [\textminus{}0.0267, 0.0294]
& 0.902 \\
EST-QCQ-CNN
& 0.9276 \textpm{} 0.0116
& 0.0084
& 0.705
& [\textminus{}0.0064, 0.0232]
& 0.192 \\
ResNet-18
& 0.9221 \textpm{} 0.0111
& 0.0139
& 1.984
& [0.0052, 0.0226]
& \textbf{0.0114} \\
ResNet-50
& 0.9408 \textpm{} 0.0027
& \textminus{}0.0048
& \textminus{}0.602
& [\textminus{}0.0147, 0.0051]
& 0.2496 \\
EfficientNet-B0
& 0.6481 \textpm{} 0.0546
& 0.2879
& 4.628
& [0.2106, 0.3651]
& \textbf{4.92e-4} \\
MobileNetV3-Small
& 0.7672 \textpm{} 0.0293
& 0.1688
& 4.684
& [0.1241, 0.2136]
& \textbf{4.70e-4} \\
\cmidrule(lr){1-6}
\textbf{FAF-HQNN}
& \textbf{0.9360 \textpm{} 0.0099}
& --
& --
& --
& -- \\
\bottomrule
\end{tabular}
\end{adjustbox}

\par\smallskip
\noindent
\begin{minipage}{\textwidth}
\scriptsize
\raggedright
Note: Mean Corr. ACC averages four non-clean levels per seed: Gaussian SDs of 0.005, 0.010, 0.020, and 0.030, or Poisson peaks of 3200, 1600, 800, and 400. Values are mean \textpm{} sample SD over five matched seeds. ACC Diff. denotes the mean paired difference (FAF-HQNN minus comparator); positive values favor FAF-HQNN. Cohen's \textit{d}\textsubscript{z} divides this difference by the sample SD of paired differences. The 95\% CIs and two-sided \textit{p}-values use paired \textit{t}-procedures (df = 4), without multiplicity adjustment; bold \textit{p}-values indicate nominal significance (\textit{p} \textless{} 0.05)
in either direction. Tests are exploratory and concern aggregate ACC only; nonsignificance does not imply equivalence. Bold FAF-HQNN entries mark the reference, not column maxima; dashes indicate self-comparisons. All checkpoints were selected by the highest validation accuracy. Analyses use per-seed ACC recorded to four decimals and precede final rounding; displayed means may not reproduce reported differences exactly.
\end{minipage}
\endgroup
\end{table*}

\section{Discussion}
\subsection{Discussion of Results}

The results on PathMNIST and BloodMNIST suggest that FAF-HQNN is effective for biomedical image classification within the evaluated simulation-based setting. On both datasets, FAF-HQNN achieves competitive clean-test accuracy compared to internal variants and external hybrid quantum--classical baselines. On PathMNIST, it achieves higher aggregate mean accuracy than all evaluated baselines under both Gaussian and Poisson noise. These findings suggest that feature-adaptive fusion may help improve classification performance, although its benefits depend on the dataset, evaluation metric, and corruption setting.

The aggregate advantages observed in these settings do not imply that FAF-HQNN performs best at every individual corruption
severity. H-Fix and H-Scalar remain competitive under some stronger corruptions and achieve higher mean ACC under the
stronger Poisson noise conditions on BloodMNIST. Thus, the relative performance of fusion strategies depends on both
corruption type and severity, and feature-adaptive fusion does not provide uniform superiority across all evaluated perturbation settings.

\paragraph{Interpretation of the quantum-branch contribution.}
The branch-contribution analyses on PathMNIST distinguish the role of the quantum-output term in the trained fusion computation from evidence of sample-specific quantum complementarity. Under fixed gating, removing the quantum outputs reduces mean accuracy, whereas shuffling complete quantum-output vectors across samples produces little change in mean accuracy. These findings support a measurable role for the quantum-output term under the evaluated inference-time interventions, but do not establish that sample-specific complementary information from the quantum branch explains the observed fusion gains. Accordingly, we do not interpret the aggregate performance improvements as evidence of such a mechanism or of an advantage attributable to quantum computation itself.

\paragraph{Circuit configuration and implementation implications.}
The quantum-module analysis shows that model performance varies with circuit layout, measurement basis, and circuit depth, although the differences across the evaluated configurations are moderate. In the present simulation setting, the single-layer circuit achieves the highest mean accuracy, and none of the deeper circuits evaluated achieves a higher mean accuracy. These results support the use of a shallow circuit in the current implementation, but do not establish a general relationship between circuit depth and predictive performance.

These findings motivate further investigation of shallow circuits for resource-constrained quantum implementations. However, classical simulation alone does not establish the practical performance of the model on noisy intermediate-scale quantum (NISQ) hardware. Evaluation on real quantum devices is needed to assess whether the observed predictive performance can be maintained under hardware noise and implementation constraints.

\paragraph{Scope of the robustness evaluation.}
The robustness findings are limited to the current train-clean/test-corrupted protocol and the selected synthetic perturbations. Gaussian and Poisson noise do not capture the full range of clinically relevant distribution shifts, including staining variation, blur, contrast changes, scanner-dependent differences, and acquisition artefacts. These factors may alter diagnostically relevant image characteristics, co-occur, and vary across clinical centers. Because such shifts are not evaluated here, the observed robustness advantages cannot be assumed to generalize to clinical workflows. The experiments therefore provide an initial controlled assessment rather than a comprehensive validation of clinical robustness.

Taken together, the findings support FAF-HQNN as a competitive predictive model under the evaluated MedMNIST conditions. Establishing quantum-specific benefits, practical hardware feasibility, and clinical applicability requires further controlled comparisons and validation beyond the present simulation-based experiments.

\subsection{Limitations}

While the results on PathMNIST and BloodMNIST provide encouraging empirical evidence for the proposed FAF-HQNN model, several limitations should be acknowledged when interpreting the present findings.

\begin{enumerate}
    \item \textbf{Simulation-based evaluation.}
    All experiments were conducted using classical simulation rather than real quantum hardware. Consequently, the present results do not establish the practical performance or implementation feasibility of the proposed framework on physical quantum devices. Hardware experiments are needed to assess whether the observed predictive performance can be maintained under real-device operating conditions.

\item \textbf{Limited training-data scale and modality diversity.}
The PathMNIST experiments use a class-balanced subset of 9,000 training images rather than the complete official training set because of the high computational cost of sample-wise quantum circuit simulation. The observed performance benefits are therefore established only within this reduced-data training setting. Additional training data may benefit both FAF-HQNN and the classical baselines, potentially altering their relative performance and the estimated contributions of individual model components. Extrapolation to the complete training set must also account for possible differences in class distribution introduced by class-balanced subsampling. Whether, and to what extent, the observed benefits persist requires direct evaluation on the complete training set, which remains an important direction for future work.

PathMNIST and BloodMNIST are low-resolution, standardized biomedical image benchmarks covering only two imaging modalities. Accordingly, it remains unclear whether the observed advantages of FAF-HQNN extend to higher-resolution inputs, other imaging modalities, or more heterogeneous clinical datasets.

\item \textbf{Restricted robustness protocol.}
The image robustness analysis is limited to a train-clean/test-corrupted setting using selected synthetic perturbations, including Gaussian and Poisson noise. Although this protocol enables a controlled assessment of sensitivity to test-time image corruption, it does not fully capture clinically relevant distribution shifts. In particular, the current experiments do not evaluate staining variation, image blur, contrast changes, scanner-dependent differences, or acquisition artefacts. These shifts may arise from differences in specimen preparation, staining protocols, scanner hardware, image focus, and acquisition or processing conditions, and may co-occur across clinical settings. Consequently, the current results should be interpreted as an initial robustness assessment under controlled synthetic perturbations rather than a comprehensive validation of clinical robustness. Whether these results generalize to clinically relevant distribution shifts and across clinical centers remains to be established.

\item \textbf{Configuration dependence and limited mechanistic understanding.}
The observed performance differences are evaluated within the specific branch architectures and fusion configurations considered in this study. On PathMNIST, fixed-gating interventions show that removing the quantum-output term reduces accuracy, whereas shuffling quantum-output vectors across samples produces little change in mean accuracy. These results do not establish that sample-specific quantum representations explain the fusion gains, nor do they establish an advantage attributable to quantum computation itself. Because these interventions retain the trained parameters and original gating coefficients without retraining or recomputing the gates, possible contributions through gate learning or joint optimization remain unresolved. Moreover, these direct branch-contribution analyses are limited to PathMNIST. Further investigation of representation structure, optimization behavior, and corruption sensitivity is needed to clarify the role of the quantum component and the conditions under which adaptive fusion is most effective.

\end{enumerate}

\subsection{Future Work}
Several directions may help further establish the practical relevance and scientific understanding of the FAF-HQNN model for biomedical image analysis.

First, an important next step is to move beyond classical simulation toward evaluation on real quantum hardware. Implementing the proposed framework on physical quantum devices would enable an assessment of its predictive performance, computational cost, and practical feasibility. Such studies will be essential for determining whether the trends observed in the present simulation-based experiments extend to real-device implementations.

Second, broader validation on more demanding biomedical imaging datasets is needed. While MedMNIST provides a useful controlled benchmark, future studies should investigate larger-scale, higher-resolution, and more heterogeneous clinical datasets that better reflect real-world visual complexity, anatomical variability, and acquisition diversity. This will be important for assessing how far the present observations generalize beyond standardized benchmark conditions.

Third, given the dataset- and corruption-dependent robustness gains observed in this study, future work should extend evaluation beyond synthetic noise to clinically relevant distribution shifts, including staining variation, image blur, contrast changes, scanner and acquisition variability, and imaging artifacts. External multi-center datasets and scanner-aware evaluation protocols would enable a more realistic assessment of generalization, particularly relative to internal variants, hybrid quantum--classical baselines, and conventional CNNs. Stain normalization and domain generalization strategies could also be investigated to improve robustness under these conditions.

Finally, the design principles and underlying mechanisms of fusion strategies warrant deeper investigation. The present results suggest that the effectiveness of the FAF-HQNN model depends strongly on how quantum and classical information are integrated, and that the preferred fusion strategy may vary across datasets and perturbation regimes. Future research should therefore pursue more principled adaptive fusion methods, while also examining representation structure, branch complementarity, optimization dynamics, and corruption sensitivity to better understand when and why feature-adaptive fusion yields practical advantages.

\section{Conclusion}

The results on PathMNIST and BloodMNIST suggest that FAF-HQNN is effective for biomedical image classification within the evaluated simulation-based setting. On both datasets, FAF-HQNN achieves competitive clean-test accuracy compared to internal variants and external hybrid quantum--classical baselines. On PathMNIST, it achieves higher aggregate mean accuracy than all evaluated baselines under both Gaussian and Poisson noise. These findings suggest that feature-adaptive fusion may help improve classification performance, although its benefits depend on the dataset, evaluation metric, and corruption setting.

Under image corruptions, FAF-HQNN achieves higher aggregate mean corrupted-test accuracy than the internal variants and external hybrid quantum--classical baselines under several evaluated noise conditions, with gains varying across datasets and corruption types. The results indicate favorable performance under Gaussian and Poisson noise on PathMNIST and under Gaussian noise on BloodMNIST, while the benefits are less consistent under Poisson noise on BloodMNIST. Overall, these findings suggest that feature-adaptive fusion may improve corrupted-test performance under specific conditions, without implying a universal robustness advantage.

Analyses of circuit layout, measurement basis, and depth further suggest that shallow VQC configurations can be effective within the evaluated adaptive-fusion framework. Overall, these findings suggest that feature-adaptive fusion can improve classification performance within the evaluated hybrid quantum–classical architectures and provide robustness benefits under specific corruption conditions; they do not establish a general quantum advantage. The robustness evaluation concerns classical image corruptions rather than quantum hardware noise, and the evidence is limited to simulation-based experiments on two MedMNIST benchmarks. Further validation on quantum hardware and external multi-center datasets under clinically relevant distribution shifts is therefore needed to assess practical applicability.

\backmatter
\bmhead{Acknowledgements}
The authors would like to thank Chaoyang Li from Zhengzhou University of Light Industry for valuable discussions and technical support.

\section*{Declarations}
\begin{itemize}
\item Funding

This work was supported by the Shandong Provincial Natural Science Foundation (Grant No. ZR2025MS1063), the Scientific Research Fund of Zaozhuang University (Grant No. 102061901), the National Natural Science Foundation of China (Grant No. 62501523), and the Zhejiang Province Selected Funding for Postdoctoral Research Projects (Grant No. ZJ2025023).

\item Competing interests

The authors declare that they have no competing interests.

\item Ethics approval and consent to participate

Not applicable.

\item Consent for publication

Not applicable.

\item Data availability

The data that support the findings of this study are available from the corresponding author upon reasonable request.

\item Materials availability

The materials that support the findings of this study are available from the corresponding author upon reasonable request.

\item Code availability

The code that supports the findings of this study is available from the corresponding author upon reasonable request.

\item Author contributions

Yan-Yan Hou wrote the manuscript and conducted the experiments, ensuring the methodology was rigorously followed and that the data collected were reliable and valid. Jian Li conceived the experimental design, providing critical insights into the research framework and guiding the overall direction of the study. Chongqiang Ye analyzed the results, employing statistical methods to interpret the data accurately and draw meaningful conclusions. Hengji Li reviewed the manuscript, offering valuable feedback on the clarity and coherence of the writing, as well as ensuring that the findings were presented in a logical manner. Zhuo Wang validated the experiments, confirming the reproducibility of the results and providing additional verification of the experimental protocols. Qinghui Liu supervised the project, contributed to the interpretation of the results, and provided overall guidance on the revision of the manuscript.

\end{itemize}

\begin{appendices}

\section{Additional Results for Classical Image Corruptions}
\label{app:classical_corruption_auc}

\subsection{AUC Results on PathMNIST and BloodMNIST}
\label{app:path_auc}

Tables~\ref{tab:path_gaussian_auc} and~\ref{tab:path_poisson_auc} report the AUC results on PathMNIST under Gaussian and Poisson noise, respectively. Tables~\ref{tab:blood_gaussian_auc} and~\ref{tab:blood_poisson_auc}
report the AUC results on BloodMNIST under Gaussian and Poisson
noise, respectively.

\begin{table*}[!htbp]
\centering
\caption{AUC comparison of model variants and external CNN baselines on PathMNIST under Gaussian noise with varying corruption intensity. Results are reported as mean \ensuremath{\pm} sample standard deviation. Bold values indicate the highest mean AUC in each column.}
\label{tab:path_gaussian_auc}
\scriptsize
\setlength{\tabcolsep}{3pt}
\renewcommand{\arraystretch}{1}
\resizebox{\textwidth}{!}{%
\begin{tabular}{lccccc}
\toprule
\textbf{Method}
& \textbf{\ensuremath{\sigma=0.00}}
& \textbf{\ensuremath{\sigma=0.005}}
& \textbf{\ensuremath{\sigma=0.01}}
& \textbf{\ensuremath{\sigma=0.02}}
& \textbf{\ensuremath{\sigma=0.03}} \\
\midrule

Classical
& 0.9754 \textpm{} 0.0029
& 0.9728 \textpm{} 0.0070
& 0.9635 \textpm{} 0.0064
& 0.9189 \textpm{} 0.0055
& 0.8581 \textpm{} 0.0197 \\

Q-Head
& \ensuremath{0.9490\pm0.0294}
& \ensuremath{0.9469\pm0.0308}
& \ensuremath{0.9383\pm0.0307}
& \ensuremath{0.8713\pm0.0288}
& \ensuremath{0.7712\pm0.0401} \\

C-AF
& \ensuremath{0.9796\pm0.0034}
& \ensuremath{0.9771\pm0.0045}
& \ensuremath{0.9670\pm0.0092}
& \ensuremath{0.9048\pm0.0304}
& \ensuremath{0.8076\pm0.0521} \\

H-Fix
& \ensuremath{0.9760\pm0.0058}
& \ensuremath{0.9738\pm0.0057}
& \ensuremath{0.9637\pm0.0081}
& \ensuremath{0.9198\pm0.0238}
& \ensuremath{0.8655\pm0.0390} \\

H-Scalar
& \ensuremath{0.9777\pm0.0048}
& \ensuremath{0.9753\pm0.0050}
& \ensuremath{0.9654\pm0.0070}
& \ensuremath{0.8966\pm0.0419}
& \ensuremath{0.7859\pm0.1030} \\
H-Sample
& \ensuremath{0.9738\pm0.0157}
& \ensuremath{0.9719\pm0.0167}
& \ensuremath{0.9645\pm0.0195}
& \ensuremath{0.9126\pm0.0259}
& \ensuremath{0.8198\pm0.0620} \\
H-Class
& \ensuremath{0.9792\pm0.0064}
& \ensuremath{0.9779\pm0.0064}
& \ensuremath{0.9717\pm0.0082}
& \ensuremath{0.9265\pm0.0377}
& \ensuremath{0.8401\pm0.0867} \\
EST-QCNN
& \ensuremath{0.9490\pm0.0114}
& \ensuremath{0.9461\pm0.0111}
& \ensuremath{0.9354\pm0.0127}
& \ensuremath{0.8719\pm0.0364}
& \ensuremath{0.7780\pm0.0584} \\
EST-RUQNN
& \ensuremath{0.9571\pm0.0132}
& \ensuremath{0.9551\pm0.0137}
& \ensuremath{0.9455\pm0.0154}
& \ensuremath{0.8766\pm0.0345}
& \ensuremath{0.7799\pm0.0534} \\


EST-QCQ-CNN
& \ensuremath{0.9590\pm0.0082}
& \ensuremath{0.9568\pm0.0093}
& \ensuremath{0.9453\pm0.0101}
& \ensuremath{0.8741\pm0.0248}
& \ensuremath{0.7949\pm0.0418} \\

\midrule

\textbf{FAF-HQNN}
& \ensuremath{0.9825\pm0.0027}
& \ensuremath{0.9805\pm0.0032}
& \ensuremath{0.9713\pm0.0050}
& \ensuremath{0.9230\pm0.0144}
& \ensuremath{0.8667\pm0.0226} \\

\midrule

ResNet-18
& \ensuremath{0.9832\pm0.0058}
& \ensuremath{0.9818\pm0.0062}
& \ensuremath{\mathbf{0.9765\pm0.0077}}
& \ensuremath{\mathbf{0.9467\pm0.0144}}
& \ensuremath{0.8985\pm0.0302} \\

ResNet-50
& \ensuremath{\mathbf{0.9867\pm0.0021}}
& \ensuremath{\mathbf{0.9847\pm0.0020}}
& \ensuremath{0.9755\pm0.0037}
& \ensuremath{0.9414\pm0.0097}
& \ensuremath{\mathbf{0.8997\pm0.0129}} \\

EfficientNet-B0
& \ensuremath{0.9822\pm0.0039}
& \ensuremath{0.9793\pm0.0053}
& \ensuremath{0.9675\pm0.0107}
& \ensuremath{0.9171\pm0.0244}
& \ensuremath{0.7565\pm0.0570} \\

MobileNetV3-Small
& \ensuremath{0.9836\pm0.0026}
& \ensuremath{0.9778\pm0.0032}
& \ensuremath{0.9631\pm0.0046}
& \ensuremath{0.9195\pm0.0203}
& \ensuremath{0.8779\pm0.0362} \\

\bottomrule
\end{tabular}%
}
\end{table*}

\begin{table*}[!htbp]
\centering
\caption{AUC comparison of different model variants and external CNN baselines on PathMNIST under Poisson noise with varying corruption intensity. Results are reported as mean \ensuremath{\pm} sample standard deviation. 
Bold values indicate the highest available mean AUC in each column.}
\label{tab:path_poisson_auc}
\scriptsize
\setlength{\tabcolsep}{3.0pt}
\renewcommand{\arraystretch}{1}
\resizebox{\textwidth}{!}{%
\begin{tabular}{lccccc}
\toprule
\textbf{Method}
& \textbf{Clean}
& \textbf{Peak=3200}
& \textbf{Peak=1600}
& \textbf{Peak=800}
& \textbf{Peak=400} \\
\midrule

Classical
& 0.9754 \textpm{} 0.0029
& 0.9475 \textpm{} 0.0049
& 0.9128 \textpm{} 0.0103
& 0.8626 \textpm{} 0.0189
& 0.8236 \textpm{} 0.0242 \\

Q-Head
& \ensuremath{0.9490\pm0.0294}
& \ensuremath{0.9175\pm0.0329}
& \ensuremath{0.8643\pm0.0320}
& \ensuremath{0.7821\pm0.0397}
& \ensuremath{0.6994\pm0.0437} \\

C-AF
& \ensuremath{0.9796\pm0.0034}
& \ensuremath{0.9463\pm0.0178}
& \ensuremath{0.9008\pm0.0311}
& \ensuremath{0.8254\pm0.0493}
& \ensuremath{0.7402\pm0.0587} \\

H-Fix
& \ensuremath{0.9760\pm0.0058}
& \ensuremath{0.9478\pm0.0140}
& \ensuremath{0.9185\pm0.0251}
& \ensuremath{0.8767\pm0.0363}
& \ensuremath{0.8243\pm0.0460} \\

H-Scalar
& \ensuremath{0.9777\pm0.0048}
& \ensuremath{0.9433\pm0.0156}
& \ensuremath{0.8915\pm0.0460}
& \ensuremath{0.8043\pm0.0871}
& \ensuremath{0.7228\pm0.1074} \\

H-Sample
& \ensuremath{0.9738\pm0.0157}
& \ensuremath{0.9481\pm0.0224}
& \ensuremath{0.9079\pm0.0232}
& \ensuremath{0.8369\pm0.0406}
& \ensuremath{0.7659\pm0.0819} \\

H-Class
& \ensuremath{0.9792\pm0.0064}
& \ensuremath{0.9582\pm0.0171}
& \ensuremath{0.9236\pm0.0396}
& \ensuremath{0.8548\pm0.0780}
& \ensuremath{0.7788\pm0.1069} \\

EST-QCNN
& \ensuremath{0.9490\pm0.0114}
& \ensuremath{0.9126\pm0.0239}
& \ensuremath{0.8654\pm0.0368}
& \ensuremath{0.7859\pm0.0537}
& \ensuremath{0.7052\pm0.0714} \\

EST-RUQNN
& \ensuremath{0.9571\pm0.0132}
& \ensuremath{0.9230\pm0.0212}
& \ensuremath{0.8738\pm0.0322}
& \ensuremath{0.7934\pm0.0460}
& \ensuremath{0.7118\pm0.0558} \\

EST-QCQ-CNN
& \ensuremath{0.9590\pm0.0082}
& \ensuremath{0.9196\pm0.0142}
& \ensuremath{0.8692\pm0.0219}
& \ensuremath{0.8018\pm0.0319}
& \ensuremath{0.7440\pm0.0442} \\

\midrule

\textbf{FAF-HQNN}
& \ensuremath{0.9825\pm0.0027}
& \ensuremath{0.9546\pm0.0081}
& \ensuremath{0.9211\pm0.0150}
& \ensuremath{0.8749\pm0.0217}
& \ensuremath{0.8264\pm0.0296} \\

\midrule

ResNet-18
& \ensuremath{0.9832\pm0.0058}
& \ensuremath{\mathbf{0.9676\pm0.0090}}
& \ensuremath{\mathbf{0.9477\pm0.0124}}
& \ensuremath{\mathbf{0.9088\pm0.0232}}
& \ensuremath{0.8541\pm0.0413} \\

ResNet-50
& \ensuremath{\mathbf{0.9867\pm0.0021}}
& \ensuremath{0.9621\pm0.0061}
& \ensuremath{0.9386\pm0.0098}
& \ensuremath{0.9019\pm0.0129}
& \ensuremath{0.8504\pm0.0140} \\

EfficientNet-B0
& \ensuremath{0.9822\pm0.0039}
& \ensuremath{0.9518\pm0.0156}
& \ensuremath{0.9171\pm0.0251}
& \ensuremath{0.8133\pm0.0359}
& \ensuremath{0.6537\pm0.0542} \\

MobileNetV3-Small
& \ensuremath{0.9836\pm0.0026}
& \ensuremath{0.9502\pm0.0077}
& \ensuremath{0.9252\pm0.0165}
& \ensuremath{0.8907\pm0.0287}
& \ensuremath{0.8572\pm0.0418} \\

\bottomrule
\end{tabular}%
}
\end{table*}

\begin{table*}[!htbp]
\centering
\caption{AUC comparison of internal model variants and external
baselines on BloodMNIST under different Gaussian noise levels.
Results are reported as mean \textpm{} standard deviation over
five runs using checkpoints selected by validation accuracy.
Bold entries indicate the highest mean AUC in each column.}
\label{tab:blood_gaussian_auc}
\scriptsize
\setlength{\tabcolsep}{3pt}
\renewcommand{\arraystretch}{1}

\resizebox{\textwidth}{!}{%
\begin{tabular}{lccccc}
\toprule
\textbf{Method}
& \textbf{Clean}
& \textbf{Noise=0.005}
& \textbf{Noise=0.010}
& \textbf{Noise=0.020}
& \textbf{Noise=0.030} \\
\midrule

Classical
& 0.9949 \textpm{} 0.0007
& 0.9948 \textpm{} 0.0007
& 0.9947 \textpm{} 0.0007
& 0.9932 \textpm{} 0.0003
& 0.9893 \textpm{} 0.0015 \\

Q-Head
& 0.9864 \textpm{} 0.0022
& 0.9862 \textpm{} 0.0022
& 0.9855 \textpm{} 0.0019
& 0.9835 \textpm{} 0.0031
& 0.9758 \textpm{} 0.0089 \\

C-AF
& 0.9950 \textpm{} 0.0011
& 0.9949 \textpm{} 0.0013
& 0.9946 \textpm{} 0.0013
& 0.9926 \textpm{} 0.0010
& 0.9870 \textpm{} 0.0022 \\

H-Fix
& 0.9938 \textpm{} 0.0013
& 0.9937 \textpm{} 0.0013
& 0.9936 \textpm{} 0.0013
& 0.9927 \textpm{} 0.0006
& 0.9893 \textpm{} 0.0021 \\

H-Scalar
& 0.9934 \textpm{} 0.0010
& 0.9934 \textpm{} 0.0010
& 0.9935 \textpm{} 0.0010
& 0.9924 \textpm{} 0.0008
& 0.9895 \textpm{} 0.0011 \\

H-Sample
& 0.9947 \textpm{} 0.0008
& 0.9947 \textpm{} 0.0009
& 0.9946 \textpm{} 0.0008
& 0.9942 \textpm{} 0.0010
& 0.9906 \textpm{} 0.0017 \\

H-Class
& 0.9948 \textpm{} 0.0006
& 0.9948 \textpm{} 0.0006
& 0.9946 \textpm{} 0.0006
& 0.9934 \textpm{} 0.0007
& 0.9894 \textpm{} 0.0012 \\
EST-QCNN
& 0.9879 \textpm{} 0.0010
& 0.9875 \textpm{} 0.0014
& 0.9869 \textpm{} 0.0013
& 0.9831 \textpm{} 0.0022
& 0.9737 \textpm{} 0.0062 \\

EST-RUQNN
& 0.9869 \textpm{} 0.0012
& 0.9870 \textpm{} 0.0014
& 0.9868 \textpm{} 0.0022
& 0.9857 \textpm{} 0.0035
& 0.9820 \textpm{} 0.0063 \\

EST-QCQ-CNN
& 0.9857 \textpm{} 0.0027
& 0.9857 \textpm{} 0.0033
& 0.9852 \textpm{} 0.0037
& 0.9820 \textpm{} 0.0044
& 0.9733 \textpm{} 0.0052 \\

\midrule

\textbf{FAF-HQNN}
& 0.9956 \textpm{} 0.0004
& 0.9955 \textpm{} 0.0003
& 0.9953 \textpm{} 0.0003
& 0.9927 \textpm{} 0.0007
& 0.9863 \textpm{} 0.0021 \\

\midrule

ResNet-18
& 0.9974 \textpm{} 0.0002
& 0.9974 \textpm{} 0.0002
& \textbf{0.9973} \textpm{} 0.0002
& 0.9964 \textpm{} 0.0004
& 0.9929 \textpm{} 0.0011 \\

ResNet-50
& 0.9972 \textpm{} 0.0004
& 0.9972 \textpm{} 0.0004
& 0.9971 \textpm{} 0.0003
& \textbf{0.9967} \textpm{} 0.0004
& \textbf{0.9948} \textpm{} 0.0005 \\

EfficientNet-B0
& \textbf{0.9976} \textpm{} 0.0001
& \textbf{0.9976} \textpm{} 0.0001
& 0.9970 \textpm{} 0.0005
& 0.9854 \textpm{} 0.0033
& 0.8302 \textpm{} 0.1361 \\

MobileNetV3-Small
& 0.9964 \textpm{} 0.0005
& 0.9954 \textpm{} 0.0009
& 0.9912 \textpm{} 0.0025
& 0.9766 \textpm{} 0.0092
& 0.9617 \textpm{} 0.0106 \\

\bottomrule
\end{tabular}%
}
\end{table*}

\begin{table*}[!htbp]
\centering
\caption{AUC comparison of internal model variants and external baselines on BloodMNIST under different Poisson noise levels. Results are reported as mean \textpm{} standard deviation over five runs using checkpoints selected by validation accuracy. Bold entries indicate the highest mean AUC in each column, including ties at the reported precision.}
\label{tab:blood_poisson_auc}
\scriptsize
\setlength{\tabcolsep}{3pt}
\renewcommand{\arraystretch}{1}

\resizebox{\textwidth}{!}{%
\begin{tabular}{lccccc}
\toprule
\textbf{Method}
& \textbf{Clean}
& \textbf{Peak=3200}
& \textbf{Peak=1600}
& \textbf{Peak=800}
& \textbf{Peak=400} \\
\midrule

Classical
& 0.9949 \textpm{} 0.0007
& 0.9944 \textpm{} 0.0007
& 0.9935 \textpm{} 0.0005
& 0.9911 \textpm{} 0.0013
& 0.9841 \textpm{} 0.0051 \\

Q-Head
& 0.9864 \textpm{} 0.0022
& 0.9856 \textpm{} 0.0025
& 0.9840 \textpm{} 0.0027
& 0.9804 \textpm{} 0.0045
& 0.9679 \textpm{} 0.0137 \\

C-AF
& 0.9950 \textpm{} 0.0011
& 0.9941 \textpm{} 0.0012
& 0.9929 \textpm{} 0.0010
& 0.9899 \textpm{} 0.0011
& 0.9819 \textpm{} 0.0035 \\

H-Fix
& 0.9938 \textpm{} 0.0013
& 0.9936 \textpm{} 0.0009
& 0.9930 \textpm{} 0.0009
& 0.9912 \textpm{} 0.0009
& 0.9853 \textpm{} 0.0044 \\

H-Scalar
& 0.9934 \textpm{} 0.0010
& 0.9932 \textpm{} 0.0010
& 0.9929 \textpm{} 0.0006
& 0.9912 \textpm{} 0.0005
& 0.9854 \textpm{} 0.0018 \\

H-Sample
& 0.9947 \textpm{} 0.0008
& 0.9946 \textpm{} 0.0008
& 0.9942 \textpm{} 0.0011
& 0.9923 \textpm{} 0.0010
& 0.9861 \textpm{} 0.0016 \\

H-Class
& 0.9948 \textpm{} 0.0006
& 0.9944 \textpm{} 0.0007
& 0.9938 \textpm{} 0.0005
& 0.9917 \textpm{} 0.0005
& 0.9844 \textpm{} 0.0022 \\
EST-QCNN
& 0.9879 \textpm{} 0.0010
& 0.9862 \textpm{} 0.0007
& 0.9844 \textpm{} 0.0015
& 0.9794 \textpm{} 0.0033
& 0.9670 \textpm{} 0.0087 \\
EST-RUQNN
& 0.9869 \textpm{} 0.0012
& 0.9872 \textpm{} 0.0019
& 0.9856 \textpm{} 0.0032
& 0.9838 \textpm{} 0.0053
& 0.9764 \textpm{} 0.0091 \\

EST-QCQ-CNN
& 0.9857 \textpm{} 0.0027
& 0.9850 \textpm{} 0.0038
& 0.9830 \textpm{} 0.0038
& 0.9789 \textpm{} 0.0046
& 0.9655 \textpm{} 0.0073 \\

\midrule

\textbf{FAF-HQNN}
& 0.9956 \textpm{} 0.0004
& 0.9945 \textpm{} 0.0004
& 0.9923 \textpm{} 0.0007
& 0.9877 \textpm{} 0.0020
& 0.9775 \textpm{} 0.0037 \\

\midrule

ResNet-18
& 0.9974 \textpm{} 0.0002
& \textbf{0.9971} \textpm{} 0.0004
& 0.9965 \textpm{} 0.0005
& 0.9945 \textpm{} 0.0013
& 0.9873 \textpm{} 0.0032 \\

ResNet-50
& 0.9972 \textpm{} 0.0004
& \textbf{0.9971} \textpm{} 0.0003
& \textbf{0.9967} \textpm{} 0.0003
& \textbf{0.9957} \textpm{} 0.0003
& \textbf{0.9922} \textpm{} 0.0010 \\

EfficientNet-B0
& \textbf{0.9976} \textpm{} 0.0001
& 0.9954 \textpm{} 0.0009
& 0.9868 \textpm{} 0.0038
& 0.8730 \textpm{} 0.1583
& 0.7287 \textpm{} 0.1343 \\

MobileNetV3-Small
& 0.9964 \textpm{} 0.0005
& 0.9852 \textpm{} 0.0068
& 0.9760 \textpm{} 0.0115
& 0.9643 \textpm{} 0.0119
& 0.9459 \textpm{} 0.0167 \\

\bottomrule
\end{tabular}%
}
\end{table*}

\clearpage
\subsection{Seed-level ACC and AUC}
\label{app:path_auc}

\noindent
\begin{minipage}{\columnwidth}
\centering
\scriptsize

\setlength{\tabcolsep}{2.5pt}
\renewcommand{\arraystretch}{0.85}
\setlength{\aboverulesep}{0.3ex}
\setlength{\belowrulesep}{0.3ex}
\setlength{\abovecaptionskip}{2pt}
\setlength{\belowcaptionskip}{3pt}

\captionof{table}{Seed-level ACC/AUC on PathMNIST under clean conditions and Gaussian and Poisson noise.}
\label{tab:bloodmnist_seed_level}
\begin{tabular*}{\columnwidth}{@{\extracolsep{\fill}}ccccc@{}}
\toprule
\textbf{Level} & \textbf{Seed}
& \textbf{Classical} & \textbf{Q-Head} & \textbf{FAF-HQNN} \\
\midrule
\multicolumn{5}{c}{\textit{Clean}} \\
\midrule
-- & 0 & 0.8795 / 0.9796 & 0.8744 / 0.9525 & 0.9050 / 0.9843 \\
-- & 1 & 0.8841 / 0.9750 & 0.8903 / 0.9673 & 0.8928 / 0.9845 \\
-- & 2 & 0.8751 / 0.9717 & 0.8866 / 0.9723 & 0.8877 / 0.9788 \\
-- & 3 & 0.8847 / 0.9761 & 0.8497 / 0.8986 & 0.8884 / 0.9843 \\
-- & 4 & 0.8766 / 0.9743 & 0.8877 / 0.9541 & 0.8854 / 0.9804 \\
\midrule
\multicolumn{5}{c}{\textit{Gaussian noise (standard deviation)}} \\
\midrule
0.005 & 0 & 0.8712 / 0.9768 & 0.8652 / 0.9504 & 0.8990 / 0.9818 \\
      & 1 & 0.8731 / 0.9728 & 0.8833 / 0.9669 & 0.8897 / 0.9834 \\
      & 2 & 0.8681 / 0.9688 & 0.8809 / 0.9705 & 0.8805 / 0.9758 \\
      & 3 & 0.8791 / 0.9743 & 0.8416 / 0.8940 & 0.8794 / 0.9828 \\
      & 4 & 0.8607 / 0.9712 & 0.8823 / 0.9525 & 0.8759 / 0.9788 \\
\addlinespace[0.5pt]
0.010 & 0 & 0.8426 / 0.9665 & 0.8273 / 0.9392 & 0.8624 / 0.9716 \\
      & 1 & 0.8430 / 0.9645 & 0.8546 / 0.9576 & 0.8648 / 0.9773 \\
      & 2 & 0.8333 / 0.9590 & 0.8620 / 0.9626 & 0.8444 / 0.9644 \\
      & 3 & 0.8556 / 0.9676 & 0.8152 / 0.8858 & 0.8444 / 0.9746 \\
      & 4 & 0.8106 / 0.9600 & 0.8681 / 0.9461 & 0.8330 / 0.9686 \\
\addlinespace[0.5pt]
0.020 & 0 & 0.6819 / 0.9094 & 0.6340 / 0.8650 & 0.6948 / 0.9175 \\
      & 1 & 0.7007 / 0.9206 & 0.6309 / 0.8685 & 0.7077 / 0.9359 \\
      & 2 & 0.6887 / 0.9110 & 0.6779 / 0.8920 & 0.6632 / 0.9031 \\
      & 3 & 0.7467 / 0.9324 & 0.6671 / 0.8283 & 0.7217 / 0.9383 \\
      & 4 & 0.6422 / 0.9115 & 0.7596 / 0.9029 & 0.6706 / 0.9203 \\
\addlinespace[0.5pt]
0.030 & 0 & 0.4837 / 0.8322 & 0.4124 / 0.7576 & 0.5336 / 0.8499 \\
      & 1 & 0.5142 / 0.8557 & 0.3659 / 0.7420 & 0.5284 / 0.8717 \\
      & 2 & 0.5454 / 0.8505 & 0.4571 / 0.7904 & 0.4837 / 0.8372 \\
      & 3 & 0.6348 / 0.8968 & 0.4469 / 0.7344 & 0.5978 / 0.8860 \\
      & 4 & 0.4713 / 0.8555 & 0.6001 / 0.8317 & 0.5659 / 0.8888 \\
\midrule
\multicolumn{5}{c}{\textit{Poisson noise (peak)}} \\
\midrule
3200 & 0 & 0.7811 / 0.9486 & 0.7506 / 0.9155 & 0.8028 / 0.9540 \\
      & 1 & 0.7792 / 0.9475 & 0.7811 / 0.9349 & 0.8021 / 0.9636 \\
      & 2 & 0.7688 / 0.9420 & 0.7975 / 0.9428 & 0.7728 / 0.9426 \\
      & 3 & 0.8104 / 0.9548 & 0.7657 / 0.8614 & 0.7926 / 0.9605 \\
      & 4 & 0.7485 / 0.9443 & 0.8251 / 0.9328 & 0.7650 / 0.9523 \\
\addlinespace[0.5pt]
1600 & 0 & 0.6465 / 0.9059 & 0.6228 / 0.8620 & 0.6933 / 0.9173 \\
      & 1 & 0.6669 / 0.9109 & 0.6146 / 0.8633 & 0.6857 / 0.9325 \\
      & 2 & 0.6696 / 0.9062 & 0.6471 / 0.8869 & 0.6411 / 0.8978 \\
      & 3 & 0.7357 / 0.9306 & 0.6465 / 0.8136 & 0.7091 / 0.9358 \\
      & 4 & 0.6329 / 0.9102 & 0.7405 / 0.8960 & 0.6717 / 0.9223 \\
\addlinespace[0.5pt]
800 & 0 & 0.5021 / 0.8476 & 0.4272 / 0.7745 & 0.5609 / 0.8598 \\
      & 1 & 0.5220 / 0.8594 & 0.3882 / 0.7519 & 0.5355 / 0.8847 \\
      & 2 & 0.5443 / 0.8509 & 0.4786 / 0.8080 & 0.4960 / 0.8445 \\
      & 3 & 0.6329 / 0.8951 & 0.4681 / 0.7399 & 0.6032 / 0.8934 \\
      & 4 & 0.4781 / 0.8598 & 0.6056 / 0.8359 & 0.5844 / 0.8919 \\
\addlinespace[0.5pt]
400 & 0 & 0.4068 / 0.8074 & 0.2557 / 0.6760 & 0.4210 / 0.7970 \\
      & 1 & 0.3923 / 0.8162 & 0.2355 / 0.6708 & 0.3982 / 0.8204 \\
      & 2 & 0.4173 / 0.8049 & 0.3359 / 0.7228 & 0.3797 / 0.8067 \\
      & 3 & 0.4983 / 0.8646 & 0.3189 / 0.6621 & 0.4812 / 0.8354 \\
      & 4 & 0.3558 / 0.8249 & 0.4316 / 0.7653 & 0.5123 / 0.8725 \\

\bottomrule
\end{tabular*}

\par\vspace{2pt}
{\raggedright
Notes: Entries report ACC / AUC (0--1). Level follows the noise
protocol. Clean results are shared across noise types and listed once.
\par}

\end{minipage}
\par

\par\medskip
\noindent
\begin{minipage}{\columnwidth}
\centering
\scriptsize

\setlength{\tabcolsep}{2.5pt}
\renewcommand{\arraystretch}{0.85}
\setlength{\aboverulesep}{0.3ex}
\setlength{\belowrulesep}{0.3ex}
\setlength{\abovecaptionskip}{2pt}
\setlength{\belowcaptionskip}{3pt}

\captionsetup{hypcap=false}
\captionof{table}{Seed-level ACC/AUC on BloodMNIST under clean conditions and Gaussian and Poisson noise.}
\label{tab:bloodmnist_seed_level}

\begin{tabular*}{\columnwidth}{@{\extracolsep{\fill}}ccccc@{}}
\toprule
\textbf{Level} & \textbf{Seed}
& \textbf{Classical} & \textbf{Q-Head} & \textbf{FAF-HQNN} \\
\midrule
\multicolumn{5}{c}{\textit{Clean}} \\
\midrule
-- & 0 & 0.9559 / 0.9943 & 0.9497 / 0.9852 & 0.9640 / 0.9957 \\
-- & 1 & 0.9588 / 0.9940 & 0.9459 / 0.9854 & 0.9664 / 0.9952 \\
-- & 2 & 0.9541 / 0.9948 & 0.9562 / 0.9859 & 0.9646 / 0.9954 \\
-- & 3 & 0.9544 / 0.9952 & 0.9582 / 0.9903 & 0.9629 / 0.9955 \\
-- & 4 & 0.9611 / 0.9959 & 0.9483 / 0.9852 & 0.9690 / 0.9962 \\
\midrule
\multicolumn{5}{c}{\textit{Gaussian noise (standard deviation)}} \\
\midrule
0.005 & 0 & 0.9570 / 0.9942 & 0.9480 / 0.9847 & 0.9635 / 0.9954 \\
      & 1 & 0.9588 / 0.9941 & 0.9456 / 0.9856 & 0.9664 / 0.9953 \\
      & 2 & 0.9518 / 0.9947 & 0.9573 / 0.9855 & 0.9632 / 0.9955 \\
      & 3 & 0.9556 / 0.9950 & 0.9579 / 0.9900 & 0.9626 / 0.9954 \\
      & 4 & 0.9602 / 0.9959 & 0.9474 / 0.9850 & 0.9705 / 0.9961 \\
\addlinespace[0.5pt]
0.010 & 0 & 0.9535 / 0.9939 & 0.9471 / 0.9845 & 0.9640 / 0.9950 \\
      & 1 & 0.9605 / 0.9942 & 0.9462 / 0.9848 & 0.9643 / 0.9953 \\
      & 2 & 0.9515 / 0.9946 & 0.9515 / 0.9847 & 0.9629 / 0.9952 \\
      & 3 & 0.9518 / 0.9947 & 0.9579 / 0.9890 & 0.9617 / 0.9952 \\
      & 4 & 0.9564 / 0.9958 & 0.9494 / 0.9847 & 0.9690 / 0.9959 \\
\addlinespace[0.5pt]
0.020 & 0 & 0.9436 / 0.9927 & 0.9433 / 0.9831 & 0.9515 / 0.9923 \\
      & 1 & 0.9541 / 0.9934 & 0.9459 / 0.9826 & 0.9600 / 0.9937 \\
      & 2 & 0.9386 / 0.9931 & 0.9369 / 0.9805 & 0.9427 / 0.9920 \\
      & 3 & 0.9418 / 0.9936 & 0.9573 / 0.9887 & 0.9544 / 0.9932 \\
      & 4 & 0.9386 / 0.9934 & 0.9462 / 0.9826 & 0.9544 / 0.9923 \\
\addlinespace[0.5pt]
0.030 & 0 & 0.9284 / 0.9899 & 0.9363 / 0.9803 & 0.9228 / 0.9856 \\
      & 1 & 0.9298 / 0.9907 & 0.9374 / 0.9770 & 0.9412 / 0.9897 \\
      & 2 & 0.9097 / 0.9873 & 0.9015 / 0.9609 & 0.9167 / 0.9857 \\
      & 3 & 0.9170 / 0.9906 & 0.9450 / 0.9844 & 0.9331 / 0.9865 \\
      & 4 & 0.9123 / 0.9882 & 0.9354 / 0.9765 & 0.9319 / 0.9842 \\
\midrule
\multicolumn{5}{c}{\textit{Poisson noise (peak)}} \\
\midrule
3200 & 0 & 0.9512 / 0.9935 & 0.9462 / 0.9840 & 0.9591 / 0.9942 \\
      & 1 & 0.9547 / 0.9940 & 0.9477 / 0.9842 & 0.9629 / 0.9941 \\
      & 2 & 0.9486 / 0.9944 & 0.9509 / 0.9845 & 0.9556 / 0.9945 \\
      & 3 & 0.9506 / 0.9946 & 0.9591 / 0.9899 & 0.9617 / 0.9946 \\
      & 4 & 0.9556 / 0.9953 & 0.9450 / 0.9853 & 0.9640 / 0.9950 \\
\addlinespace[0.5pt]
1600 & 0 & 0.9430 / 0.9930 & 0.9433 / 0.9836 & 0.9506 / 0.9928 \\
      & 1 & 0.9541 / 0.9931 & 0.9453 / 0.9818 & 0.9579 / 0.9925 \\
      & 2 & 0.9389 / 0.9933 & 0.9436 / 0.9820 & 0.9369 / 0.9915 \\
      & 3 & 0.9430 / 0.9942 & 0.9564 / 0.9884 & 0.9544 / 0.9930 \\
      & 4 & 0.9448 / 0.9937 & 0.9471 / 0.9844 & 0.9538 / 0.9915 \\
\addlinespace[0.5pt]
800 & 0 & 0.9313 / 0.9911 & 0.9389 / 0.9824 & 0.9366 / 0.9887 \\
      & 1 & 0.9436 / 0.9922 & 0.9404 / 0.9787 & 0.9442 / 0.9899 \\
      & 2 & 0.9167 / 0.9892 & 0.9246 / 0.9743 & 0.9129 / 0.9853 \\
      & 3 & 0.9319 / 0.9925 & 0.9506 / 0.9864 & 0.9380 / 0.9887 \\
      & 4 & 0.9272 / 0.9904 & 0.9398 / 0.9802 & 0.9372 / 0.9860 \\
\addlinespace[0.5pt]
400 & 0 & 0.9114 / 0.9871 & 0.9310 / 0.9777 & 0.8980 / 0.9788 \\
      & 1 & 0.9094 / 0.9878 & 0.9199 / 0.9665 & 0.9217 / 0.9834 \\
      & 2 & 0.8647 / 0.9755 & 0.8550 / 0.9453 & 0.8737 / 0.9754 \\
      & 3 & 0.8968 / 0.9868 & 0.9269 / 0.9798 & 0.8980 / 0.9760 \\
      & 4 & 0.8910 / 0.9835 & 0.9252 / 0.9702 & 0.9030 / 0.9741 \\
\bottomrule
\end{tabular*}
\par\vspace{2pt}
{\raggedright
Notes: Entries report ACC / AUC (0--1). Level follows the noise
protocol. Clean results are shared across noise types and listed once.
Checkpoints are selected by validation accuracy.
\par}

\end{minipage}
\par

\end{appendices}

\clearpage
\bibliography{sample}

\end{document}